%% file: 00_main_ieee.tex
\documentclass[journal,final]{IEEEtran}

\usepackage[utf8x]{inputenc}
\usepackage{amsmath,amsfonts,amssymb,amsthm}
\usepackage{chngcntr}
\usepackage[pdftex,
			pdfauthor={C. Steffens, M. Pesavento, M.E. Pfetsch},
			pdftitle={A Compact Formulation for the l21 Mixed-Norm Minimization Problem},
            pdfkeywords={Multiple Measurement Vectors, Joint Sparsity, Mixed-Norm Minimization, Gridless Estimation}]{hyperref}
\usepackage{cite}

\usepackage{tikz}
\usetikzlibrary{calc,positioning,fit,shapes,arrows,spy,backgrounds}
\usepackage{pgfplots}
\pgfplotsset{width=10cm,compat=newest}
\usepackage{graphics}

\input{texMacros}
\input{tikzMacros}

\IEEEoverridecommandlockouts

\title{A Compact Formulation for the \\\texorpdfstring{$\ell_{2,1}$}{l21}~Mixed-Norm Minimization Problem}
\author{Christian Steffens, Marius Pesavento, and Marc E. Pfetsch
\thanks{
This work was supported by the EXPRESS project within the DFG priority program CoSIP (DFG-SPP 1798).}
\thanks{
Christian Steffens and Marius Pesavento are with the Communication Systems Group, TU Darmstadt, Merckstr. 25, 64283 Darmstadt, Germany (e-mail: \{steffens, pesavento\}@nt.tu-darmstadt.de).}
\thanks{
Marc E. Pfetsch is with the Discrete Optimization Group, TU Darmstadt, Dolivostr. 15, 64293 Darmstadt, Germany (e-mail: pfetsch@mathematik.tu-darmstadt.de).}
}

\begin{document}

\maketitle

\input{01_abstract}
\begin{IEEEkeywords}
Multiple Measurement Vectors, Joint Sparsity, Mixed-Norm Minimization, Gridless Estimation 
\end{IEEEkeywords}

\input{02_motivation}

\input{03_model}

\input{04_mixedNorm}

\input{05_magRec}

\input{06_algorithms}
\input{07_relatedWork}

\input{08_simulations}

\input{09_conclusion}
\input{10_appendix}

% \nocite{*}
\bibliographystyle{IEEEtran}
\bibliography{11_refferences}

\end{document}

%% file: texMacros.tex
\newcommand{\tj}{\text{\sf j}}
\newcommand{\tT}{\text{\sf T}}
\newcommand{\tH}{\text{\sf H}}
\newcommand{\tF}{\text{\sf F}}

\newcommand{\telse}{\text{ else }}
\newcommand{\tif}{\text{ if }}
\newcommand{\tst}{{\rm s.t.}}
\newcommand{\tsnr}{{\rm SNR}}
\newcommand{\tdB}{{\rm \,dB}}
\newcommand{\tE}{\ensuremath{\text{E}}}
\newcommand{\te}{\ensuremath{\text{e}}}

\newcommand{\diag}{{\rm diag}}

\newcommand{\tr}{{\rm Tr}}
\newcommand{\rank}{{\rm rank}}

\newcommand{\mb}[1]{\ensuremath{\boldsymbol{#1}}}
\newcommand{\bmtx}{\ensuremath{\begin{bmatrix}}}
\newcommand{\emtx}{\ensuremath{\end{bmatrix}}}
\newcommand{\mtx}[1]{\ensuremath{\begin{bmatrix} #1 \end{bmatrix}}}
\newcommand{\toep}{{\rm Toep}}
\newcommand{\opt}[1]{\ensuremath{\smash{\hat{#1}}}}

\newcommand{\pdMat}{\mathbb{D}_{+}}
\newcommand{\dMat}{\mathbb{D}}

\def\myName{SPARROW}
\def\myNameB{Sparrow}
\newcommand{\mybar}{}

\newtheorem{theorem}{Theorem}

\newtheorem{corollary}{Corollary}

%% file: tikzMacros.tex
\pgfplotsset{every axis legend/.append style={row sep=-3pt, inner ysep=0pt, inner xsep=2pt}}

\pgfplotsset{colorbar/draw/.code={
      \begin{axis}[
        colormap/jet,
        colorbar horizontal,
        every colorbar,
		at={(0.5,1.1)},
		anchor=south,
		height=5pt,
		xticklabel pos=upper,
        colorbar shift,
        colorbar=false,
      ]
        \pgfkeysvalueof{/pgfplots/colorbar addplot}
      \end{axis}
    }]
}

\tikzset{every picture/.append style={baseline=(current bounding box.north west)}}

\colorlet{cs0}{black!25!orange}
\colorlet{cs1}{black!10!red}
\colorlet{cs2}{white!15!blue}
\colorlet{cs3}{black!25!green}

\def\NSen{6}
\def\NSnp{4}
\def\NGrd{12}
\def\dA{3}
\def\dB{5}
\def\dC{10}
\def\seedA{107}
\def\seedB{198}
\def\cRange{60}
\def\bSizeMod{1.7mm}

\def\seedA{20}
\def\seedB{500}

%% file: 01_abstract.tex
% 150-250 words

\begin{abstract}
Parameter estimation from multiple measurement vectors (MMVs) is a fundamental problem in many signal processing applications, e.g., spectral analysis and direction-of-arrival estimation. Recently, this problem has been address using prior information in form of a jointly sparse signal structure. A prominent approach for exploiting joint sparsity considers mixed-norm minimization in which, however, the problem size grows with the number of measurements and the desired resolution, respectively. In this work we derive an equivalent, compact reformulation of the \texorpdfstring{$\ell_{2,1}$}{l21} mixed-norm minimization problem which provides new insights on the relation between different existing approaches for jointly sparse signal reconstruction. The reformulation builds upon a compact parameterization, which models the row-norms of the sparse signal representation as parameters of interest, resulting in a significant reduction of the MMV problem size.
Given the sparse vector of row-norms, the jointly sparse signal can be computed from the MMVs in closed form. For the special case of uniform linear sampling, we present an extension of the compact formulation for gridless parameter estimation by means of semidefinite programming. Furthermore, we derive in this case from our compact problem formulation the exact equivalence between the $\ell_{2,1}$ mixed-norm minimization and the atomic-norm minimization. Additionally, for the case of irregular sampling or a large number of samples, we present a low complexity, grid-based implementation based on the coordinate descent method.
\end{abstract}

%% file: 02_motivation.tex
\section{Introduction}
% Storyline
% \begin{itemize}
%   \item Many snapshots are good for estimation
%   \item Subspace methods are asymptotically good
%   \item Sparse Regression better for extreme case
%   \item extensive research and advances on single measurement vector problem (LARS, continuous)
%   \item Joint sparsity problems suffer from computational complexity, research focus on reduction of complexity
%   \item Different approaches like covariance matching or l1-svd simplify signal or model
%   \item we have the solution to ALL problems
% \end{itemize}

% finding sparse solutions to vastly underdetermined systems of linear equations
% sparse inverse problems

Sparse Signal Reconstruction (SSR) techniques have gained a considerable research interest over the last decades \cite{Tibshirani:Lasso, Chen98atomicdecomposition, 1614066, 1580791, 1542412, candes:2688127, candes:12380704, Donoho02optimallysparse}. Traditionally, SSR considers the problem of reconstructing a high-dimensional sparse signal vector from a low-dimensional Single Measurement Vector (SMV), which is characterized by an underdetermined system of linear equations. It has been shown that exploiting prior knowledge on the sparsity structure of the signal admits a unique solution to the underdetermined system. In the signal processing context, this implies that far fewer samples than postulated by the Shannon-Nyquist sampling theorem for bandlimited signals are required for perfect signal reconstruction \cite{tropp2010}, whereas, in the parameter estimation context,  this indicates that SSR methods exhibit the superresolution property \cite{donoho1992}.

% Applications of spare signal reconstruction arise in various fields such as spectral analysis, direction-of-arrival (DOA) estimation, image processing, geophysics, tomography and magnetic resonance imaging, or machine learning.

While SSR under the classical $\ell_0$ formulation constitutes a combinatorial and NP-complete optimization problem, several heuristics exist to approximately solve the SSR problem. Most prominent heuristics are based on convex relaxation in terms of $\ell_1$ norm minimization, which makes the SSR problem computationally tractable while providing sufficient conditions for exact recovery \cite{Tibshirani:Lasso, Chen98atomicdecomposition, 1614066, 1580791, 1542412,candes:2688127, candes:12380704, Donoho02optimallysparse}, or greedy methods, such as OMP \cite{258082,4385788} and CoSaMP \cite{Needell2009301}, which have low computational complexity but provide reduced recovery guarantees. In the context of parameter estimation, e.g., in Direction-Of-Arrival (DOA) estimation, the  SSR problem has been extended to an infinite-dimensional vector space by means of total variation norm and atomic norm minimization \cite{candes2012b, candes2012, chandrasekaran2012, gongguo2013, bhaskar2011, 6552292}, leading to
gridless parameter estimation methods.

Besides the aforementioned SMV problem, many practical applications deal with the problem of finding a jointly sparse signal representation from Multiple Measurement Vectors (MMVs), also referred to as the multiple snapshot estimation problem. Similar to the SMV case, heuristics for the MMV-based SSR problem include convex relaxation by means of mixed-norm minimization \cite{10.2307/3647556, Tropp2006589,turlach2005simultaneous, Kowalski2009303}, and greedy methods \cite{tropp2006algorithms, 1453780}. Recovery guarantees for the MMV case have been established in \cite{6408167, 4014378, Lai2011402}, and it has been shown that rank awareness in MMV-based SSR can further enhance the recovery performance as compared to the SMV case \cite{6145474}. An extension to the infinite-dimensional vector space for MMV-based SSR, using atomic norm minimization, has been proposed in \cite{7313018, yang2014a, yang2014b}.

Apart from SSR, MMV-based parameter estimation is a classical problem in array signal processing \cite{Krim:TwoDecades, vanTrees2002}. Prominent applications in array processing include beamforming and DOA estimation. Beamforming considers the problem of signal reconstruction in the presence of noise and interference while DOA estimation falls within the concept of parameter estimation and is addressed, e.g., by the subspace-based MUSIC method \cite{1143830}. The MUSIC method has been shown to perform asymptotically optimal \cite{17564} and offers the super-resolution property at tractable computational complexity. On the other hand, in the non-asymptotic case of low number of MMVs or correlated source signals, the performance of subspace-based estimation methods can drastically deteriorate such that SSR techniques provide an attractive alternative for these scenarios \cite{Malioutov:LassoDoa, 5466152, kim2010compressive}. In fact, due
to similar objectives in SSR and array signal processing, strong links between the two fields of research have been established in literature. The OMP has an array processing equivalent in the CLEAN method \cite{clean} for source localization in radio astronomy, i.e., both methods rely on the same greedy estimation approach. In \cite{558475, 1453780} the authors present the FOCUSS method, which provides sparse estimates by iterative weighted norm minimization, with application to DOA estimation. SSR based on an $\ell_{2,0}$ mixed-norm approximation has been considered in \cite{5466152}, while a convex relaxation approach based on the $\ell_{2,1}$ mixed-norm has been proposed in \cite{Malioutov:LassoDoa}. DOA estimation based on second-order signal statistics has been addressed in \cite{1687-6180-2012-111, 6494328}, where a sparse covariance matrix representation is exploited by application of a sparsity prior on the source covariance matrix, leading to an SMV-like sparse minimization
problem. In \cite{5599897, 5617289, 2014arXiv1406.7698S} the authors propose the SPICE method, which is based on weighted covariance matching and constitutes a sparse estimation problem which does not require the assumption of a sparsity prior. Links between SPICE and SSR formulations have been established in \cite{5617289,6553252, babu2014connection, 2014arXiv1406.7698S, yang2014b}, which show that SPICE can be reformulated as an $\ell_{2,1}$ mixed-norm minimization problem.

In this paper we consider jointly sparse signal reconstruction from MMVs by means of the classical $\ell_{2,1}$ mixed-norm minimization problem, with application to DOA estimation in array signal processing. Compared to recently presented sparse methods such as SPICE \cite{5599897, 5617289, 2014arXiv1406.7698S} and atomic norm minimization \cite{7313018, yang2014a, yang2014b}, the classical $\ell_{2,1}$ formulation has the general shortcoming that its problem size grows with the number of measurements and
the resolution requirement, respectively. Heuristic approaches to deal with the aforementioned problems have been presented, e.g., in \cite{Malioutov:LassoDoa}.
% Furthermore, there are gaps in understanding the relation between the different approaches and formulations.

While the classical $\ell_{2,1}$ mixed-norm minimization problem has a large number of variables in the jointly sparse signal representation, in this paper we derive an equivalent problem reformulation based on a compact parameterization in which the optimization parameters represent the row-norms of the signal representation, rather then the signal matrix itself. We refer to this formulation as the SPARse ROW-norm reconstruction ({\myName}). Given the sparse signal row-norms, the jointly sparse signal matrix is reconstructed from the MMVs in closed-form. We point out that support recovery is determined by the sparse vector of row-norms and only relies on the sample covariance matrix instead of the MMVs themselves. In this sense we achieve a concentration of the optimization variables as well as the measurements, leading to a significantly reduced problem size in the case of a large number of MMVs. Regarding the implementation of the {\myName} problem, we present a gridless estimation approach based on 
semidefinite
programming as well as a grid-based, low complexity implementation in form of a coordinate descent method. Due to the large variety of competing approaches for SSR in the MMV context, it is of fundamental interest to explore similarities and differences between different techniques and to develop new links among different approaches. We compare our new problem formulation to existing alternative approaches for the MMV problem, viz. atomic norm minimization and SPICE, and establish new links and equivalences in terms of problem formulation as well as implementation. Specifically, we prove from our gridless, compact reformulation the exact equivalence between the classical $\ell_{2,1}$ mixed-norm minimization problem \cite{10.2307/3647556, Malioutov:LassoDoa} and the recently proposed atomic norm minimization formulation for MMV scenarios \cite{7313018, yang2014a, yang2014b}. We conclude our presentation by a short numerical analysis of the parameter estimation performance and the computation time of our proposed {\myName} formulation which shows a significant reduction in the computational complexity of our proposed 
reformulation as compared to both equivalent formulations, the classical $\ell_{2,1}$ mixed-norm \cite{10.2307/3647556, Malioutov:LassoDoa} and the atomic norm \cite{7313018, yang2014a, yang2014b} problem formulations.

In summary, our main contributions are the following:
\begin{itemize}
  \item We derive an equivalent, compact reformulation of the classical $\ell_{2,1}$ mixed-norm minimization problem \cite{10.2307/3647556, Malioutov:LassoDoa}, named {\myName}, with significantly reduced computational complexity.
  \item We provide a gridless and a low complexity implementation of the {\myName} formulation.
  \item We proof the equivalence of the gridless {\myName} formulation and the atomic norm minimization problem \cite{7313018, yang2014a, yang2014b}.
  \item We show theoretical links between the {\myName} formulation and the SPICE method \cite{5599897, 5617289, 2014arXiv1406.7698S}.
\end{itemize}

The paper is organized as followed: In Section \ref{sec:model} we present the sensor array signal model. A short review of the classical $\ell_{2,1}$ mixed-norm minimization problem is provided in Section \ref{sec:StateOfTheArtl21} before the equivalent, compact {\myName} formulation is introduced in Section \ref{sec:magrec} and for which an efficient implementation is discussed in Section \ref{sec:algorithms}. Section \ref{sec:relatedWork}  provides a theoretical comparison of the {\myName} formulation and related methods for jointly sparse recovery. Simulation results regarding estimation performance and computational complexity of the various formulations are presented in Section \ref{sec:simulations}. Conclusions are provided in Section \ref{sec:conclusion}.

\vspace{0.1em}
\textbf{Notation:} Boldface uppercase letters $\mb{X}$ denote matrices, boldface lowercase letters $\mb{x}$ denote column vectors, and regular letters $x,N$ denote scalars, with $\tj$ denoting the imaginary unit. Superscripts $\mb{X}^\tT$ and $\mb{X}^\tH$ denote transpose and conjugate transpose of a matrix $\mb{X}$, respectively. The sets of diagonal and nonnegative diagonal matrices are denoted as $\dMat$ and $\pdMat$, respectively. We write $[\mb{X}]_{m,n}$ to indicate the element in the $m$th row and $n$th column of matrix $\mb{X}$. The statistical expectation of a random variable $x$ is denoted as $\tE\{x\}$, and the trace of a matrix $\mb{X}$ is referred to as $\tr(\mb{X})$. The Frobenius norm and the $\ell_{p,q}$ mixed-norm of a matrix $\mb{X}$ are referred to as $\|\mb{X}\|_{\tF}$ and $\|\mb{X}\|_{p,q}$, respectively, while the $\ell_p$ norm of a vector $\mb{x}$ is denoted as $\|\mb{x}\|_p$. $\toep(\mb{u})$ describes a Hermitian Toeplitz matrix with $\mb{u}$ as its first column and $\diag(\mb{x})$
denotes a diagonal matrix with the elements in $\mb{x}$ on its main diagonal.
%For an integer scalar $N$ we define $\intSet{N}=1,\ldots,N$ as the set of integers from 1 to $N$.

%% file: 03_model.tex
\section{Signal Model} \label{sec:model}

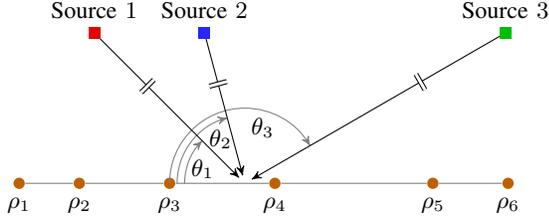
\begin{figure}[t]
\begin{center}
  \small    
  \input{p01_modelImg.tex}
\end{center}
\vspace{-0.08cm}
\caption{Exemplary setup for a linear array of $M=6$ sensors and $L=3$ source signals}
\label{fig:SigModel}
\vspace{0.1cm}
\end{figure}

Consider a linear array of $M$ omnidirectional sensors, as depicted in Figure \ref{fig:SigModel}. Further, assume a set of $L$ narrowband far-field sources in angular directions $\theta_1, \ldots, \theta_L$, summarized as $\mb{\theta} = [ \theta_1, \ldots, \theta_L ]^\tT$. The spatial frequencies are defined as
\begin{align}
  \mu_l = \cos \theta_l \in [-1,1), 
\end{align}
for $l = 1,\ldots,L$, comprised in the vector $\mb{\mu} = [\mu_1, \ldots, \mu_L]^\tT$. The array output provides measurement vectors, also referred to as snapshots, which are recorded over $N$ time instants where we assume that the sources transmit time-varying signals while the frequencies in $\mb{\mu}$ remain constant within the entire observation time. The measurement vectors are collected in the multiple measurement vector (MMV) matrix $\mb{Y} \in \mathbb{C}^{M \times N}$, where $[\mb{Y}]_{m,n}$ denotes the output at sensor $m$ at time instant $n$. The MMV matrix is modeled as 
\begin{equation}  
  \mb{Y} = \mb{A}( \mb{\mu} ) \mb{\varPsi} + \mb{N},
  \label{eq:sigModelMmv}
\end{equation}
where $\mb{\varPsi} \in \mathbb{C}^{L \times N} $ is the source signal matrix, with $[\mb{\varPsi}]_{l,n}$ denoting the signal transmitted by source $l$ in time instant $n$, and $\mb{N} \in \mathbb{C}^{M \times N}$ represents circular and spatio-temporal white Gaussian sensor noise with covariance matrix $\tE \{\mb{N} \mb{N}^\tH \}/N = \sigma^2 \mb{I}_M$, where $\mb{I}_M$ and $\sigma^2$ denote the $M \times M$ identity matrix and the noise power, respectively. The $M \times L$ array steering matrix $\mb{A}(\mb{\mu})$ in \eqref{eq:sigModelMmv} is given by
\begin{align}
 \mb{A}(\mb{\mu}) = [ \mb{a} ( \mu_1 ), \ldots, \mb{a} ( \mu_L ) ], 
 \label{eq:sigModelA}
\end{align}
where 
\vspace{-0.1cm}
\begin{equation}
  \mb{a} (\mu) = [1, \te^{-\tj \pi \mu \rho_2 }, \ldots, \te^{-\tj \pi \mu \rho_M } ]^\tT
  \label{eq:bk}
\end{equation}
is the array manifold vector with $\rho_{m} \in \mathbb{R}$, for $m=1, \ldots, M$, denoting the position of the $m$th sensor in half signal wavelength, relative to the first sensor in the array, hence $\rho_{1} = 0$. 

% We assume that the sensor noise and source signals are uncorrelated such that the receive covariance matrix is given by
% \begin{align}
%   \mb{R} =& \E \{ \mb{Y} \mb{Y}^\tH \} / N \nonumber \\
% 		 =& \mb{A}(\mb{\mu}) \mb{\Phi} \mb{A}^\tH(\mb{\mu}) + \sigma^2 \mb{I}
%   \label{eq:covar}
% \end{align}
% where 
% \begin{align}
%   \mb{\Phi} = \E \{ \mb{\varPsi} \mb{\varPsi}^\tH \} / N
%   \label{eq:SrcCovar}
% \end{align}
% denotes the source covariance matrix. In the case of uncorrelated source signals the source covariance matrix $\mb{\Phi}$ is a nonnegative diagonal matrix, i.e., $\mb{\Phi} \in \pdMat$.
% 
% Given the MMV-matrix $\mb{Y}$, the sample covariance matrix is computed as
% \begin{align}
%   \hat{\mb{R}} % &= \frac{1}{N} \sum_{n=1}^N \mb{y}(n) \mb{y}^\tH(n) \nonumber \\
% 			   = \mb{Y} \mb{Y}^\tH / N.
%   \label{eq:smpCovar}
% \end{align}
% The MMV model \eqref{eq:sigModelMmv} forms the basis for various DoA estimation methods such as subspace-based methods or sparse signal reconstruction methods. 

%% file: p01_modelImg.tex
\tikzstyle{sensor}=[circle, fill=cs0, inner sep=0pt, minimum height=1.5mm]
\tikzstyle{source}=[fill=red, inner sep=0pt, minimum height=1.5mm, minimum width=1.5mm]

\begin{tikzpicture}
          
     \tikzset{>=stealth'}
%     \tikzset{arrow style mul/.style={draw,sloped,midway,fill=white}}  
    
    % coordinates of array 1    
    % draw nodes
%     \foreach \x in {1,...,5} {
% 	  \node[sensor, fill=cs0, label=below:$\rho_\x$] (s\x) at (\x-3,0) {};
%     }

	\node[sensor, fill=cs0, label=below:$\rho_1$] (s1) at (-3.0,0) {};
    \node[sensor, fill=cs0, label=below:$\rho_2$] (s2) at (-2.2,0) {};
    \node[sensor, fill=cs0, label=below:$\rho_3$] (s3) at (-1.0,0) {};
    \node[sensor, fill=cs0, label=below:$\rho_4$] (s4) at ( 0.4,0) {};
    \node[sensor, fill=cs0, label=below:$\rho_5$] (s5) at ( 2.5,0) {};
    \node[sensor, fill=cs0, label=below:$\rho_6$] (s6) at ( 3.5,0) {};

    % draw connections
    \draw[gray] (s1) -- (s2) -- (s3) -- (s4) -- (s5) -- (s6); % A <-> C
       
    %%% draw coordinate system
    \node[circle, inner sep=2pt] (cs) at (0,0) {};
        
    %%% draw sources
    \def\angA{180-\dA/\NGrd*180}
    \def\angB{180-\dB/\NGrd*180}
    \def\angC{180-\dC/\NGrd*180}
    
    \coordinate (src1) at (intersection cs: first line={(cs)--($(cs)+(\angA:6)$)}, second line={(0,2)--(3,2)}) {};
    \coordinate (src2) at (intersection cs: first line={(cs)--($(cs)+(\angB:6)$)}, second line={(0,2)--(3,2)}) {};
    \coordinate (src3) at (intersection cs: first line={(cs)--($(cs)+(\angC:6)$)}, second line={(0,2)--(3,2)}) {};
    
%     \node at ($(src1)-(2.5,0)$) {$\cdots$};
%     \node at ($(src2)+(2.5,0)$) {$\cdots$};
	
    \draw[gray,<-] ($(cs)+(\angA:0.8)$) arc (\angA:180:0.8cm);
    \node[anchor=south east] at ($(cs)+(-0.3,-0.05)$) {\textcolor{black}{$\theta_1$}};
    \draw[gray,->] ($(cs)+(180:0.9)$) arc (180:\angB:0.9cm);
    \node[anchor=south east] at ($(cs)+(-0.06,0.35)$){\textcolor{black}{$\theta_2$}};
    \draw[gray,->] ($(cs)+(180:1)$) arc (180:\angC:1cm);
    \node[anchor=south east] at ($(cs)+(0.5,0.5)$){\textcolor{black}{$\theta_3$}};
    
    \draw[->] (src1) -- (cs); % A <-> C
    \draw[->] (src2) -- (cs); % A <-> C
    \draw[->] (src3) -- (cs); % A <-> C
   
	\def\wcut{1.2mm}
	\def\hcut{.4mm}
    \coordinate (cut1) at ($(src1)!1cm!(cs)$);
    \draw[double distance=\hcut] ($ (cut1)!\wcut!90:(cs) $) -- ($ (cut1)!\wcut!-90:(cs) $);
    \draw[white] ($ (cut1)!\wcut!90:(cs) $) -- ($ (cut1)!\wcut!-90:(cs) $);
    \coordinate (cut2) at ($(src2)!0.7cm!(cs)$);
    \draw[double distance=\hcut] ($ (cut2)!\wcut!90:(cs) $) -- ($ (cut2)!\wcut!-90:(cs) $);
    \draw[white] ($ (cut2)!\wcut!90:(cs) $) -- ($ (cut2)!\wcut!-90:(cs) $);
    \coordinate (cut3) at ($(src3)!1.3cm!(cs)$);
    \draw[double distance=\hcut] ($ (cut3)!\wcut!90:(cs) $) -- ($ (cut3)!\wcut!-90:(cs) $);
    \draw[white] ($ (cut3)!\wcut!90:(cs) $) -- ($ (cut3)!\wcut!-90:(cs) $);
   
	\node[source, fill=cs1, label={above:Source $1$}] at (src1) {};
    \node[source, fill=cs2, label={above:Source $2$}] at (src2) {};
    \node[source, fill=cs3, label={above:Source $3$}] at (src3) {};
       
\end{tikzpicture}

%% file: 04_mixedNorm.tex
\section{Sparse Representation and Mixed-Norm~Minimization} \label{sec:StateOfTheArtl21}

For the application of SSR to DOA estimation we define a sparse representation of the MMV model in \eqref{eq:sigModelMmv} as
\begin{align}
  \mb{Y} = \mb{A}(\mb{\nu}) \mb{X} + \mb{N} ,
  \label{eq:sparseSigModelMmv}
\end{align}
with $\mb{X}$ denoting a $K \times N$ row-sparse signal matrix, and the $M \times K$ overcomplete dictionary matrix $\mb{A}(\mb{\nu})$  is defined in correspondence to \eqref{eq:sigModelA}, where the vector $\mb{\nu} = [ \nu_1, \ldots,  \nu_K]^\tT$ is obtained by sampling the spatial frequencies in $K \gg L$ points $\nu_1, \ldots,  \nu_K$. For ease of notation we will drop the argument in the remainder of the paper an refer to the dictionary matrix as $\mb{A}=\mb{A}(\mb{\nu})$. We assume that the frequency grid is sufficiently fine, such that the true frequencies in $\mb{\mu}$ are contained in the frequency grid $\mb{\nu}$, i.e., 
\begin{align}
  \{\mu_l\}_{l=1}^L \subset \{\nu_k\}_{k=1}^K .
  \label{eq:onGrid}
\end{align}
Since the true frequencies in $\mb{\mu}$ are not known in advance and the grid-size is limited in practice, the on-grid assumption \eqref{eq:onGrid} is usually not fulfilled, leading to spectral leakage effects and basis mismatch \cite{yuejie2011, herman2010}. In section \ref{sec:SDP} we present an extension of our proposed formulation which does not rely on the on-grid assumption. However, elsewhere we assume \eqref{eq:onGrid} to hold true for ease of presentation. 

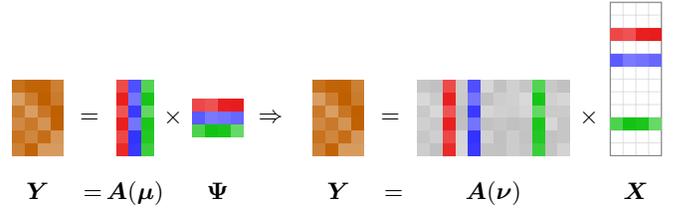
\begin{figure}[t]
\begin{center}
  \small    
  \input{p02_sparseRepMmv.tex}
\end{center}
% \vspace{-0.7cm}
\caption{Signal model and sparse representation (neglecting additive noise  and basis mismatch) for $M=6$ sensors, $L=3$ source signals and $K=12$ grid points }
\label{fig:sparseRepMmv}
\vspace{-0.3cm}
\end{figure}

The $K \times N$ sparse signal matrix $\mb{X}$ in \eqref{eq:sparseSigModelMmv} contains elements
\begin{align}
  [\mb{X}]_{k,n} =&
  \begin{cases}
	[\mb{\varPsi}]_{l,n} \quad &\text{if } \nu_k = \mu_l \\
	0		  \quad &\text{else,}
  \end{cases}
  \label{eq:rowSparseStruct}
\end{align}
for $k=1,\ldots,K$, $l=1,\ldots,L$. Thus $\mb{X}$ exhibits a row-sparse structure, i.e., the elements in a row of $\mb{X}$ are either jointly zero or primarily non-zero, as illustrated in Figure \ref{fig:sparseRepMmv}. To exploit the joint sparsity assumption in the estimation problem, it was proposed, e.g., in \cite{10.2307/3647556, Tropp2006589, turlach2005simultaneous, Kowalski2009303, Malioutov:LassoDoa, 5466152}, to utilize a mixed-norm formulation leading to the classical $\ell_{p,q}$ mixed-norm minimization problem
\begin{align}  
  \min_{\mb{X}} 
  \frac{1}{2} \left\| \mb{A} \mb{X} - \mb{Y} \right\|_\tF^2 + 
%   \lambda \sum_{i=1}^{N} \left( \sum_{j=1}^\tT \left| s_{ij} \right|^p \right)^{1/p}.
%   \lambda \sum_{n=1}^{N} \left\| \mybar{\mb{x}}_n \right\|_p .
  \lambda \| \mb{X} \|_{p,q}.
  \label{eq:mixedVectorNorm}
\end{align}
In \eqref{eq:mixedVectorNorm}, the data fitting $\| \mb{A} \mb{X} - \mb{Y} \|_\tF^2$ is performed by means of the Frobenius norm to ideally match the reconstructed measurements $\mb{A} \mb{X}$ in the presence of additive white Gaussian noise. The regularization parameter $\lambda > 0$ admits balancing the data fitting fidelity versus the sparsity level in $\mb{X}$, where the choice of a small $\lambda$ in \eqref{eq:mixedVectorNorm} tends to result in a large number of non-zero rows, whereas a large vlaue of $\lambda$ tends to result in a small number of non-zero rows. Joint sparsity in $\mb{X}$ is induced by the $\ell_{p,q}$ mixed-norm, which is defined as 
\begin{align}
  \| \mb{X} \|_{p,q} = \left( \sum_{k=1}^{K} \left\| \mybar{\mb{x}}_k \right\|_p^q \right)^{1/q} ,
%   \| \mb{X} \|_{p,q} = \left( \sum_{k=1}^{K} \left( \sum_{n=1}^{N} \big| [\mb{X}]_{k,n} \big|^p \right)^{q/p} \right)^{1/q} ,
  \label{eq:lpqNorm}
\end{align}
applying an \textit{inner} $\ell_p$ norm on the rows $\mb{x}_k$, for $k=1,\ldots,K$, in $\mb{X}=[\mybar{\mb{x}}_1, \ldots, \mybar{\mb{x}}_K]^\tT$ and an \textit{outer} $\ell_q$ norm on the $\ell_p$ row-norms. The \textit{inner} $\ell_p$ norm provides a nonlinear coupling among the elements in a row, leading to the desired row-sparse structure of the signal matrix $\mb{X}$. Ideally, considering the original signal model in \eqref{eq:mixedVectorNorm}, we desire a problem formulation containing an $\ell_{p,0}$ pseudo-norm, leading, however, to an NP-complete problem, such that convex relaxation in form of $\ell_{p,1}$ mixed-norm is considered in practice to obtain computationally tractable problems. In the SMV case, i.e., $N=1$, the $\ell_{p,1}$ mixed-norm reduces to the $\ell_1$ norm, such that $\ell_{p,1}$ mixed-norm minimization can be considered as a generalization of the classical $\ell_1$ norm minimization problem \cite{Tibshirani:Lasso, Chen98atomicdecomposition} to the MMV case with $N>1$. Common 
choices of mixed-norms are the $\ell_{2,1}$ norm \cite{10.2307/3647556, Malioutov:LassoDoa} and 
the $\ell_{\infty,1}$ norm \cite{Tropp2006589, turlach2005simultaneous}. Similar to the SMV case, 
recovery guarantees for the MMV-based joint SSR problem have been derived \cite{6408167, 4014378, Lai2011402}, providing conditions for the noiseless case under which the sparse signal matrix $\mb{X}$ can be perfectly reconstructed. Moreover, it has been shown that rank-awareness in the signal reconstruction can additionally improve the reconstruction performance \cite{6145474}.

Given a row-sparse minimizer $\opt{\mb{X}}$ for \eqref{eq:mixedVectorNorm}, the DOA estimation problem reduces to identifying the union support set, i.e., the indices of the non-zero rows, from which the set of estimated spatial frequencies can be obtained as
\begin{align}
  \{ \hat{\mu}_l \}_{l=1}^{\hat{L}} = \left\{ \nu_k \left| \; \| \opt{\mybar{\mb{x}}}_k\|_p > 0, \, k = 1, \ldots, K \right. \right\}
  \label{eq:sparseDoA}
\end{align}
where $\opt{\mybar{\mb{x}}}_k$ corresponds to the $n$th row of the signal matrix $\opt{\mb{X}} = [\opt{\mybar{\mb{x}}}_1, \ldots, \opt{\mybar{\mb{x}}}_K]^\tT$ and $\hat{L}$ denotes the number of non-zero rows in $\opt{\mb{X}}$, i.e., the estimated model order. 

One major drawback of the mixed-norm minimization problem in \eqref{eq:mixedVectorNorm} lies in its computational complexity, which is determined by the size of the $K \times N$ source signal matrix $\mb{X}$. A large number of grid points $K$ is desired to improve the frequency resolution, while a large number of measurement vectors $N$ is desired to improve the estimation performance. However, the choice of too large values $K$ and $N$ makes the problem computationally intractable. To reduce the computational complexity in the MMV problem it was suggested in \cite{Malioutov:LassoDoa} to reduce the dimension of the measurement matrix by matching only the signal subspace of $\mb{Y}$, leading to the prominent $\ell_1$-SVD method. To achieve high frequency resolution it was further suggested in \cite{Malioutov:LassoDoa} to perform an adaptive grid refinement. For the special case of uniform linear arrays (ULAs) and ULAs with missing sensors the authors in \cite{7313018, 
yang2014a, 
yang2014b} proposed an extension of the mixed-norm minimization problem in \eqref{eq:mixedVectorNorm} to the infinite-dimensional vector space, i.e., gridless signal reconstruction, in terms of atomic norm minimization.

%% file: p02_sparseRepMmv.tex
\begin{tikzpicture}

  \begin{scope}[xshift=-1cm]
   
  \node (Y) at (0,0) {

	\begin{tikzpicture}
	  \pgfmathsetseed{\seedA}
	  \foreach \x in {1,...,\NSnp}{
		\foreach \y in {1,...,\NSen}{			
		  \pgfmathrandominteger{\r}{\cRange}{100};
		  \fill[cs0!\r] (\x*\bSizeMod, \y*\bSizeMod) rectangle +(\bSizeMod,\bSizeMod);
		}
	  }
	\end{tikzpicture}	  
  };

  \node[anchor=west] (eq) at ($(Y.east)+(0mm,0mm)$) {$=$};	
  \node[anchor=west] (A) at ($(eq.east)+(0mm,0mm)$) {	
	\begin{tikzpicture}
	  \pgfmathsetseed{\seedB}
	  \foreach \y in {1,...,\NSen}{			
		  \pgfmathrandominteger{\r}{\cRange}{100};
		  \fill[cs1!\r] (1*\bSizeMod, \y*\bSizeMod) rectangle +(\bSizeMod,\bSizeMod);
	  }
	  \foreach \y in {1,...,\NSen}{			
		  \pgfmathrandominteger{\r}{\cRange}{100};
		  \fill[cs2!\r] (2*\bSizeMod, \y*\bSizeMod) rectangle +(\bSizeMod,\bSizeMod);
	  }
	  \foreach \y in {1,...,\NSen}{			
		  \pgfmathrandominteger{\r}{\cRange}{100};
		  \fill[cs3!\r] (3*\bSizeMod, \y*\bSizeMod) rectangle +(\bSizeMod,\bSizeMod);
	  }
	\end{tikzpicture}	  
  };
  
  \node[anchor=west] (times) at ($(A.east)+(-1mm,0mm)$) {$\times$};	

  \node[anchor=west] (X) at ($(times.east)+(-1mm,0mm)$) {
	\begin{tikzpicture}
	\foreach \x in {1,...,\NSnp}{			
		\pgfmathrandominteger{\r}{\cRange}{100};
		\fill[cs1!\r] (\x*\bSizeMod, 2*\bSizeMod) rectangle +(\bSizeMod,\bSizeMod);
		\pgfmathrandominteger{\r}{\cRange}{100};
		\fill[cs2!\r] (\x*\bSizeMod, 1*\bSizeMod) rectangle +(\bSizeMod,\bSizeMod);
		\pgfmathrandominteger{\r}{\cRange}{100};
		\fill[cs3!\r] (\x*\bSizeMod, 0) rectangle +(\bSizeMod,\bSizeMod);	  	
	}
	\end{tikzpicture}	
  };	  

  \node[anchor=north] (Y2) at ($(Y.south)+(0mm,-1mm)$) {$\strut\mb{Y}$};
  \node[anchor=center, baseline = (Y2.base)] (eq2) at (Y2.center-|eq.center) {$\strut=$};
  \node[anchor=center, baseline = (Y2.base)] (A2) at (Y2.center-|A.center) {$\strut\mb{A}(\mb{\mu})$};
  \node[anchor=center, baseline = (Y2.base)] (X2) at (A2.center-|X.center) {$\strut{\mb{\Psi}}$};

  \end{scope}

  \node at (2.1,0) {$\Rightarrow$};
  
  \begin{scope}[xshift=3cm]
   
  \node (Y) at (0,0) {

	\begin{tikzpicture}
	  \pgfmathsetseed{\seedA}
	  \foreach \x in {1,...,\NSnp}{
		\foreach \y in {1,...,\NSen}{			
		  \pgfmathrandominteger{\r}{\cRange}{100};
		  \fill[cs0!\r] (\x*\bSizeMod, \y*\bSizeMod) rectangle +(\bSizeMod,\bSizeMod);
		}
	  }
	\end{tikzpicture}	  
  };

  \node[anchor=west] (eq) at ($(Y.east)+(0mm,0mm)$) {$=$};	
  \node[anchor=west] (A) at ($(eq.east)+(0mm,0mm)$) {	
  % 	\drawMtxB{red}{6}{15}{\bSizeMod}
	\begin{tikzpicture}
	  \pgfmathsetseed{\seedA}
	  \foreach \x in {1,...,\NGrd}{
	  \foreach \y in {1,...,\NSen}{			
		  \pgfmathrandominteger{\r}{\cRange}{100};
		  \fill[lightgray!\r] (\x*\bSizeMod, \y*\bSizeMod) rectangle +(\bSizeMod,\bSizeMod);
	  }
	  }
	  \pgfmathsetseed{\seedB}
	  \foreach \y in {1,...,\NSen}{			
		  \pgfmathrandominteger{\r}{\cRange}{100};
		  \fill[cs1!\r] (\dA*\bSizeMod, \y*\bSizeMod) rectangle +(\bSizeMod,\bSizeMod);
	  }
	  \foreach \y in {1,...,\NSen}{			
		  \pgfmathrandominteger{\r}{\cRange}{100};
		  \fill[cs2!\r] (\dB*\bSizeMod, \y*\bSizeMod) rectangle +(\bSizeMod,\bSizeMod);
	  }
	  \foreach \y in {1,...,\NSen}{			
		  \pgfmathrandominteger{\r}{\cRange}{100};
		  \fill[cs3!\r] (\dC*\bSizeMod, \y*\bSizeMod) rectangle +(\bSizeMod,\bSizeMod);
	  }
	\end{tikzpicture}	  
  };
  \node[anchor=west] (times) at ($(A.east)+(-1mm,0mm)$) {$\times$};	
  
  \node[anchor=south west] (X) at ($(A.south east)+(3mm,0mm)$) {
	\begin{tikzpicture}
		\draw[step=\bSizeMod,lightgray!50,very thin] (\bSizeMod,0) grid (\NSnp*\bSizeMod+\bSizeMod, \NGrd*\bSizeMod);
		\draw[gray] (\bSizeMod,0) rectangle (\NSnp*\bSizeMod+\bSizeMod, \NGrd*\bSizeMod);
		\foreach \x in {1,...,\NSnp}{		
		  \pgfmathrandominteger{\r}{\cRange}{100};
		  \fill[cs1!\r] (\x*\bSizeMod, \NGrd*\bSizeMod-\dA*\bSizeMod) rectangle +(\bSizeMod,\bSizeMod);
		  \pgfmathrandominteger{\r}{\cRange}{100};
		  \fill[cs2!\r] (\x*\bSizeMod, \NGrd*\bSizeMod-\dB*\bSizeMod) rectangle +(\bSizeMod,\bSizeMod);
		  \pgfmathrandominteger{\r}{\cRange}{100};
		  \fill[cs3!\r] (\x*\bSizeMod, \NGrd*\bSizeMod-\dC*\bSizeMod) rectangle +(\bSizeMod,\bSizeMod);	  	
		}
	\end{tikzpicture}	
  };	  

  \node[anchor=north] (Y2) at ($(Y.south)+(0mm,-1mm)$) {$\strut\mb{Y}$};
  \node[anchor=center, baseline = (Y2.base)] (eq2) at (Y2.center-|eq.center) {$\strut=$};
  \node[anchor=center, baseline = (Y2.base)] (A2) at (Y2.center-|A.center) {$\strut\mb{A}(\mb{\nu})$};
  \node[anchor=center, baseline = (Y2.base)] (X2) at (A2.center-|X.center) {$\strut\mb{X}$};

  \end{scope}
  
\end{tikzpicture}
  

%% file: 05_magRec.tex
\section{{\myNameB}: A Reformulation of the \texorpdfstring{$\ell_{21}$}{l21}~Mixed-Norm Minimization Problem} \label{sec:magrec}

As discussed in Section \ref{sec:StateOfTheArtl21}, the MMV-based $\ell_{2,1}$ mixed-norm minimization problem can be considered as a generalization of the prominent $\ell_1$ norm minimization problem for SMVs \cite{Tibshirani:Lasso, Chen98atomicdecomposition}. In this context, one of our main results is given by the following theorem:

\begin{theorem} \label{th:equivalence}
The row-sparsity inducing $\ell_{2,1}$ mixed-norm minimization problem 
\begin{equation} 
  \min_{\mb{X}} 
  \frac{1}{2} \left\| \mb{A} \mb{X} - \mb{Y} \right\|_\tF^2 + 
%   \lambda \sum_{i=1}^{N} \left( \sum_{j=1}^\tT \left| s_{ij} \right|^p \right)^{1/p}.
%   \lambda \sum_{n=1}^{N} \left\| \mybar{\mb{x}}_k \right\|_2
  \lambda \sqrt{N} \left\| \mb{X} \right\|_{2,1}
  \label{eq:mixedVectorNorm_v2}
\end{equation}
is equivalent to the convex problem
\begin{align} 
  \min_{\mb{S} \in \pdMat} &\;
  \tr \big( (\mb{A} \mb{S} \mb{A}^\tH  + \lambda \mb{I}_M)^{-1} \hat{\mb{R}} \big) + \tr(\mb{S}), \label{eq:smr1}
\end{align}
with $\hat{\mb{R}} = \mb{Y} \mb{Y}^\tH /N$ denoting the sample covariance matrix and $\pdMat$ describing the set of nonnegative diagonal matrices, in the sense that minimizers $\opt{\mb{X}}$ and $\opt{\mb{S}}$ for problems \eqref{eq:mixedVectorNorm_v2} and \eqref{eq:smr1}, respectively, are related by 
\begin{align}
  \opt{\mb{X}} =&  \opt{\mb{S}} \! \mb{A}^\tH  ( \mb{A} \opt{\mb{S}} \! \mb{A}^\tH + \lambda \mb{I}_M )^{-1} \mb{Y} .
  \label{eq:smr2}
\end{align}
\end{theorem}

A proof of the equivalence is provided at the end of this section, while a proof of the convexity of \eqref{eq:smr1} is provided in Section \ref{sec:SDP} by establishing equivalence to a semidefinite program. 

In addition to \eqref{eq:smr2}, we observe that the matrix $\opt{\mb{S}}=\diag(\opt{s}_1, \ldots, \opt{s}_K)$ contains the row-norms of the sparse signal matrix $\opt{\mb{X}} = [\opt{\mybar{\mb{x}}}_1, \ldots, \opt{\mybar{\mb{x}}}_K]^\tT$ on its diagonal according to
\begin{align}
  \opt{s}_k = \frac{1}{\sqrt{N}} \| \opt{\mybar{\mb{x}}}_k \|_2,
  \label{eq:magIdentity}
\end{align}
for $k=1,\ldots,K$, such that the union support of $\opt{\mb{X}}$ is equivalently represented by the support of the sparse vector of row-norms $[\opt{s}_1, \ldots, \opt{s}_K]$. We will refer to \eqref{eq:smr1} as SPARse ROW-norm reconstruction ({\myName}). In this regard, we emphasize that $\opt{\mb{S}}$ should not be mistaken for a sparse representation of the source covariance matrix, i.e., $\opt{\mb{S}} \neq \tE\{\opt{\mb{X}} \opt{\mb{X}}^\tH\}/N$. While the mixed-norm minimization problem in \eqref{eq:mixedVectorNorm_v2} has $NK$ complex variables in ${\mb{X}}$, the {\myName} problem in \eqref{eq:smr1} provides a reduction to only $K$ nonnegative variables in the diagonal matrix ${\mb{S}}$. However, the union support of $\opt{\mb{X}}$ is similarly provided by $\opt{\mb{S}}$. Moreover, the {\myName} problem in \eqref{eq:smr1} only relies on the sample covariance matrix $\hat{\mb{R}}$ instead of the MMVs in $\mb{Y}$ themselves, leading to a reduction in problem 
size, especially in the case of large number of MMVs $N$. Interestingly, this indicates that the union support of the signal matrix $\opt{\mb{X}}$ is fully encoded in the sample covariance~$\hat{\mb{R}}$, rather than the instantaneous MMVs in~$\mb{Y}$, as may be concluded from the $\ell_{2,1}$ formulation in \eqref{eq:mixedVectorNorm_v2}. As seen from \eqref{eq:smr2}, the instantaneous MMVs in $\mb{Y}$ are only required for the signal reconstruction, which, in the context of array signal processing, can be interpreted as a form of beamforming \cite{vanTrees2002}, where the row-sparse structure in $\opt{\mb{X}}$ is induced by premultiplication with the sparse diagonal matrix~$\opt{\mb{S}}$.

\vspace{.3em}
\begin{IEEEproof}[Proof of Theorem \ref{th:equivalence}]
A key component in establishing the equivalence in equations \eqref{eq:mixedVectorNorm_v2} and \eqref{eq:smr1} is the observation that the $\ell_2$ norm of a vector $\mybar{\mb{x}}_k$ can be rewritten as
\begin{subequations}
\label{eq:normDecomp}
\begin{align}
  \left\| \mybar{\mb{x}}_k \right\|_2 = \min_{\gamma_k, \mybar{\mb{g}}_k} &\; \frac{1}{2} (|\gamma_k|^2 + \| \mybar{\mb{g}}_k\|_2^2) \\
  \text{s.t. } &\; \gamma_k \mybar{\mb{g}}_k = \mybar{\mb{x}}_k ,
\end{align}
\end{subequations}
where $\gamma_k$ is a complex scalar and $\mybar{\mb{g}}_k$ is a complex vector of dimension $N \times 1$, similar to $\mybar{\mb{x}}_k$. For the optimal solution of \eqref{eq:normDecomp}, it holds that
\begin{align}
  \left\| \mybar{\mb{x}}_k \right\|_2 = |\gamma_k|^2 = \| \mybar{\mb{g}}_k\|_2^2 .
  \label{eq:s_xl2_eq}
\end{align}
To see this, consider that any feasible solution must fulfill 
\begin{align}
  \left\| \mybar{\mb{x}}_k \right\|_2 = \sqrt{|\gamma_k|^2 \| \mybar{\mb{g}}_k\|_2^2} \leq \frac{1}{2} (|\gamma_k|^2 + \| \mybar{\mb{g}}_k\|_2^2)
\end{align}
which constitutes the inequality of arithmetic and geometric means, with equality holding if and only if $|\gamma_k| = \| \mybar{\mb{g}}_k\|_2$.

We can extend the idea in \eqref{eq:normDecomp} to the $\ell_{2,1}$ mixed-norm of the source signal matrix $\mb{X} = [\mybar{\mb{x}}_1, \ldots, \mybar{\mb{x}}_K]^\tT$ composed of rows $\mybar{\mb{x}}_k$, for $k=1,\ldots,K$, by 
\begin{subequations}
\label{eq:l21Fac}
\begin{align}
  \| \mb{X} \|_{2,1} = \sum_{k=1}^{K} \left\| \mybar{\mb{x}}_k \right\|_2 = 
  \min_{\substack{\mb{\varGamma} \in \dMat, \mb{G}}} &\; 
  \frac{1}{2} (\| \mb{\varGamma} \|_\tF^2 + \| \mb{G}\|_\tF^2)  \\
  \text{s.t. } & \mb{X} = \mb{\varGamma} \mb{G} \label{eq:l21FacConst}
\end{align}
\end{subequations}
where $\mb{\varGamma} = \diag(\gamma_1, \ldots, \gamma_K)$ is a $K \times K$ complex diagonal matrix and $\mb{G}=[\mybar{\mb{g}}_1, \ldots, \mybar{\mb{g}}_K]^\tT$ is a $K \times N$ complex matrix with rows $\mybar{\mb{g}}_k$, for $k=1,\ldots,K$. After inserting \eqref{eq:l21Fac} into the $\ell_{2,1}$ mixed-norm minimization problem in \eqref{eq:mixedVectorNorm_v2}, we formulate the minimization problem
\begin{equation}  
  \min_{\substack{\mb{\varGamma} \in \dMat, \mb{G}}}
  \frac{1}{2} \left\| \mb{A}\mb{\varGamma} \mb{G} - \mb{Y} \right\|_\tF^2 + \frac{\lambda \sqrt{N}}{2} (\| \mb{\varGamma} \|_\tF^2 + \| \mb{G}\|_\tF^2) 
  \label{eq:bilinOpt} .
\end{equation}
For a fixed matrix $\mb{\varGamma}$, the minimizer $\opt{\mb{G}}$ of problem \eqref{eq:bilinOpt} admits the closed form expression
\begin{align}
  \opt{\mb{G}} &=
%   \argmin_{\mb{G}} 
%   \frac{1}{2} \left\| \mb{A}\mb{\varGamma} \mb{G} - \mb{Y} \right\|_\tF^2 + \frac{\lambda \sqrt{N}}{2} (\| \mb{\varGamma} \|_\tF^2 + \| \mb{G}\|_\tF^2)
%   \nonumber \\ &= 
  \big( \mb{\varGamma}^\tH \mb{A}^\tH \mb{A} \mb{\varGamma} + \lambda \sqrt{N} \mb{I}_K \big)^{-1} \mb{\varGamma}^\tH \mb{A}^\tH \mb{Y} 
  \nonumber \\ &= 
  \mb{\varGamma}^\tH \mb{A}^\tH  \big( \mb{A} \mb{\varGamma} \mb{\varGamma}^\tH \mb{A}^\tH + \lambda \sqrt{N} \mb{I}_M \big)^{-1} \mb{Y}
  \label{eq:optG}
\end{align}
where the last identity is derived from the matrix inversion lemma. Reinserting the optimal matrix $\opt{\mb{G}}$ into equation \eqref{eq:bilinOpt} and performing basic reformulations of the objective function results in the compact minimization problem
\begin{align}
  \min_{\mb{\varGamma} \in \dMat} \frac{\lambda \sqrt{\!N} }{2} \! \Big( \!
  \text{Tr}\big( ( \mb{A} \mb{\varGamma} \mb{\varGamma}^\tH \! \mb{A}^\tH \!+\! \lambda \sqrt{N} \mb{I}_M )^{-1} \mb{Y} \mb{Y}^\tH \big)
  \!+\! \text{Tr} \big( \mb{\varGamma} \mb{\varGamma}^\tH \big) \! \Big) .
  \label{eq:equivalenceStep3}
\end{align}
Upon substituting $\mb{Y} \mb{Y}^\tH = N \hat{\mb{R}}$ and defining the nonnegative diagonal matrix 
\begin{align}
  \mb{S}=\mb{\varGamma} \mb{\varGamma}^\tH / \sqrt{N} \in \pdMat
  \label{eq:SMatDef}
\end{align}
we can rewrite \eqref{eq:equivalenceStep3} as the problem
\begin{align}
  \min_{\mb{S} \in \pdMat} \;
  \frac{\lambda N}{2} \Big( \text{Tr}\big( ( \mb{A} \mb{S} \mb{A}^\tH + \lambda \mb{I}_M )^{-1} \hat{\mb{R}} \big) 
  & + \text{Tr} \big( \mb{S} \big) \Big) .
  \label{eq:equivalenceStep4}
\end{align}
Neglecting the factor $\lambda N/2$ in \eqref{eq:equivalenceStep4}, we arrive at formulation \eqref{eq:smr1}. From equation \eqref{eq:s_xl2_eq} and the definition of $\mb{S}~=~\diag(s_1, \ldots,s_K)$ in \eqref{eq:SMatDef} we furthermore conclude that
\begin{align}
  s_k = \frac{1}{\sqrt{N}} \| \mybar{\mb{x}}_k \|_2,
\end{align}
for $k=1,\ldots,K$, as given by \eqref{eq:magIdentity}. Making further use of the factorization in \eqref{eq:l21FacConst} we obtain
\begin{align}
  \opt{\mb{X}} =& \opt{\mb{\varGamma}} \opt{\mb{G}} \nonumber \\
			   =&  \opt{\mb{\varGamma}} \opt{\mb{\varGamma}}^\tH \mb{A}^\tH  ( \mb{A} \opt{\mb{\varGamma}} \opt{\mb{\varGamma}}^\tH \mb{A}^\tH + \lambda \sqrt{N} \mb{I}_M )^{-1} \mb{Y} \nonumber \\
			   =&  \opt{\mb{S}} \mb{A}^\tH  ( \mb{A} \opt{\mb{S}} \mb{A}^\tH + \lambda \mb{I}_M )^{-1} \mb{Y}
\end{align}
which is \eqref{eq:smr2}. 
\end{IEEEproof}

%% file: 06_algorithms.tex
\section{Implementation of the {\myNameB} Problem} \label{sec:algorithms}

In this section we provide a simple implementation of the {\myName} problem via SemiDefinite Programming (SDP), which further admits gridless frequency estimation in the case of a uniform linear array. Additionally, for arbitrary array geometries we present a grid-based, low complexity implementation of problem \eqref{eq:smr1} in terms of the coordinate descent method for application with a large number of sensors~$M$.

\subsection{SDP Implementation and Gridless {\myName}} \label{sec:SDP}

To show convexity of the {\myName} formulation \eqref{eq:smr1} and for implementation with standard convex solvers, such as SeDuMi \cite{S98guide}, consider the following corollaries:
\begin{corollary} \label{col:sdp1}
The {\myName} problem in \eqref{eq:smr1} is equivalent to the semidefinite program (SDP)
\begin{subequations}
\label{eq:sdp1}
\begin{align}
  \min_{\mb{S}, \mb{U}_N} \; & \;\; \frac{1}{N} \tr( \mb{U}_N ) + \tr( \mb{S} ) \\
  \tst & \; \mtx{ \mb{U}_N & \mb{Y}^\tH \\ \mb{Y} & 
  \mb{A} \mb{S} \mb{A}^\tH + \lambda \mb{I}_M } \succeq \mb{0}  \label{eq:sdp1Con1} \\
  & \; \; \mb{S} \in \pdMat 
\end{align}	
\end{subequations}
where $\mb{U}_N$ is a Hermitian matrix of size $N \times N$.
\end{corollary}
To see the equivalence of the two problems, note that in \eqref{eq:sdp1} $\mb{A} \mb{S} \mb{A}^\tH + \lambda \mb{I}_M \succ \mb{0}$ is positive definite, since $\mb{S} \succeq \mb{0}$ and $\lambda > 0$. Further consider the Schur complement of the constraint \eqref{eq:sdp1Con1}
\begin{align}
  \mb{U}_N \succeq \mb{Y}^\tH (\mb{A} \mb{S} \mb{A}^\tH + \lambda \mb{I}_M)^{-1} \mb{Y} ,
  \label{eq:sdpEquivalence1}
\end{align}
which implies
\begin{align}
  \frac{1}{N} \tr(\mb{U}_N) &\geq \frac{1}{N} \tr( \mb{Y}^\tH (\mb{A} \mb{S} \mb{A}^\tH + \lambda \mb{I}_M)^{-1} \mb{Y}) \nonumber \\
						  &= \tr( (\mb{A} \mb{S} \mb{A}^\tH + \lambda \mb{I}_M)^{-1} \hat{\mb{R}}) .
  \label{eq:sdpEquivalence2}
\end{align}
For any optimal point $\opt{\mb{S}}$ of \eqref{eq:smr1} we can construct a feasible point of \eqref{eq:sdp1} with the same objective function value by choosing $\mb{U}_N = \mb{Y}^\tH (\mb{A} \opt{\mb{S}} \mb{A}^\tH + \lambda \mb{I}_M)^{-1} \mb{Y}$.
Reversely, any optimal solution pair $\opt{\mb{U}}_N,\opt{\mb{S}}$ of \eqref{eq:sdp1} is also feasible for \eqref{eq:smr1}.
\begin{corollary} \label{col:sdp2}
The {\myName} formulation in \eqref{eq:smr1} admits the equivalent problem formulation
\begin{subequations}
\label{eq:sdp1b}
\begin{align}
  \min_{\mb{S}, \mb{U}_M} \; & \;\; \tr( \mb{U}_M \hat{\mb{R}} ) + \tr( \mb{S} ) \\
  \tst & \; \mtx{ \mb{U}_M & \mb{I}_M \\ \mb{I}_M & \mb{A} \mb{S} \mb{A}^\tH + \lambda \mb{I}_M } \succeq \mb{0} \label{eq:sdp1bCon1} \\
  & \; \; \mb{S} \in \pdMat 
\end{align}	
\end{subequations}
where $\mb{U}_M$ is a Hermitian matrix of size $M \times M$.
\end{corollary}
The proof of Corollary \ref{col:sdp2} follows the same line of arguments as in the proof of Corollary \ref{col:sdp1}. In contrast to the constraint \eqref{eq:sdp1Con1}, the dimension of the semidefinite constraint \eqref{eq:sdp1bCon1} is independent of the number of MMVs $N$. It follows that either problem formulation \eqref{eq:sdp1} or \eqref{eq:sdp1b} can be selected to solve the {\myName} problem in \eqref{eq:smr1}, depending on the number of MMVs $N$ and the resulting dimension of the semidefinite constraint, i.e., \eqref{eq:sdp1} is preferable for $N \leq M$ and \eqref{eq:sdp1b} is preferable otherwise.

While the above SDP implementations are applicable to arbitrary array geometries, we consider next the special case of a uniform linear array (ULA) with sensor positions $\rho_m = m-1$, for $m=1,\ldots,M$, such that $\mb{A}=[\mb{a}(\nu_1), \ldots, \mb{a}(\nu_K)]$ is a Vandermonde matrix of size $M \times K$. In contrast to previous considerations, we further assume that $K \leq M$ and that the frequencies $\nu_1,\ldots,\nu_K$ for the signal representation are arbitrary, i.e., not confined to lie on a fixed grid. Under the given assumptions, the matrix product $\mb{A} \mb{S} \mb{A}^\tH$ exhibits a Toeplitz structure according to
\begin{align}
  \toep(\mb{u}) = \mb{A} \mb{S} \mb{A}^\tH = \sum_{k=1}^K s_k \mb{a}(\nu_k) \mb{a}^\tH (\nu_k) ,
  \label{eq:VanDec}
\end{align}
where $\toep(\mb{u})$ denotes a Hermitian Toeplitz matrix with $\mb{u}$ as its first column. As discussed in \cite{gongguo2013}, by the Caratheodory theorem \cite{Cara1, Cara2, Toeplitz}, any Toeplitz matrix $\toep(\mb{u})$ can be represented by a Vandermonde decomposition according to \eqref{eq:VanDec} for any distinct frequencies $\nu_1, \ldots, \nu_K$ and corresponding magnitudes $s_1,\ldots,s_K > 0$, with $\rank(\toep(\mb{u})) = K \leq M$. Given a Toeplitz matrix $\toep(\mb{u})$, the Vandermonde decomposition according to \eqref{eq:VanDec} can be obtained by first recovering the frequencies $\nu_k$, e.g., by Prony's method \cite{prony1795}, the matrix pencil approach \cite{56027} or linear prediction methods \cite{1456696}, where the frequency recovery is performed in a gridless fashion. The corresponding signal magnitudes in $\mb{s}=[s_1,\ldots, s_K]^\tT$ can be reconstructed by solving the linear system 
\begin{align}
  \mb{A} \, \mb{s} = \mb{u} ,
  \label{eq:GlMagRec}
\end{align}
i.e., by exploiting that $[\mb{a}(\nu)]_1 = 1$, for all $\nu \in [-1,1)$, and considering the first column in the representation \eqref{eq:VanDec}. Based on \eqref{eq:VanDec}, we rewrite problem \eqref{eq:sdp1} in a gridless version as 
\begin{subequations}
\label{eq:GL_smr}
\begin{align}
  \min_{\mb{u}, \mb{U}_N} \; & \;\; \frac{1}{N} \tr\big( \mb{U}_N \big) + \frac{1}{M} \tr\big( \toep(\mb{u}) \big) \\
  \tst & \; \mtx{ \mb{U}_N & \mb{Y}^\tH \\ \mb{Y} & \toep(\mb{u}) + \lambda \mb{I}_M } \succeq \mb{0}  \label{eq:GL_smr0b} \\
  & \; \; \toep(\mb{u})\succeq \mb{0} ,
\end{align}	
\end{subequations}
where we additionally make use of the identity
\begin{align}
  \tr( \mb{S} ) = \frac{1}{M} \tr(\mb{A} \mb{S} \mb{A}^\tH ) = \frac{1}{M} \tr\big( \toep(\mb{u}) \big) ,
\end{align}
with the factor $1/M$ resulting from $\| \mb{a}(\nu) \|_2^2 = M$, for all $\nu \in [-1,1)$. Alternatively, using the formulation \eqref{eq:sdp1b}, we can define the gridless estimation problem
\begin{subequations}
\label{eq:GL_smrb}
\begin{align}
  \min_{\mb{u}, \mb{U}_M} \; & \;\; \tr\big( \mb{U}_N \hat{\mb{R}} \big) + \frac{1}{M} \tr\big( \toep(\mb{u}) \big) \\
  \tst & \; \mtx{ \mb{U}_M & \mb{I}_M \\ \mb{I}_M & \toep(\mb{u}) + \lambda \mb{I}_M } \succeq \mb{0}  \\
  & \; \; \toep(\mb{u})\succeq \mb{0} .
\end{align}	
\end{subequations}
Given a minimizer $\opt{\mb{u}}$ of problem \eqref{eq:GL_smr} or \eqref{eq:GL_smrb}, the number of sources, i.e., the model order, can be directly estimated as
\begin{align}
  \hat{L} = \rank \big(\toep(\opt{\mb{u}})\big) ,
  \label{eq:GL_smr_NSrc0}
\end{align}
while the frequencies $\{ \hat{\mu}_l \}_l^{\hat{L}}$ and corresponding magnitudes $\{ \hat{s}_l \}_l^{\hat{L}}$ can be estimated by Vandermonde decomposition according to \eqref{eq:VanDec}, as discussed above. With the frequencies in $\{ \hat{\mu}_l \}_l^{\hat{L}}$ and signal magnitudes in $\{ \hat{s}_l \}_l^{\hat{L}}$, the corresponding signal matrix $\hat{\mb{X}}$ can be reconstructed by application of \eqref{eq:smr2}. 

We remark that unique Vandermonde decomposition requires that $\hat{L} = \rank \big(\toep(\opt{\mb{u}})\big) < M$. The rank $\hat{L}$ can be interpreted as the counterpart of the number of non-zero elements in the minimizer $\hat{\mb{S}}$ in the grid-based problems \eqref{eq:sdp1} and \eqref{eq:sdp1b}. Similarly as the regularization parameter $\lambda$ determines the number of non-zero elements, i.e., the sparsity level of $\hat{\mb{S}}$, there always exists a value $\lambda$ which yields a minimizer $\opt{\mb{u}}$ of the gridless formulations \eqref{eq:GL_smr} and \eqref{eq:GL_smrb} which fulfills $\hat{L} = \rank \big(\toep(\opt{\mb{u}})\big) < M$ such that a unique Vandermonde decomposition is obtained. We provide a description for the appropriate choice of the regularization parameter $\lambda$ in Section~\ref{sec:simulations}.

\subsection{Implementation by the Coordinate Descent Method} \label{sec:CD}
For sensor arrays with a large number of sensors $M$, the SDP implementation in the previous section may become computationally intractable, due to the large dimension of the semidefinite matrix constraints. Similar observations have been made for the gridless atomic norm minimization problem, which likewise relies on an SDP implementation, such that in \cite{bhaskar2011,6810450} it was suggested to avoid gridless estimation in the case of large sensor arrays and to return to a grid-based implementation of SSR that avoids SDP instead.

A particularly simple algorithm for solving the $\ell_{2,1}$ formulation \eqref{eq:mixedVectorNorm_v2} is the coordinate descent (CD) method \cite{qin2013efficient,wright2015}. Its simplicity mainly lies in the closed-form and low-complexity solutions for the coordinate updates. However, the computational complexity of the CD implementation of the conventional $\ell_{2,1}$ mixed norm minimization problem \eqref{eq:mixedVectorNorm_v2} increases with the number of MMVs $N$. On the other hand, the computational complexity of the {\myName} formulation in \eqref{eq:smr1} is independent of the number of MMVs $N$ and, as we will show in this section, a simple CD implementation also exists for the {\myName} formulation which can be implemented without expensive matrix inversions. 

Consider a function $f(\mb{S})$ which is jointly convex in the variables $s_1, \ldots, s_K$. To be consistent with previous notation we summarize the variables in the diagonal matrix $\mb{S}=\diag(s_1, \ldots, s_K)$. Furthermore, consider uncoupled constraints of the form $s_k \geq 0$, for $k=1,\ldots,K$. The CD method provides sequential and iterative coordinate updates, where coordinate $s_k^{(\tau)}$ in iteration $\tau$ is updated with the optimal stepsize $\opt{d}_{k}^{(\tau)}$, computed as
\begin{subequations}
\label{eq:cd2}
\begin{align}
  \opt{d}_{k}^{(\tau)} \!\! = 
  \arg \min_{d}&\; f(\mb{S}_{k,\tau} \! + d \, \mb{E}_k) \\
  \text{s.t.} &\; s_{k}^{(\tau)}+d \geq 0 \label{eq_cd2con}  .
\end{align}
\end{subequations}
In \eqref{eq:cd2}, the diagonal matrix
\begin{align}
  \mb{S}_{k,\tau} = \diag \Big( s_{1}^{(\tau+1)}\!, \ldots, s_{k-1}^{(\tau+1)}, s_{k}^{(\tau)}, \ldots, s_{K}^{(\tau)} \Big)
\end{align}
denotes the approximate solution for the minimizer of $f(\mb{S})$ in iteration $\tau$, before updating coordinate $k$, and matrix $\mb{E}_k$ with elements
\begin{align}
  [\mb{E}_k]_{m,n} = 
  \begin{cases}
	1 \quad & \tif m=n=k \\
	0 \quad & \telse
  \end{cases}
\end{align}
denotes a selection matrix. Given the update stepsize $\opt{d}_{k}^{(\tau)}$, the coordinate update is performed according to
\begin{align}
  \mb{S}_{k,\tau+1} = \mb{S}_{k,\tau} \! + \opt{d}_k^{(\tau)} \, \mb{E}_k .
  \label{eq:cdUpS}
\end{align}
Regarding the {\myName} problem in \eqref{eq:smr1}, the objective function of the subproblem in \eqref{eq:cd2} is given as
\begin{align}
  f(\mb{S}_{k,\tau} \! + d \, \mb{E}_k) = \!
  \tr \big( (\mb{U}_{k,\tau} \! + d \, \mb{a}_k \mb{a}_k^\tH)^{-1} \hat{\mb{R}} \big) \!+ \tr\big( \mb{S}_{k,\tau} \big) \!+ d,
  \label{eq:cdObj}
\end{align}
with $\mb{a}_k = \mb{a}(\nu_k)$ denoting the $k$th column of the $M \times K$ dictionary matrix $\mb{A}$, computed from a fixed grid of frequencies $\nu_1, \ldots, \nu_K$ as discussed in Section~\ref{sec:StateOfTheArtl21}, and $\mb{U}_{k,\tau}~=~\mb{A} \mb{S}_{k,\tau} \mb{A}^\tH + \lambda \mb{I}_M$. Upon application of the matrix inversion lemma 
\begin{align}
  & (\mb{U}_{k,\tau} \! + d \, \mb{a}_k \mb{a}_k^\tH)^{-1} = \;  
  \mb{U}_{k,\tau}^{-1} - 
  \frac{d \, \mb{U}_{k,\tau}^{-1} \mb{a}_k \mb{a}_k^\tH \mb{U}_{k,\tau}^{-1} }{1 + d \, \mb{a}_k^\tH \mb{U}_{k,\tau}^{-1} \mb{a}_k} 
  \label{eq:cdQInv}
\end{align}
and by exploiting the cyclic property of the trace, equation \eqref{eq:cdObj} can be rewritten as
\begin{align}
  & f(\mb{S}_{k,\tau} \! + d \, \mb{E}_k) \nonumber \\ = \;
  & \tr \big( \mb{U}_{k,\tau}^{-1} \hat{\mb{R}} \big) - 
	\frac{d \, \mb{a}_k^\tH \mb{U}_{k,\tau}^{-1} \hat{\mb{R}} \mb{U}_{k,\tau}^{-1} \mb{a}_k  }
		 {1 + d \, \mb{a}_k^\tH \mb{U}_{k,\tau}^{-1} \mb{a}_k }  + \tr\big( \mb{S}_{k,\tau} \big) + d .
  \label{eq:cdObj2}
\end{align}
The function $f(\mb{S}_{k,\tau} \! + d \, \mb{E}_k)$ in \eqref{eq:cdObj2} behaves asymptotically linear in $d$ and has stationary points in
\begin{align}
  \tilde{d}_{1,2} = \frac{ \pm \sqrt{ \mb{a}_k^\tH \mb{U}_{k,\tau}^{-1} \hat{\mb{R}} \mb{U}_{k,\tau}^{-1} \mb{a}_k}-1}
				  {\mb{a}_k^\tH \mb{U}_{k,\tau}^{-1} \mb{a}_k} ,
\end{align}
symmetrically located around the simple pole in
\begin{align}
  \tilde{d}_0 = 
  -\frac{1}{\mb{a}_k^\tH \mb{U}_{k,\tau}^{-1} \mb{a}_k} = 
  -\frac{1 + s_{k}^{(\tau)} \, \mb{a}_k^\tH \mb{U}_{-k,\tau}^{-1} \mb{a}_k }{\mb{a}_k^\tH \mb{U}_{-k,\tau}^{-1} \mb{a}_k} ,
  \label{eq:cdPole}
\end{align}
where the last identity in \eqref{eq:cdPole} follows from the matrix inversion lemma applied to $\mb{U}_{k,\tau}^{-1} = (\mb{U}_{-k,\tau} + s_{k}^{(\tau)} \, \mb{a}_k \mb{a}_k^\tH )^{-1}$, with $\mb{U}_{-k,\tau}~=~\mb{A}_{-k} \mb{S}_{-k,\tau} \mb{A}_{-k}^\tH + \lambda \mb{I}_M$, where $\mb{A}_{-k} = [\mb{a}_1 , \ldots , \mb{a}_{k-1} , \mb{a}_{k+1}, \ldots, \mb{a}_K]$ and $\mb{S}_{-k,\tau} = \diag(s_{1}^{(\tau)}, \ldots, s_{k-1}^{(\tau)}, s_{k+1}^{(\tau)}, \ldots, s_{K}^{(\tau)})$. By taking account of the constraint $ s_{k}^{(\tau)}+d \geq 0$ in \eqref{eq_cd2con}, it can easily be verified that the optimal stepsize must fulfill $\opt{d}_k^{(\tau)} \geq -s_{k}^{(\tau)} > \tilde{d}_0$, i.e., it must be located on the right hand side of the pole $\tilde{d}$, such that the optimal stepsize according to \eqref{eq:cd2} is computed as
\begin{align}
  \opt{d}_k^{(\tau)} 
%   =& \max \left( \frac{\sqrt{\alpha_{k,\tau}}-1}{\beta_{k,\tau}}, -s_{k}^{(\tau)} \right) \nonumber \\
  =& \max \left( \frac{ \sqrt{ \mb{a}_k^\tH \mb{U}_{k,\tau}^{-1} \hat{\mb{R}} \mb{U}_{k,\tau}^{-1} \mb{a}_k }-1} 
		  {\mb{a}_k^\tH \mb{U}_{k,\tau}^{-1} \mb{a}_k}, -s_{k}^{(\tau)} \right) \label{eq:cdOptStep} .
\end{align}
Given the stepsize $\opt{d}_k^{(\tau)}$, the variable update is performed according to \eqref{eq:cdUpS}. The matrix inverse $\mb{U}_{k+1,\tau}^{-1}$, including the updated coordinate $s_{k}^{(\tau+1)} = s_{k}^{(\tau)}+d_k^{(\tau)}$ as required for updating the next coordinate $s_{k+1}^{(\tau)}$, can be computed by the matrix inversion lemma as shown in \eqref{eq:cdQInv}, such that computationally expensive explicit matrix inversion can be avoided. We remark that the computation time of the CD method can be drastically reduced if the sparsity in $\mb{S}_{k,\tau}$ is exploited, by excluding zero elements in $\mb{S}_{k,\tau}$ from the computation. 
% Alternatively, the coordinate updates in \eqref{eq:cdOptStep} can be employed for parallel methods such as FISTA \cite{fista} and STELA \cite{stela}.

%% file: 07_relatedWork.tex
\section{Relation to Existing Algorithms} \label{sec:relatedWork}

In recent years, numerous publications have considered SSR from MMVs. In this section we provide a comparison of the $\ell_{2,1}$ mixed-norm minimization problem, and our compact reformulations, with two prominent alternative approaches which show particular similarities to our proposed {\myName} formulation, namely the atomic norm minimization approach \cite{7313018, yang2014a, yang2014b} and the SPICE method \cite{5599897, 5617289, 2014arXiv1406.7698S}.

\subsection{Atomic Norm Minimization} \label{sec:Anm}
The concept of Atomic Norm Minimization (ANM) has been introduced in \cite{chandrasekaran2012} as a unifying framework for different types of sparse recovery methods, such as $\ell_1$ norm minimization for sparse vector reconstruction or nuclear norm minimization for low-rank matrix completion. In \cite{gongguo2013,bhaskar2011,6552292} ANM was introduced for gridless line spectral estimation from SMVs in ULAs. The extension of ANM to MMVs under this setup was studied in \cite{7313018, yang2014a, yang2014b}, which will be revised in the following. Consider the noise-free MMV matrix $\mb{Y}_0 = \sum_{l=1}^L \mb{a}(\mu_l) \mb{\psi}_l^\tT$, obtained at the output of a ULA for $L$ impinging source signals with spatial frequencies $\mu_1, \ldots, \mu_L$, where the $l$th source signal is contained in the $N \times 1$ vector $\mb{\psi}_l$. In the ANM framework \cite{7313018, yang2014a, yang2014b}, the MMV matrix $\mb{Y}_0$ is considered as a weighted superposition of atoms $\mb{a}(\nu) \mb{b}^\tH$ with $\nu \in [-1,
1)$, $\mb{b} \in \mathbb{C}^{N}$ and $\| \mb{b} \|_2=1$. The atomic norm of $\mb{Y}_0$ is defined as
\begin{align}
  \| \mb{Y}_0 \|_{\mathcal{A}} 
%   &= \inf \big\{ t > 0 : \mb{Y}_0 \in t \, \text{conv}(\mathcal{A}) \big\} \\
%   &= \inf_{\substack{c_k \geq 0 \\ \mb{b} \in \mathbb{C}^N,\\ -1 \leq \nu < 1}}
	&= \smash{\inf_{\substack{\{c_k, \mb{b}_k, \\\nu_k\}}}} 
	   \Big\{ \sum_{k} c_k : \mb{Y}_0 \!=\! \sum_{k} c_k \mb{a}(\nu_k) \mb{b}_k^\tH, \, c_k \geq 0 \Big\}, 
  \label{eq:atomicDef}
\end{align}
and computed by the SDP \cite{chandrasekaran2012,gongguo2013,bhaskar2011,6552292,7313018, yang2014a, yang2014b}
\begin{subequations}
\label{eq:anm}
\begin{align}
  \| \mb{Y}_0 \|_{\mathcal{A}}
  = \inf_{\substack{\mb{v}, \mb{V}_N }}
	  & \, \frac{1}{2} \tr\big( \mb{V}_N \big)  + \frac{1}{2M} \tr \big( \toep(\mb{v}) \big) \label{eq:anm1}\\
	  \text{s.t. } & \, \mtx{ \mb{V}_N & \mb{Y}_0^\tH \\ \mb{Y}_0 & \toep(\mb{v}) } \succeq \mb{0}  \label{eq:anm2} \\
					& \toep(\mb{v}) \succeq \mb{0}  
\end{align}
\end{subequations}
where the Toeplitz matrix representation in the constraint \eqref{eq:anm2} relies on the assumption of a ULA\footnote{An interesting extension of the ANM problem in \eqref{eq:anm} considers the application of missing sensors. Although not treated here, the {\myName} formulations in \eqref{eq:GL_smr} and \eqref{eq:GL_smrb} can similarly deal with this application, e.g., by replacing $\toep(\mb{u})$ in \eqref{eq:GL_smr} and \eqref{eq:GL_smrb} by $\mb{J} \toep(\mb{u}) \mb{J}^\tT$, where $\mb{J}$ denotes a selection matrix representing the missing sensors.
% or by considering the dual problem
}, following similar arguments as for the gridless GL-{\myName} implementation discussed in Section \ref{sec:SDP}. Correspondingly, the frequency estimates $\hat{\mb{\mu}}$ can be recovered by Vandermonde decomposition \eqref{eq:VanDec}.
As proposed in \cite{7313018, yang2014a, yang2014b}, given a noise-corrupted MMV matrix $\mb{Y}$ as defined in \eqref{eq:sigModelMmv}, jointly sparse recovery from MMVs can be performed by using \eqref{eq:atomicDef} as
\begin{align}
  \min_{\mb{Y}_0} & \; \frac{1}{2} \| \mb{Y} - \mb{Y}_0 \|_{\tF}^2 + \lambda \sqrt{N} \| \mb{Y}_0 \|_{\mathcal{A}} 
\end{align}
or, equivalently, by using the SDP formulation in \eqref{eq:anm}, as
\begin{subequations}
\label{eq:relSdp2}
\begin{align}
  \smash{\min_{\substack{\mb{v}, \mb{V}_N, \\\mb{Y}_0 }}}
  & \, \frac{1}{2} \| \mb{Y} - \mb{Y}_0 \|_{\tF}^2 + 
		\frac{\lambda \sqrt{N}}{2} \Big( \tr \big( \mb{V}_N \big) + \frac{1}{M} \tr \big( \toep(\mb{v}) \big) \Big) \\
  \text{s.t. } & \, \mtx{ \mb{V}_N & \mb{Y}_0^\tH \\ \mb{Y}_0 & \toep(\mb{v}) } \succeq \mb{0} \\
					& \toep(\mb{v}) \succeq \mb{0} .
\end{align}
\end{subequations}
Problem \eqref{eq:relSdp2} and the GL-{\myName} formulation \eqref{eq:GL_smr} exhibit a similar structure in the objective functions and semidefinite constraints. In fact, both problems are equivalent in the sense that minimizers are related by
\begin{align}
  \opt{\mb{u}} = \opt{\mb{v}} / \sqrt{N}, 
\end{align}
where the factor $\sqrt{N}$ results from the definition in \eqref{eq:SMatDef}. The spatial frequencies of interest $\mb{\nu}$ are encoded in the vectors $\opt{\mb{u}}$ and $\opt{\mb{v}}$, as found by Vandermonde decomposition \eqref{eq:VanDec}, such that the GL-{\myName} problem in \eqref{eq:sdp1} and the ANM problem in \eqref{eq:relSdp2} both provide the same estimates. A proof of the equivalence is given in the appendix. 

However, from a computational viewpoint, in contrast to the GL-{\myName} problem in \eqref{eq:GL_smr}, the ANM problem in \eqref{eq:relSdp2} has additional $M N$ variables in the matrix $\mb{Y}_0$, which need to be matched to the MMV matrix $\mb{Y}$ by an additional quadratic term in the objective function. Moreover, the size of the ANM problem \eqref{eq:relSdp2} scales with the number of MMVs $N$. In contrast to that, the GL-{\myName} problem \eqref{eq:GL_smr} can be equivalently formulated as \eqref{eq:GL_smrb}, which is independent of the number of MMVs $N$. In this context the GL-{\myName} formulations \eqref{eq:GL_smr} and \eqref{eq:GL_smrb} admit significantly reduced computational complexity as compared to the ANM formulation \eqref{eq:relSdp2}. 

\subsection{SPICE} \label{sec:spice}

The SParse Iterative Covariance-based Estimation (SPICE) method \cite{5599897,5617289,2014arXiv1406.7698S} seeks to match the sample covariance matrix $\hat{\mb{R}} = \mb{Y} \mb{Y}^\tH / N$ with a sparse representation of the covariance matrix $\mb{R}_0$, as shortly reviewed in the following. 

The signal model $\mb{Y} = \mb{A}( \mb{\mu} ) \mb{\varPsi} + \mb{N}$, as defined in \eqref{eq:sigModelMmv}, admits the covariance matrix
\begin{align}
  \mb{R} = \tE \{ \mb{Y} \mb{Y}^\tH \}/N = \mb{A}(\mb{\mu}) \mb{\varPhi} \mb{A}^\tH(\mb{\mu}) + \sigma^2 \mb{I}_M .
  \label{eq:covar}
\end{align}
In contrast to our consideration the authors in \cite{5599897,5617289,2014arXiv1406.7698S} explicitly assume that the signals in $\mb{\varPsi}$ are uncorrelated, such that the source covariance matrix
\begin{align}
  \mb{\varPhi} = \tE\{ \mb{\varPsi} \mb{\varPsi}^\tH \} / N
  \label{eq:SrcCovar}
\end{align}
has a diagonal structure, i.e., $\mb{\varPhi} = \diag( \phi_1, \ldots, \phi_L )$. The sparse representation $\mb{R}_0$ of the covariance matrix in \eqref{eq:covar} is introduced as
\begin{align}
  \mb{R}_0 = \mb{A} \mb{P} \mb{A}^\tH + \epsilon \mb{I}_M ,
  \label{eq:SparseCovar}
\end{align}
where $\mb{A}$ denotes the dictionary matrix computed for a fixed grid of frequencies $\nu_1, \ldots, \nu_K ,$ as used in \eqref{eq:sparseSigModelMmv}, $\epsilon=\sigma^2$ denotes the noise power and the elements of the sparse diagonal source covariance matrix $\mb{P} = \diag( p_1, \ldots,  p_K) \in \pdMat$ are given as
\begin{align}
  p_k =& 
  \begin{cases}
	\phi_l \quad &\text{if } \nu_k = \mu_l \\
	0		  \quad &\text{else,}
  \end{cases}
  \label{eq:SparseSrcCovar}
\end{align}
for $k=1,\ldots,K$ and $l=1,\ldots,L$, with $\phi_l$ denoting the diagonal elements of the source covariance as defined \eqref{eq:SrcCovar}. 

Two types of weighted covariance matching functions have been proposed in \cite{5599897,5617289,2014arXiv1406.7698S}. The undersampled case, with $N < M$, is treated by minimization of a weighted covariance matching function according to
\begin{alignat}{2}
  & \min_{\substack{\mb{P} \in \pdMat, \\ \epsilon \geq 0}}
  \Big\{ \big\| \mb{R}_0^{-1/2} ( \hat{\mb{R}} - \mb{R}_0 ) \big\|_\tF^2 
  \,:\, \eqref{eq:SparseCovar} \Big\}
  \nonumber  \\	
  = 
  & {\min_{\substack{\mb{P} \in \pdMat, \\ \epsilon \geq 0}}}
  \Big\{ \tr \big(\mb{R}_0^{-1} \smash{\hat{\mb{R}}}^{\,2} \,\big) + \tr \big( \mb{R}_0 \big) - 2 \tr \big( \hat{\mb{R}} \big) 
  \,:\, \eqref{eq:SparseCovar}	\Big\} ,
\label{eq:spice2} 
\end{alignat}
where sparsity in $\mb{P}$ is induced in the objective of \eqref{eq:spice2} in form of the trace penalty term $\tr \big( \mb{R}_0 \big)$ as can be observed from the following identity:
\begin{align}
  \tr \big( \mb{R}_0 \big) &= \epsilon M + \sum_{k=1}^K \|\mb{a}_k \|_2^2 \cdot p_k 
  = M ( \epsilon + \sum_{k=1}^K p_k ).
\end{align}

The oversampled case, with $N \geq M$ where the sample covariance matrix $\hat{\mb{R}}$ is non-singular, is treated by the minimization of the weighted covariance matching function according to
\begin{alignat}{2}
  & \min_{\substack{\mb{P} \in \pdMat,\\ \epsilon \geq 0}} 
    \Big\{ \big\| \mb{R}_0^{-1/2} \big( \hat{\mb{R}} - \mb{R}_0 \big) \smash{\hat{\mb{R}}}^{-1/2} \big\|_\tF^2 
    \,:\, \eqref{eq:SparseCovar} \Big\}
  \nonumber \\ =
  & {\min_{\substack{\mb{P} \in \pdMat,\\ \epsilon \geq 0}}} 
    \Big\{ \tr \big( \mb{R}_0^{-1} \hat{\mb{R}} \big) + \tr \big(\mb{R}_0 \smash{\hat{\mb{R}}}^{-1} \big) - 2M
    \,:\, \eqref{eq:SparseCovar} \Big\},
  \label{eq:spice1}
\end{alignat}
where sparsity in $\mb{P}$ is induced by summation of its diagonal elements with data dependent weights according to
\begin{align}
  \tr \big(\mb{R}_0 \smash{\hat{\mb{R}}}^{-1} \big) = 
  \epsilon \tr \big(\smash{\hat{\mb{R}}}^{-1} \big) + \sum_{k=1}^K \mb{a}^\tH_k \smash{\hat{\mb{R}}}^{-1} \mb{a}_k \cdot p_k .
  \label{eq:spice1b}
\end{align}

We remark that our proposed {\myName} formulation in \eqref{eq:smr1} exhibits similarities with both SPICE formulations \eqref{eq:spice2} and \eqref{eq:spice1}. While the {\myName} formulation shares the uniformly weighted summation of its variables in $\tr(\mb{S})$ with the SPICE formulation in \eqref{eq:spice2}, it shares the structure of the data fitting function $\tr \big( (\mb{A} \mb{S} \mb{A}^\tH  + \lambda \mb{I}_M)^{-1} \hat{\mb{R}} \big)$ with the SPICE formulation in \eqref{eq:spice1}. There is, however, a fundamental difference between the {\myName} formulation and the SPICE formulations in the fact that the variables in $\mb{S}$ correspond to the normalized row-norms of the signal matrix, i.e., $\opt{s}_k = \frac{1}{\sqrt{N}} \| \opt{\mybar{\mb{x}}}_k \|_2$, for $k=1,\ldots,K$, as seen from \eqref{eq:magIdentity}, while the variables in $\mb{P}$ correspond to the signal powers, i.e., $\opt{p}_k = \frac{1}{\sqrt{N}} \tE \{ \| \opt{\mybar{\mb{x}}}_k \|_2^2 \}$, for $k=1,\ldots,K$, as seen 
from \eqref{eq:SrcCovar} and \eqref{eq:SparseSrcCovar}. 

Moreover, the SPICE formulations make assumptions on the second-order signal statistics in form of the covariance matrix in \eqref{eq:SparseCovar}, namely, the sparse source covariance matrix $\mb{P}$ is modeled as a diagonal matrix, which involves the assumption of uncorrelated source signals. In contrast to that, the {\myName} problem in \eqref{eq:smr1} does not rely on any such assumptions.

% An advantage of the SPICE formulations is that it is hyperparameter-free, i.e., the noise power $\epsilon$ is included as an estimation parameter. Regarding the structural similarities, the counterpart of the noise power $\epsilon$ in the SPICE formulations is given by the regularization parameter $\lambda$ in the {\myName} formulation. While the common approach would be to determine $\lambda$ from a-priori knowledge, it follows that $\lambda$ could as well be included as an optimization variable in the {\myName} formulation, similar to the SPICE formulations. Conversely, the noise power $\epsilon$ could also be provided as a-priori knowledge to the SPICE formulations. A detailed investigation of these different approaches is, however, beyond the scope of this paper and left for future research.

An extension of SPICE to the GridLess Spice (GLS) method for ULAs was proposed in \cite{yang2014b}, which relies on an SDP formulation of the SPICE problems \eqref{eq:spice2} and \eqref{eq:spice1}, and Vandermonde decomposition of Toeplitz matrices, similar to the {\myName} and ANM problems discussed in Sections \ref{sec:SDP} and \ref{sec:Anm}.

%% file: 08_simulations.tex
\section{Numerical Experiments} \label{sec:simulations}

The parameter estimation performance of the $\ell_{2,1}$ mixed-norm minimization, ANM and SPICE has been numerically investigated in various publications, e.g., \cite{Malioutov:LassoDoa, 5466152, 5599897, 5617289, 2014arXiv1406.7698S, 7313018, yang2014a, yang2014b}. In this paper we extend the existing simulation results by a numerical analysis of the parameter estimation performance in terms of estimation bias, standard deviation and root-mean-square error, for varying frequency separation as well as varying number of MMVs. In our experiments we compare gridless {\myName} \eqref{eq:GL_smrb} (referred to as GL-{\myName}), under- and oversampled SPICE, i.e., \eqref{eq:spice2} and \eqref{eq:spice1}, (referred to as US-SPICE and OS-SPICE, respectively), under- and oversampled GridLess Spice \cite{yang2014b} (referred to as US-GLS and OS-GLS, respectively), spectral MUSIC \cite{1143830}, root-MUSIC \cite{Krim:TwoDecades, vanTrees2002}, and the stochastic Cramer-Rao Bound (CRB) \cite{stoica2001stochastic}. We 
remind the reader, that the {\myName} formulation is equivalent to $\ell_{2,1}$ mixed-norm 
minimization and ANM, as discussed in Sections \ref{sec:magrec} and \ref{sec:Anm}, such that the latter two methods are not included in the performance analysis. Instead we provide a comparison of computation time for the equivalent approaches.

Optimal selection of a regularization parameter for the $\ell_{2,1}$ mixed-norm minimization, and correspondingly for the {\myName} problem in \eqref{eq:GL_smrb}, is an open problem in SSR research and beyond the scope of this paper. In this work, we follow a heuristic approach which provides good estimation performance in our investigated scenarios. For this, we consider problem \eqref{eq:mixedVectorNorm_v2} as a normalized combination of multiple SMV problems. Given a single SMV problem, in \cite{bhaskar2011} it was suggested to select the regularization parameter as
\begin{align}
  \lambda = \sqrt{\sigma^2 M \log M} ,
  \label{eq:regSelect}
\end{align}
for a large number of sensors $M$. We also apply the regularization parameter selection \eqref{eq:regSelect} to our {\myName} formulation \eqref{eq:smr1}. We remark that other approaches of regularization parameter selection can be used. The study of this is, however, not a subject of investigation in this work.

Note that {\myName}, SPICE and MUSIC all make different assumptions on the availability of a-priori knowledge. While SPICE does not require any a-priori knowledge, we assume perfect knowledge of the noise power $\sigma^2$ for the regularization parameter selection of {\myName}, and perfect knowledge of the number of source signals $L$ for the MUSIC method. 

% We remark once more, that different approaches for {\myName} and SPICE would be possible, but that such investigation is beyond the scope of this paper, as discussed in Section \ref{sec:spice}.

\def\pw{\columnwidth}
\def\ph{.66\columnwidth}

\subsection{Bias and Resolution Capability}

% use 'mark repeat=<number>' 'mark phase=<number>' or 'mark indices={<index list>}' do change marker behavior

\begin{figure}[t]
\begin{center}
  \scriptsize
  \input{p03_sep_bias_log}
  \vspace{-0.3cm}
\end{center}
\caption{Bias of estimated frequencies for two $L=2$ source signals for $N=50$ MMVs and $\tsnr=10\,\text{dB}$}
\label{fig:SepBias1}

\begin{center}
  \scriptsize
  \input{p04_sep_var_log}
  \vspace{-0.3cm}
\end{center}
\caption{Variance of estimated frequencies for two $L=2$ source signals for $N=50$ MMVs and $\tsnr=10\,\text{dB}$}
\label{fig:SepStd1}
\end{figure}

\begin{figure}[t]
\begin{center}
  \scriptsize
  \input{p05_sep_mse_log_10dB}
  \vspace{-0.3cm}
\end{center}
\caption{RMSE of estimated frequencies for two $L=2$ source signals for $N=50$ MMVs and $\tsnr=10\,\text{dB}$}
\label{fig:mse10dB}

\begin{center}
  \scriptsize
  \input{p06_sep_mse_log_3dB}
  \vspace{-0.3cm}
\end{center}
\caption{RMSE of estimated frequencies for two $L=2$ source signals for $N=20$ MMVs and $\tsnr=3\,\text{dB}$}
\label{fig:mse3dB}
\end{figure}

As discussed in \cite{Malioutov:LassoDoa}, $\ell_{2,1}$ mixed-norm minimization provides biased frequency estimates in the case of sources with closely separated frequencies. To the best of our knowledge, no such bias investigation has been performed for SPICE. For our first experiment on estimation bias and resolution capability we consider a uniform linear array of $M=6$ sensors with half signal wavelength spacing and fix the Signal-to-Noise Ratio ($\tsnr$), defined as $\tsnr = 1/\sigma^2$, to $\tsnr=10 \tdB$ and the number of MMVs to $N=50$. We perform $T=1000$ Monte-Carlo trials and for each trial we consider two independent complex Gaussian sources with static spatial frequencies. The first source signal has fixed spatial frequency $\mu_1=0.5$ while the spatial frequency $\mu_2$ of the second source is selected from the interval $[-0.5,\, 0.499]$ for each trial. For all grid-based estimation methods we make use of a uniform grid of $K=1000$ points. The estimation bias is computed as
\begin{align}
  \text{Bias} (\hat{\mb{\mu}}) = \sqrt{\frac{1}{L} \sum_{l=1}^L \left( \mu_l - {\rm Mean} (\hat{\mu}_l) \right)^2 }, 
  \label{eq:bias}
\end{align}
where the mean estimate for frequency $\mu_l$ is computed as
\begin{align}
  {\rm Mean} (\hat{\mu}_l) = \frac{1}{T} \sum_{t=1}^T \hat{\mu}_l(t), 
\end{align}
with $\hat{\mu}_l(t)$ denoting the estimate of the $l$th frequency $\mu_l$ in Monte Carlo trial $t$. Since the bias computation \eqref{eq:bias} requires the number of estimated source signals $\hat{L}$ to be equal to the true number of source signals $L$, we have to consider two special cases: in the case of overestimation of the model order, $\hat{L} > L$, we select the $L$ frequency estimates with the largest corresponding magnitudes, whereas we select $L-\hat{L}$ additional random spatial frequencies in the case of underestimation $\hat{L} < L$. Furthermore, we compute the standard deviation as
\begin{align}
  \text{Std}(\hat{\mb{\mu}}) = \sqrt{\frac{1}{T L} \sum_{l=1}^L 
					   \sum_{t=1}^T \big|  {\rm Mean} (\hat{\mu}_\ell) - \hat{\mu}_\ell(t) \big|_{\rm wa}^2 }  ,
\end{align}
where $|\opt{\mu}_1-\opt{\mu}_2|_{\rm wa}=\min_{i \in \mathbb{Z}}|\opt{\mu}_1-\opt{\mu}_2+2i|$ denotes the wrap-around distance for frequencies $\opt{\mu}_1,\opt{\mu}_2 \in [-1, 1)$. 

Figures \ref{fig:SepBias1} and \ref{fig:SepStd1} show the resulting bias and standard deviation versus the frequency separation $\Delta \mu = |\mu_2-\mu_1|_{\rm wa}$. As can be observed from the figures, our proposed GL-{\myName} method provides a relatively large bias in the case of closely spaced frequencies, with $\Delta \mu \leq 0.33$, but provides source resolution performance, i.e., thresholding performance, slightly superior to that of root-MUSIC, with successful source resolution for $\Delta \mu \geq 0.05$. For frequency separation $\Delta \mu \geq 0.33$ the estimation bias reduces significantly and becomes negligible with respect to the standard deviation. 

Similar to GL-{\myName}, US-SPICE and OS-SPICE show an estimation bias for closely spaced source signals with $\Delta \mu <0.44$, but provide degraded source resolution performance for $\Delta < 0.14$, similar to spectral MUSIC. In contrast to that, the US- and OS-GLS versions display negligible estimation bias (not shown here), while exhibiting a reduced estimation performance in terms of standard deviation. 
% Regarding the difference in performance, we found in our experiments that the main performance difference in the SPICE and GLS versions is the estimation of the noise power $\epsilon$, which in consequence affects the estimation of the spatial spectrum and the spatial frequencies.

Figure \ref{fig:mse10dB} shows the root-mean-square error (RMSE) of the schemes under investigation, which is computed according to
\begin{align}
  \text{RMSE}(\hat{\mb{\mu}})= \sqrt{\frac{1}{L T} \sum_{t=1}^T \sum_{l=1}^L \big| \mu_l - \hat{\mu}_l(t) \big|_{\rm wa}^2}.
\end{align}
As can be seen, GL-{\myName} does not reach the CRB for frequency separations $0.05 \leq \Delta \mu \leq 0.3$, due to the large estimation bias as compared to the CRB. The RMSE performance of the remaining schemes is comparable to the performance in terms of standard deviation, since the estimation bias for these schemes is negligible as compared to the CRB. Figure \ref{fig:mse3dB} shows the RMSE performance for a modified scenario with $\tsnr=3 \tdB$ and $N=20$ MMVs. In this case, the estimation bias of GL-{\myName} is negligible compared to the CRB, such that the RMSE approaches the CRB even for low frequency separation. Figure \ref{fig:mse3dB} also shows an improved threshold performance of the gridless sparse estimation methods GL-{\myName}, US-GLS and OS-GLS  as compared to the root-MUSIC method, such that for the given scenario sparse methods can be considered as a viable supplement to subspace-based methods.

We remark that in the gridless implementation for the case of ULAs, the estimation bias is inherent in the estimation method and independent of grid effects, and can be countered by bias mitigation techniques \cite{6375850} or a final maximum likelihood (ML) estimation step \cite{Krim:TwoDecades,vanTrees2002}. For instance, a combination of the SPICE and ML estimation has been proposed in \cite{Stoica20121580} in form of the LIKES method.

\subsection{Varying Number of Measurement Vectors}

\begin{figure}[t]
% \begin{center}
%   \scriptsize
%   \input{./plots/snp_var}
%   \vspace{-0.3cm}
% \end{center}
% \caption{Standard deviation of estimated DoAs for two $L=2$ source signals with frequency separation $\Delta \mu = 0.15$ and $\tsnr=3\tdB$}
% \label{fig:SnpStd2}

\begin{center}
  \scriptsize
  \input{p07_snp_mse}
  \vspace{-0.3cm}
\end{center}
\caption{RMSE of estimated frequencies for two $L=2$ source signals with frequency separation $\Delta \mu = 0.15$ and $\tsnr=3\tdB$}
\label{fig:SnpMse}

\begin{center}
  \scriptsize
  \input{p08_snp_res}
  \vspace{-0.3cm}
\end{center}
\caption{Resolution percentage of estimated frequencies for two $L=2$ source signals with frequency separation $\Delta \mu = 0.15$ and $\tsnr=3\tdB$}
\label{fig:SnpRes}

\end{figure}

\begin{figure}[t]

\begin{center}
  \scriptsize
  \input{p09_runtime_gb}
  \vspace{-0.3cm}
\end{center}
\caption{Average computation time for grid-based methods; $M=6$ sensors, $\Delta \mu = 0.15$ and $\tsnr=10\tdB$ }
\label{fig:runtimeGb}

\begin{center}
  \scriptsize
  \input{p10_runtime_gl}
  \vspace{-0.3cm}
\end{center}
\caption{Average computation time for gridless methods; $M=6$ sensors, $\Delta \mu = 0.15$ and $\tsnr=10\tdB$ }
\label{fig:runtimeGl}

\end{figure}

In our second experiment we investigate the estimation performance of the various methods for a varying number of MMVs. We consider two independent complex Gaussian sources with static spatial frequencies $\mu_1=0.35$ and $\mu_2=0.5$ and a ULA with $M=6$ sensors. The $\tsnr$ is fixed at $3\tdB$. Figure \ref{fig:SnpMse} shows the RMSE of the schemes under investigation from which we observe that GL-{\myName} clearly outperforms all other methods in terms of threshold performance. However, for large number of MMVs, the RMSE of GL-{\myName} saturates due to the estimation bias. GLS shows slightly worse RMSE performance compared to GL-{\myName} for a low number of MMVs $N$ and also does not reach the CRB for a large number of MMVs. In contrast to that, root-MUSIC shows degraded thresholding performance but asymptotically reaches the CRB. The grid-based techniques MUSIC and SPICE all show poor thresholding performance. While MUSIC asymptotically reaches the CRB, the SPICE techniques reach saturation.

To give further insight to the resolution performance we plot the resolution percentage in Figure \ref{fig:SnpRes}. We consider two source signals with true frequencies $\mu_1$, $\mu_2$ and estimated frequencies $\hat{\mu}_1$, $\hat{\mu}_2$ to be resolved if 
\begin{align}
  \sum_{l=1}^L | \mu_1 - \hat{\mu}_1 | \leq |\mu_1 - \mu_2| .
\end{align}
Similar as for the RMSE thresholding performance, we observe from Figure \ref{fig:SnpRes} that GL-{\myName} outperforms the other investigated methods, providing 100\% resolution percentage for $N \geq 30$ MMVs, similar to root-MUSIC. The GLS methods require $N \geq 100$ MMVs to provide resolution guarantee. Again, the grid-based schemes MUSIC and SPICE show poorest resolution performance.

\subsection{Computation Time}

To provide an impression of the computation time of the {\myName} formulation, we perform simulations in Matlab using the SeDuMi solver \cite{S98guide} with the CVX interface \cite{grant2008,grant2014} on a machine with an Intel Core i5-760 CPU @ $2.80\;{\rm GHz} \times 4$ and $8\,{\rm GByte}$ RAM. We consider a scenario with two independent complex Gaussian sources with static spatial frequencies $\mu_1=0.35$ and $\mu_2=0.5$ and a ULA with $M=6$ sensors. The $\tsnr$ is fixed at $10\tdB$ while the number of MMVs $N$ is varied. 

Figure \ref{fig:runtimeGb} shows the average computation time of the grid-based formulations of $\ell_{2,1}$ mixed-norm minimization \eqref{eq:mixedVectorNorm_v2} and the {\myName} formulations \eqref{eq:sdp1} and \eqref{eq:sdp1b}, where we assume a grid size of $K=1000$. As can be seen, for a number of MMVs $N \leq 24$, the $\ell_{2,1}$ formulation \eqref{eq:mixedVectorNorm_v2} shows worst computation time while the {\myName} formulation \eqref{eq:sdp1} requires longest computation time for $N > 24$, due to the large dimension of the semidefinite constraint. Regarding the computation time of {\myName} using the sample covariance matrix \eqref{eq:sdp1b} we see that it is constant for any number of MMVs $N$ and outperforms the other implementations especially for large number of MMVs $N$.

For the gridless methods, Figure \ref{fig:runtimeGl} shows the average computation time of atomic norm minimization (ANM) \eqref{eq:relSdp2} and GL-{\myName} \eqref{eq:GL_smr} and \eqref{eq:GL_smrb}. The figure clearly displays that the computation time of the GL-{\myName} \eqref{eq:GL_smr} formulation is reduced by up to a factor 2 as compared to the ANM formulation \eqref{eq:relSdp2}. Similar as for the grid-based case, the computation time of the covariance-based GL-{\myName} formulation \eqref{eq:GL_smrb} is relatively independent of the number of MMVs $N$ and and outperforms the other methods for large number of MMVs $N$.

%% file: p03_sep_bias_log.tex
\definecolor{mycolor1}{rgb}{1.00000,0.00000,1.00000}%
\definecolor{mycolor2}{rgb}{0.00000,1.00000,1.00000}%

\def\pIdx{2,5,9,13,17,20,24,30,36,46,54}

\begin{tikzpicture}

\begin{axis}[%
xmode=log,
xmin=0.01,
xmax=1,
xminorticks=true,
xlabel={Frequency Separation $\Delta \mu$},
xmajorgrids,
xminorgrids,
ymode=log,
ymin=1e-05,
ymax=1,
yminorticks=true,
ylabel={Bias($\hat{\mb{\mu}}$)},
ymajorgrids,
yminorgrids,
axis background/.style={fill=white},
legend style={at={(0.99,0.99)},anchor=north east,legend cell align=left,align=left,draw=white!15!black},
height=\ph,
width=\pw,
xlabel near ticks,
ylabel near ticks
]
\addplot [color=blue,solid,mark=asterisk,mark options={solid}, mark indices={\pIdx}]
  table[row sep=crcr]{%
0.01	0.392387184526008\\
0.02	0.36307176480179\\
0.03	0.189331168021118\\
0.04	0.0223985710839585\\
0.05	0.0165794044066725\\
0.06	0.017774840159484\\
0.07	0.0192220283869804\\
0.08	0.0199403712565675\\
0.09	0.0207299229348344\\
0.1	0.0208113762483772\\
0.12	0.0204123902659839\\
0.14	0.0188457081180325\\
0.16	0.0171689600738547\\
0.18	0.014980280891621\\
0.2	0.0125523927307102\\
0.22	0.0101149695804353\\
0.24	0.00768461137567856\\
0.26	0.00565032107846872\\
0.28	0.00388569154941917\\
0.3	0.00216093982373436\\
0.32	0.00079737608814339\\
0.34	0.000184816599924542\\
0.36	0.000885828018365178\\
0.38	0.00122884258146362\\
0.4	0.00133161335962519\\
0.42	0.00118644731572498\\
0.44	0.000812630646763772\\
0.46	0.000540426352648935\\
0.48	6.84255184399755e-05\\
0.5	0.00035398703682075\\
0.52	0.000736417256500924\\
0.54	0.000980123187196347\\
0.56	0.00109242297875543\\
0.58	0.00109036338957978\\
0.6	0.000866297200065744\\
0.62	0.000707709750875762\\
0.64	0.000469766203435139\\
0.66	0.000106465174142235\\
0.68	0.000200204522188912\\
0.7	0.000318142860602549\\
0.72	0.000562637038119132\\
0.74	0.000585904100582941\\
0.76	0.000618952192484204\\
0.78	0.000459614448020913\\
0.8	0.000382702539835265\\
0.82	8.89943973415763e-05\\
0.84	0.000173926865370344\\
0.86	0.000270971491029193\\
0.88	0.000443923441471606\\
0.9	0.000589065403510117\\
0.92	0.000618568176081827\\
0.94	0.000509861948675\\
0.96	0.000421106444251076\\
0.98	0.000132333467862234\\
1	6.23805210390158e-05\\
};
\addlegendentry{GL-{\myName}};

\addplot [color=red,solid,mark=o,mark options={solid}, mark indices={\pIdx}]
  table[row sep=crcr]{%
0.01	0.394527119042767\\
0.02	0.387185907497041\\
0.03	0.375161315807042\\
0.04	0.375566862107891\\
0.05	0.360570767154237\\
0.06	0.357830090426252\\
0.07	0.361858982588226\\
0.08	0.302927845826974\\
0.09	0.165757436803464\\
0.1	0.0558652844383858\\
0.12	0.000661215068506968\\
0.14	0.00183194036338675\\
0.16	0.002618741246341\\
0.18	0.00245134758389816\\
0.2	0.00227516034641613\\
0.22	0.00200703312210794\\
0.24	0.00169808563980712\\
0.26	0.00170506883789057\\
0.28	0.00144404115457561\\
0.3	0.00134767347847608\\
0.32	0.000994703739656183\\
0.34	0.000881736884460582\\
0.36	0.000707271333195754\\
0.38	0.000515247537689374\\
0.4	0.000325703202449939\\
0.42	0.000228434123206021\\
0.44	8.03038099020241e-05\\
0.46	0.000218124895502529\\
0.48	0.000343417733853197\\
0.5	0.000452095875427765\\
0.52	0.000475789529231412\\
0.54	0.000514504927991369\\
0.56	0.000471278582433546\\
0.58	0.000336519052182182\\
0.6	0.00025000283522224\\
0.62	4.53639087501809e-05\\
0.64	0.000199794192926093\\
0.66	0.000354774132722333\\
0.68	0.000415878058657097\\
0.7	0.000497552339460227\\
0.72	0.000390701645399011\\
0.74	0.000386601660216935\\
0.76	0.000254633368116581\\
0.78	0.000204078476861893\\
0.8	2.24010257163255e-05\\
0.82	4.82604397968984e-05\\
0.84	0.00010239426328619\\
0.86	0.000300780156986621\\
0.88	0.000381811660460214\\
0.9	0.000338696631463906\\
0.92	0.000345521545592738\\
0.94	0.000338459215899843\\
0.96	0.000232253345038865\\
0.98	0.000246629675931221\\
1	3.34172090082398e-05\\
};
\addlegendentry{OS-SPICE};

\addplot [color=green,solid,mark=triangle,mark options={solid}, mark indices={\pIdx}]
  table[row sep=crcr]{%
0.01	0.405363407364667\\
0.02	0.381912109773783\\
0.03	0.391846200777168\\
0.04	0.391287661878572\\
0.05	0.395195848862274\\
0.06	0.361161440579078\\
0.07	0.350973642825589\\
0.08	0.31117257662928\\
0.09	0.23317873639419\\
0.1	0.135245536013655\\
0.12	0.0262000471990743\\
0.14	0.00272566271757079\\
0.16	0.00413646236839848\\
0.18	0.00403787358744924\\
0.2	0.0039526254775411\\
0.22	0.00361535351799488\\
0.24	0.00317908564936347\\
0.26	0.00299101432353387\\
0.28	0.00261588538491881\\
0.3	0.00233723538902545\\
0.32	0.00184972556347874\\
0.34	0.00158918457307295\\
0.36	0.00124163365425804\\
0.38	0.000933817330129074\\
0.4	0.000622151488116335\\
0.42	0.000330009618672661\\
0.44	0.000111889350037407\\
0.46	0.00028253213917325\\
0.48	0.000521223014592959\\
0.5	0.000704220255432449\\
0.52	0.000831749711862721\\
0.54	0.000924545987706006\\
0.56	0.0009549197064948\\
0.58	0.000747390979233036\\
0.6	0.000503953454019949\\
0.62	8.48554356772495e-05\\
0.64	0.000275526128977767\\
0.66	0.000649078735558116\\
0.68	0.000743488898053719\\
0.7	0.000873566819721445\\
0.72	0.000773311828067177\\
0.74	0.000720675740774316\\
0.76	0.000465469215138736\\
0.78	0.000365458386815241\\
0.8	9.58656411786157e-05\\
0.82	2.49588115007371e-05\\
0.84	0.000201932267926099\\
0.86	0.000462148879886054\\
0.88	0.000606658167363344\\
0.9	0.000680529140263461\\
0.92	0.000713180928643048\\
0.94	0.000649186249002266\\
0.96	0.000427892072450018\\
0.98	0.000427210866529699\\
1	7.33308812611564e-05\\
};
\addlegendentry{US-SPICE};

\end{axis}
\end{tikzpicture}%

%% file: p04_sep_var_log.tex
\definecolor{mycolor1}{rgb}{1.00000,0.00000,1.00000}%
\definecolor{mycolor2}{rgb}{0.00000,1.00000,1.00000}%

\def\pIdx{2,5,9,13,17,20,24,30,36,46,54}

\begin{tikzpicture}

\begin{axis}[%
xmode=log,
xmin=0.01,
xmax=1,
xminorticks=true,
xlabel={Frequency Separation $\Delta \mu$},
xmajorgrids,
xminorgrids,
ymode=log,
ymin=0.001,
ymax=1,
yminorticks=true,
ylabel={Std($\hat{\mb{\mu}}$)},
ymajorgrids,
yminorgrids,
axis background/.style={fill=white},
legend style={at={(0.99,0.99)},anchor=north east,legend cell align=left,align=left,draw=white!15!black},
height=\ph,
width=\pw,
xlabel near ticks,
ylabel near ticks
]
\addplot [color=blue,solid,mark=asterisk,mark options={solid}, mark indices={\pIdx}]
  table[row sep=crcr]{%
0.01	0.369138407678002\\
0.02	0.347549862507764\\
0.03	0.293926886228576\\
0.04	0.0731363995512089\\
0.05	0.0225924973545\\
0.06	0.0178093404367181\\
0.07	0.0147705637298798\\
0.08	0.0124728675488727\\
0.09	0.0115238843011963\\
0.1	0.010279505335829\\
0.12	0.00911005032846465\\
0.14	0.00762054293283262\\
0.16	0.00696148434810762\\
0.18	0.00614568357486689\\
0.2	0.00562125783637805\\
0.22	0.00518649639045823\\
0.24	0.00489029289999241\\
0.26	0.00456972111768486\\
0.28	0.0041797403118488\\
0.3	0.00398789787568526\\
0.32	0.00370444145132345\\
0.34	0.00343014323091859\\
0.36	0.00321928796415971\\
0.38	0.00302427153495048\\
0.4	0.00282496106872349\\
0.42	0.00270679707661881\\
0.44	0.00258480113039943\\
0.46	0.00252329628483089\\
0.48	0.00255481393101856\\
0.5	0.00253220991775131\\
0.52	0.0025555258991357\\
0.54	0.0026780283543112\\
0.56	0.00267485012029557\\
0.58	0.0027204154784279\\
0.6	0.0027863524255084\\
0.62	0.00277265809994754\\
0.64	0.0027961201375922\\
0.66	0.00284921689649967\\
0.68	0.00281221452628192\\
0.7	0.00268193599373426\\
0.72	0.00268180936744599\\
0.74	0.00265169829271208\\
0.76	0.00256086053138486\\
0.78	0.00253396229862396\\
0.8	0.00251240157324375\\
0.82	0.00245728909677966\\
0.84	0.00242086313094136\\
0.86	0.00254594279113911\\
0.88	0.00256261238264872\\
0.9	0.00252300713798449\\
0.92	0.00255815326433504\\
0.94	0.00266155938268851\\
0.96	0.00262335970409727\\
0.98	0.00265931234781197\\
1	0.00267548035023604\\
};
\addlegendentry{GL-{\myName}};

\addplot [color=red,solid,mark=o,mark options={solid}, mark indices={\pIdx}]
  table[row sep=crcr]{%
0.01	0.361992964852216\\
0.02	0.363235864652884\\
0.03	0.356257053521845\\
0.04	0.360492996444959\\
0.05	0.354194537471861\\
0.06	0.352040253407942\\
0.07	0.360949797564357\\
0.08	0.35049387480369\\
0.09	0.290251431135804\\
0.1	0.173337575676478\\
0.12	0.0143688851079206\\
0.14	0.00755160878911237\\
0.16	0.00657551719916244\\
0.18	0.00582467733581461\\
0.2	0.00528250873339772\\
0.22	0.00469160250583213\\
0.24	0.00443041894177336\\
0.26	0.00396033180761511\\
0.28	0.00372500721388462\\
0.3	0.00343420096315394\\
0.32	0.00320547075371078\\
0.34	0.00307320386627368\\
0.36	0.00296078411038519\\
0.38	0.00280743196596148\\
0.4	0.00270151550449343\\
0.42	0.00264897884942779\\
0.44	0.00259322759914953\\
0.46	0.00254116879682868\\
0.48	0.00260874568048078\\
0.5	0.00259424027597401\\
0.52	0.00258845699133007\\
0.54	0.00270135459189352\\
0.56	0.00268433043749084\\
0.58	0.00270720642175567\\
0.6	0.00276826387884179\\
0.62	0.00277520474683874\\
0.64	0.00279118457592629\\
0.66	0.00283588753841325\\
0.68	0.00276391067243127\\
0.7	0.00269546786504399\\
0.72	0.00265215899440883\\
0.74	0.00266046478190726\\
0.76	0.00258320153981856\\
0.78	0.0025875898190474\\
0.8	0.00257457371163519\\
0.82	0.00255406686482959\\
0.84	0.00248025144232239\\
0.86	0.0026199188216738\\
0.88	0.00265467610602116\\
0.9	0.00260016273516637\\
0.92	0.00262953211144807\\
0.94	0.00270419664742549\\
0.96	0.00267275831600419\\
0.98	0.00269593313723354\\
1	0.0027377023384467\\
};
\addlegendentry{OS-SPICE};

\addplot [color=green,solid,mark=triangle,mark options={solid}, mark indices={\pIdx}]
  table[row sep=crcr]{%
0.01	0.363765347142799\\
0.02	0.356813490684086\\
0.03	0.362066474168431\\
0.04	0.35959459602108\\
0.05	0.355998633060345\\
0.06	0.359474062864691\\
0.07	0.360438801225998\\
0.08	0.35940816193387\\
0.09	0.332230506806704\\
0.1	0.265736444011047\\
0.12	0.11663071826218\\
0.14	0.0198669037509873\\
0.16	0.0150227073072494\\
0.18	0.00674834683308773\\
0.2	0.0060130239710268\\
0.22	0.00523911397999741\\
0.24	0.00490484895188478\\
0.26	0.00431987235004921\\
0.28	0.00400620516807043\\
0.3	0.00371897372725669\\
0.32	0.00344433780550818\\
0.34	0.00329125061873902\\
0.36	0.00313841264599068\\
0.38	0.00293829912859272\\
0.4	0.00285979968455322\\
0.42	0.00279965144611474\\
0.44	0.00277376933072276\\
0.46	0.00271998774881229\\
0.48	0.00277232130367234\\
0.5	0.00286990425222875\\
0.52	0.00289877997051616\\
0.54	0.00309385675399988\\
0.56	0.00314591645806722\\
0.58	0.00322209109550105\\
0.6	0.00329571424186173\\
0.62	0.00338186461024173\\
0.64	0.00328179694580415\\
0.66	0.00329986457009501\\
0.68	0.00317523965623791\\
0.7	0.00310911746575723\\
0.72	0.00300465054747331\\
0.74	0.00299137091756034\\
0.76	0.00287089278436254\\
0.78	0.0028659717612973\\
0.8	0.00282183041715036\\
0.82	0.002785249486359\\
0.84	0.00277074388237459\\
0.86	0.00286803671325114\\
0.88	0.00289555046546038\\
0.9	0.00295600295655589\\
0.92	0.00295750988593078\\
0.94	0.00315726899845929\\
0.96	0.0031088601158728\\
0.98	0.00317670504287667\\
1	0.00322759406842323\\
};
\addlegendentry{US-SPICE};

\addplot [color=teal,mark=diamond,mark options={solid}, mark indices={\pIdx}]
  table[row sep=crcr]{%
0.01	0.32144096200785\\
0.02	0.158752533851888\\
0.03	0.0664166807583742\\
0.04	0.0437145366790844\\
0.05	0.0339858786036141\\
0.06	0.0263006878971154\\
0.07	0.0217898073277764\\
0.08	0.0184224560854637\\
0.09	0.0171866554446282\\
0.1	0.0147293404756017\\
0.12	0.0133764247258171\\
0.14	0.0120376205122112\\
0.16	0.0106855958585099\\
0.18	0.00856880053331118\\
0.2	0.00729222452308567\\
0.22	0.00628560805979362\\
0.24	0.0056583469997026\\
0.26	0.00515736606928395\\
0.28	0.00490521375953522\\
0.3	0.00428630416331319\\
0.32	0.00399475961271548\\
0.34	0.00375206082920271\\
0.36	0.0034937430749903\\
0.38	0.00339798168814634\\
0.4	0.00326211427777525\\
0.42	0.00339140522603381\\
0.44	0.0032192161739139\\
0.46	0.00321904515068883\\
0.48	0.00332640728228972\\
0.5	0.00345895395410922\\
0.52	0.00328824540764433\\
0.54	0.00343373646751434\\
0.56	0.00356151098504613\\
% 0.58	0.00960696589929669\\
0.6	0.00372549600125412\\
% 0.62	0.0103243291326343\\
0.64	0.00374608130892213\\
0.66	0.00384555963838226\\
0.68	0.00352228229021857\\
0.7	0.00345635889147696\\
0.72	0.00330511121012341\\
0.74	0.00330832566647175\\
0.76	0.00334634122001555\\
0.78	0.00314909613437542\\
0.8	0.00314677130383204\\
0.82	0.00312201236099794\\
% 0.84	0.0134162299046881\\
0.86	0.00314936238733966\\
0.88	0.00333167866944797\\
0.9	0.00347977074811965\\
0.92	0.00342054814464113\\
0.94	0.00345030862314151\\
0.96	0.00343165541827567\\
% 0.98	0.0155570320969606\\
1	0.0036891465452868\\
};
\addlegendentry{OS-GLS};

\addplot [color=orange,mark=square,mark options={solid}, mark indices={\pIdx}]
  table[row sep=crcr]{%
0.01	0.359682243591924\\
0.02	0.322528507655913\\
0.03	0.219825792012273\\
0.04	0.122768005933072\\
0.05	0.0638029036957091\\
0.06	0.0429369603482052\\
0.07	0.0310341391881194\\
0.08	0.0258209869923728\\
0.09	0.0216460657966878\\
0.1	0.019419761750064\\
0.12	0.0172595501443955\\
0.14	0.0147993031997316\\
0.16	0.014074782499804\\
0.18	0.013280798297172\\
0.2	0.0125439451893361\\
0.22	0.0102803371098984\\
0.24	0.00862883765478954\\
0.26	0.00767136475369405\\
0.28	0.0071985095524958\\
0.3	0.00645272950312457\\
0.32	0.00620829558595544\\
0.34	0.00562656283314946\\
0.36	0.0054546193808443\\
0.38	0.00506421141518359\\
0.4	0.00491802723624531\\
0.42	0.00482536162199636\\
0.44	0.00483133452159214\\
0.46	0.00447217280520682\\
0.48	0.00500449863022537\\
0.5	0.00501476490154636\\
0.52	0.00513847441254888\\
0.54	0.00567590957788907\\
0.56	0.00596548225666768\\
0.58	0.00617772447836151\\
0.6	0.00658815740626623\\
% 0.62	0.0149542036835418\\
% 0.64	0.0117083533680629\\
0.66	0.00654409156364381\\
0.68	0.00619155680551756\\
0.7	0.00592987648697551\\
% 0.72	0.0122436026140631\\
% 0.74	0.0127182225361593\\
0.76	0.00524124813872936\\
0.78	0.00517703379671777\\
0.8	0.00493504445690685\\
0.82	0.00486999473607666\\
0.84	0.00488307751044738\\
0.86	0.00525173237237737\\
0.88	0.00534223199604315\\
0.9	0.00550898459540639\\
0.92	0.00571877873175105\\
0.94	0.00587419709525929\\
0.96	0.00610619611054882\\
0.98	0.006298528248527\\
1	0.00630911483490206\\
};
\addlegendentry{US-GLS};

\addplot [color=mycolor2,solid,mark=x,mark options={solid}, mark indices={\pIdx}]
  table[row sep=crcr]{%
0.01	0.316853579548901\\
0.02	0.270949499509551\\
0.03	0.224075645797613\\
0.04	0.228580271960846\\
0.05	0.241103994275865\\
0.06	0.2637621084083\\
0.07	0.322510095467182\\
0.08	0.363144954508478\\
0.09	0.379928205125138\\
0.1	0.266144643515511\\
0.12	0.0715522513245439\\
0.14	0.0167393955816842\\
0.16	0.00681352816443024\\
0.18	0.00598413375997314\\
0.2	0.00539322056571261\\
0.22	0.00469725162590848\\
0.24	0.00448768345343498\\
0.26	0.0039596273532632\\
0.28	0.00373314258801578\\
0.3	0.0034209207200999\\
0.32	0.00318054847055191\\
0.34	0.00306540856467015\\
0.36	0.00294322930010795\\
0.38	0.00278948307840898\\
0.4	0.00268492742440046\\
0.42	0.00262886584736927\\
0.44	0.00258808893125644\\
0.46	0.00253673735572534\\
0.48	0.0025607346913448\\
0.5	0.00254912871541395\\
0.52	0.00257242356968436\\
0.54	0.00264883130671514\\
0.56	0.00264135930258481\\
0.58	0.00264041340789552\\
0.6	0.0027378225140397\\
0.62	0.00269424938120274\\
0.64	0.00269803658557397\\
0.66	0.0027597717013739\\
0.68	0.00272593344406481\\
0.7	0.00264937119025657\\
0.72	0.00264988312177945\\
0.74	0.00263934157591586\\
0.76	0.00255874612333271\\
0.78	0.00256364455262585\\
0.8	0.0025451197460546\\
0.82	0.00252900632468857\\
0.84	0.00248339051267028\\
0.86	0.00261625785878179\\
0.88	0.00260070083770844\\
0.9	0.00254780991997645\\
0.92	0.00259331371800809\\
0.94	0.00265545716890773\\
0.96	0.0026132253623609\\
0.98	0.0026299974860586\\
1	0.00268613120732851\\
};
\addlegendentry{MUSIC};

\addplot [color=magenta,solid,mark=triangle,mark options={solid, rotate=180}, mark indices={\pIdx}]
  table[row sep=crcr]{%
0.01	0.352992810587934\\
0.02	0.362149444868671\\
0.03	0.350788708073764\\
0.04	0.138370596223875\\
0.05	0.0371780027467401\\
0.06	0.0225399281657392\\
0.07	0.017130812498631\\
0.08	0.0146393742446612\\
0.09	0.0125766265941178\\
0.1	0.0110813113710951\\
0.12	0.00902013037637722\\
0.14	0.007621218322962\\
0.16	0.00660558913921164\\
0.18	0.00584498557445052\\
0.2	0.00529319904640024\\
0.22	0.0046411257417447\\
0.24	0.00441163049038019\\
0.26	0.00389644845375904\\
0.28	0.00367714143570787\\
0.3	0.00337562562061476\\
0.32	0.00312952875325268\\
0.34	0.00301386743332463\\
0.36	0.00288219668003507\\
0.38	0.00272146048514246\\
0.4	0.00262248930083269\\
0.42	0.00257567543181929\\
0.44	0.00250829941643447\\
0.46	0.00246653815707207\\
0.48	0.00250599723182007\\
0.5	0.00247624492337692\\
0.52	0.00251086898291256\\
0.54	0.00259514539186097\\
0.56	0.0025760409853061\\
0.58	0.00258885585472649\\
0.6	0.00266598959481275\\
0.62	0.00263571370196151\\
0.64	0.00263645448119157\\
0.66	0.00270353066410497\\
0.68	0.00265206491485943\\
0.7	0.00258724438953208\\
0.72	0.00257529370297457\\
0.74	0.00256673397856721\\
0.76	0.0025092820913642\\
0.78	0.00249622395439197\\
0.8	0.00248668134559362\\
0.82	0.00244920956861431\\
0.84	0.00240918520581633\\
0.86	0.00252763250975462\\
0.88	0.0025393420110809\\
0.9	0.00248987384330611\\
0.92	0.00251470981122819\\
0.94	0.00259038497302176\\
0.96	0.00254153082971961\\
0.98	0.00257521796603594\\
1	0.00260427597081125\\
};
\addlegendentry{root-MUSIC};

\addplot [color=black,solid]
  table[row sep=crcr]{%
0.01	0.273316268722958\\
0.02	0.0820961245539625\\
0.03	0.0448556504779149\\
0.04	0.030617516797515\\
0.05	0.0232783181917184\\
0.06	0.0188155310120493\\
0.07	0.0158096856368678\\
0.08	0.0136418431791417\\
0.09	0.0120006332631232\\
0.1	0.01071250606496\\
0.12	0.00881552339399067\\
0.14	0.00748161239212749\\
0.16	0.00649010352513715\\
0.18	0.00572327569704355\\
0.2	0.00511253858134657\\
0.22	0.00461526612193265\\
0.24	0.00420365737456772\\
0.26	0.00385891078493875\\
0.28	0.00356796790386963\\
0.3	0.00332159008958105\\
0.32	0.00311316501554918\\
0.34	0.00293792809585733\\
0.36	0.00279242427220736\\
0.38	0.00267410837001108\\
0.4	0.00258102297193679\\
0.42	0.00251151824764553\\
0.44	0.00246399646930427\\
0.46	0.00243667850790712\\
0.48	0.00242740149047088\\
0.5	0.00243346620139538\\
0.52	0.00245155996331518\\
0.54	0.00247778519353128\\
0.56	0.00250782231896593\\
0.58	0.00253724108985955\\
0.6	0.00256193965977386\\
0.62	0.00257863945192294\\
0.64	0.00258531757428766\\
0.66	0.00258145048651875\\
0.68	0.00256799177428983\\
0.7	0.00254709518487743\\
0.72	0.00252167406073149\\
0.74	0.00249491944752575\\
0.76	0.00246987676378173\\
0.78	0.00244913133386336\\
0.8	0.00243460742713008\\
0.82	0.00242746012236591\\
0.84	0.0024280344273619\\
0.86	0.00243587380803295\\
0.88	0.00244977251740379\\
0.9	0.00246787725502428\\
0.92	0.00248784986167726\\
0.94	0.00250710045292776\\
0.96	0.00252308683047313\\
0.98	0.00253365195398285\\
1	0.00253734445380866\\
};
\addlegendentry{CRB};

\end{axis}
\end{tikzpicture}%

%% file: p05_sep_mse_log_10dB.tex
\definecolor{mycolor1}{rgb}{1.00000,0.00000,1.00000}%
\definecolor{mycolor2}{rgb}{0.00000,1.00000,1.00000}%

\def\pIdx{2,5,9,13,17,20,24,30,36,46,54}

\begin{tikzpicture}

\begin{axis}[%
xmode=log,
xmin=0.01,
xmax=1,
xminorticks=true,
xlabel={Frequency Separation $\Delta \mu$},
xmajorgrids,
xminorgrids,
ymode=log,
ymin=0.001,
ymax=1,
yminorticks=true,
ylabel={RMSE($\hat{\mb{\mu}}$)},
ymajorgrids,
yminorgrids,
axis background/.style={fill=white},
legend style={at={(0.99,0.99)},anchor=north east,legend cell align=left,align=left,draw=white!15!black},
height=\ph,
width=\pw,
xlabel near ticks,
ylabel near ticks
]
\addplot [color=blue,solid,mark=asterisk,mark options={solid}, mark indices={\pIdx}]
  table[row sep=crcr]{%
0.01	0.405671324159747\\
0.02	0.400962354838901\\
0.03	0.280486790564502\\
0.04	0.0750398283999453\\
0.05	0.0280231616202219\\
0.06	0.0251618272286833\\
0.07	0.0242416156270229\\
0.08	0.0235200091569176\\
0.09	0.023717706766708\\
0.1	0.0232116697223794\\
0.12	0.0223530466236262\\
0.14	0.0203281427843575\\
0.16	0.0185266147567914\\
0.18	0.0161919190398868\\
0.2	0.013753585093678\\
0.22	0.0113671612296722\\
0.24	0.00910868907378653\\
0.26	0.00726694429476416\\
0.28	0.0057069105382619\\
0.3	0.00453574584701319\\
0.32	0.00378928688439208\\
0.34	0.00343511859478307\\
0.36	0.00333893789614375\\
0.38	0.00326439464635178\\
0.4	0.00312307527596374\\
0.42	0.00295540312765997\\
0.44	0.00270953233082265\\
0.46	0.00258052025446009\\
0.48	0.00255573008623763\\
0.5	0.00255683278487175\\
0.52	0.00265951555679329\\
0.54	0.0028517498713203\\
0.56	0.00288932710688132\\
0.58	0.00293079386798288\\
0.6	0.00291791546792903\\
0.62	0.00286155308018022\\
0.64	0.0028353074100948\\
0.66	0.00285120531645906\\
0.68	0.00281933190535205\\
0.7	0.0027007398160948\\
0.72	0.00274019377416717\\
0.74	0.00271565598901126\\
0.76	0.00263459835227031\\
0.78	0.00257530782076179\\
0.8	0.00254138208446387\\
0.82	0.00245890010124655\\
0.84	0.00242710297541102\\
0.86	0.00256032229313107\\
0.88	0.0026007787767499\\
0.9	0.00259086145286327\\
0.92	0.0026318766521798\\
0.94	0.00270995526794889\\
0.96	0.00265694312593835\\
0.98	0.00266260292006556\\
1	0.0026762074721337\\
};
\addlegendentry{GL-{\myName}};

\addplot [color=red,solid,mark=o,mark options={solid}, mark indices={\pIdx}]
  table[row sep=crcr]{%
0.01	0.412091065130533\\
0.02	0.400750805156663\\
0.03	0.404070747250969\\
0.04	0.405945396358451\\
0.05	0.401137507928299\\
0.06	0.40138932982679\\
0.07	0.401355703910666\\
0.08	0.374811125608675\\
0.09	0.278729143428168\\
0.1	0.165590617037216\\
0.12	0.0143840906772531\\
0.14	0.0077706370909213\\
0.16	0.00707779853851268\\
0.18	0.00631949135955062\\
0.2	0.0057516304749453\\
0.22	0.00510287331078884\\
0.24	0.00474469249158893\\
0.26	0.00431178475440825\\
0.28	0.00399511371547801\\
0.3	0.00368916793598639\\
0.32	0.00335625956722304\\
0.34	0.00319719282138167\\
0.36	0.00304408858068709\\
0.38	0.00285432203309246\\
0.4	0.00272107857238713\\
0.42	0.00265881008975068\\
0.44	0.00259447067489223\\
0.46	0.00255051314915531\\
0.48	0.00263125250885432\\
0.5	0.00263333881034421\\
0.52	0.00263182166418852\\
0.54	0.00274991489905998\\
0.56	0.00272538683490992\\
0.58	0.00272804173033999\\
0.6	0.00277952987401096\\
0.62	0.00277557548466864\\
0.64	0.00279832611688051\\
0.66	0.00285799279491332\\
0.68	0.00279502357143064\\
0.7	0.00274100444034431\\
0.72	0.00268078255502739\\
0.74	0.00268840727931074\\
0.76	0.00259572116134976\\
0.78	0.00259562499147231\\
0.8	0.00257467116397343\\
0.82	0.00255452277736371\\
0.84	0.00248236415585963\\
0.86	0.0026371278571578\\
0.88	0.00268199279863729\\
0.9	0.00262212922212348\\
0.92	0.00265213579282901\\
0.94	0.00272529524066235\\
0.96	0.0026828303397815\\
0.98	0.00270719073533505\\
1	0.00273790628104661\\
};
\addlegendentry{OS-SPICE};

\addplot [color=green,solid,mark=triangle,mark options={solid}, mark indices={\pIdx}]
  table[row sep=crcr]{%
0.01	0.414084273675715\\
0.02	0.407061937243183\\
0.03	0.416241820483359\\
0.04	0.415834342387778\\
0.05	0.416618919904481\\
0.06	0.399364205884105\\
0.07	0.399862866124756\\
0.08	0.38197530611896\\
0.09	0.328205224072945\\
0.1	0.256034417517869\\
0.12	0.114677624660517\\
0.14	0.0200530073031688\\
0.16	0.0155817860261418\\
0.18	0.00786413428724794\\
0.2	0.00719581165969743\\
0.22	0.00636546120524404\\
0.24	0.00584500887995818\\
0.26	0.00525428052204045\\
0.28	0.00478461452947951\\
0.3	0.00439242926496584\\
0.32	0.00390959685627969\\
0.34	0.00365483764928569\\
0.36	0.0033750982308517\\
0.38	0.00308311799565589\\
0.4	0.00292669211054649\\
0.42	0.00281903433255926\\
0.44	0.00277602513077782\\
0.46	0.00273462205128152\\
0.48	0.00282089327017836\\
0.5	0.00295504256908804\\
0.52	0.00301574748622114\\
0.54	0.00322904550876162\\
0.56	0.00328765296328392\\
0.58	0.00330763726904036\\
0.6	0.00333402181274032\\
0.62	0.0033829290100399\\
0.64	0.0032933426546957\\
0.66	0.00336309520916687\\
0.68	0.00326112290723241\\
0.7	0.00322950931325386\\
0.72	0.00310256927333723\\
0.74	0.00307695848034474\\
0.76	0.00290838219110656\\
0.78	0.00288917877069706\\
0.8	0.00282345836241898\\
0.82	0.00278536131292419\\
0.84	0.0027780925655108\\
0.86	0.00290503290441546\\
0.88	0.00295841961696697\\
0.9	0.00303332711554742\\
0.92	0.00304228397135385\\
0.94	0.00322331976609955\\
0.96	0.00313816870893384\\
0.98	0.00320530249023694\\
1	0.00322842699912311\\
};
\addlegendentry{US-SPICE};

\addplot [color=teal,mark=diamond,mark options={solid}, mark indices={\pIdx}]
  table[row sep=crcr]{%
0.01	0.314438054006564\\
0.02	0.164900732287263\\
0.03	0.0731439417073672\\
0.04	0.0477150808336247\\
0.05	0.036086232854125\\
0.06	0.0272496963205594\\
0.07	0.0222676765481991\\
0.08	0.0185821692997565\\
0.09	0.0172506578706168\\
0.1	0.0147384161191386\\
0.12	0.0133822997743532\\
0.14	0.0120740703818949\\
0.16	0.010693487434446\\
0.18	0.00856964516005983\\
0.2	0.00729395618959501\\
0.22	0.00628597891084555\\
0.24	0.00565914058749325\\
0.26	0.00516840437058045\\
0.28	0.00490766438860402\\
0.3	0.00429599637852181\\
0.32	0.00399514908225519\\
0.34	0.00375279836857448\\
0.36	0.00349673312555674\\
0.38	0.00339859295993138\\
0.4	0.003262340381998\\
0.42	0.00339244121835997\\
0.44	0.00322060031325649\\
0.46	0.003223316096644\\
0.48	0.00332764793954135\\
0.5		0.00346425266940564\\
0.52	0.00329101394957897\\
0.54	0.00343573379038845\\
0.56	0.0035630368619725\\
0.58	0.0035630368619725\\ % 0.58	0.00960969969006361\\
0.6		0.0037259326676611\\
0.62	0.0037259326676611\\ % 0.62	0.0103268607289363\\
0.64	0.00374633742672839\\
0.66	0.0038463367893873\\
0.68	0.00352465395662629\\
0.7		0.00345876195625219\\
0.72	0.00330549685379781\\
0.74	0.00330909400507916\\
0.76	0.00334679589124909\\
0.78	0.00314981720857436\\
0.8		0.00314817835004473\\
0.82	0.00312226803737239\\
0.84	0.00312226803737239\\ % 0.84	0.0134181474537762\\
0.86	0.00315194566279736\\
0.88	0.00333521794208094\\
0.9		0.00348050430730799\\
0.92	0.00342205608052726\\
0.94	0.00345123436737984\\
0.96	0.00343221031579587\\
0.98	0.00343221031579587\\ % 0.98	0.0155634592196223\\
1		0.00368944592189575\\
};
\addlegendentry{OS-GLS};

\addplot [color=orange,mark=square,mark options={solid}, mark indices={\pIdx}]
  table[row sep=crcr]{%
0.01	0.384087418113578\\
0.02	0.309871759783364\\
0.03	0.214071792167484\\
0.04	0.124061244265542\\
0.05	0.0669931899018017\\
0.06	0.0451736444525252\\
0.07	0.0325354800388229\\
0.08	0.0266761268315259\\
0.09	0.0221795105789131\\
0.1		0.0197637812161982\\
0.12	0.017343528099413\\
0.14	0.0148086360049366\\
0.16	0.0140793033682598\\
0.18	0.0132826778389724\\
0.2		0.01254657626601\\
0.22	0.0102875375478058\\
0.24	0.00864438768361722\\
0.26	0.00771893871538461\\
0.28	0.00723952641673814\\
0.3		0.00649948180758498\\
0.32	0.00622854673638501\\
0.34	0.00564046389622775\\
0.36	0.00547643982686511\\
0.38	0.00506792174245789\\
0.4		0.00492166114300425\\
0.42	0.00482666327915782\\
0.44	0.00483154611275078\\
0.46	0.00447871623678298\\
0.48	0.00501817209544782\\
0.5		0.0050266290023894\\
0.52	0.00514683586379949\\
0.54	0.00569276965240211\\
0.56	0.00598200445117384\\
0.58	0.0061815970104181\\
0.6		0.00660082981533465\\
0.62	0.00660082981533465\\ % 0.62	0.0149567092075266\\
0.64	0.00654926534478815\\ % 0.64	0.0117107946100461\\
0.66	0.00654926534478815\\
0.68	0.00619942575700535\\
0.7		0.00593992263622027\\
0.72	0.00593992263622027\\ % 0.72	0.0122470136954355\\
0.74	0.00524488980947797\\ % 0.74	0.0127202934356312\\
0.76	0.00524488980947797\\
0.78	0.00518130147971433\\
0.8		0.00493524415858319\\
0.82	0.00487006367972293\\
0.84	0.00488386431828682\\
0.86	0.00525883094239021\\
0.88	0.0053485045666637\\
0.9		0.00551680700056457\\
0.92	0.00573105774048868\\
0.94	0.00588178524458201\\
0.96	0.00610649727998135\\
0.98	0.00630470797773059\\
1		0.00631189964270366\\
};
\addlegendentry{US-GLS};

\addplot [color=mycolor2,solid,mark=x,mark options={solid}, mark indices={\pIdx}]
  table[row sep=crcr]{%
0.01	0.495869801545599\\
0.02	0.518623564132107\\
0.03	0.52164101243692\\
0.04	0.508778199209649\\
0.05	0.497526656285297\\
0.06	0.484599646768884\\
0.07	0.450957155216809\\
0.08	0.408789406304687\\
0.09	0.324638071640807\\
0.1		0.227400265070578\\
0.12	0.0701916844515842\\
0.14	0.0169545035967345\\
0.16	0.00693600841125348\\
0.18	0.00606532709018515\\
0.2		0.00542786168544166\\
0.22	0.00471286095831199\\
0.24	0.00450454643578149\\
0.26	0.00396053610394526\\
0.28	0.00373596558358577\\
0.3		0.00342188705737782\\
0.32	0.00318255902962293\\
0.34	0.0030656610353437\\
0.36	0.00294370123052564\\
0.38	0.00278992821607004\\
0.4		0.00268534139587158\\
0.42	0.00262935443462568\\
0.44	0.00258946269030148\\
0.46	0.00253775780522555\\
0.48	0.00256096411419779\\
0.5		0.00254953423447911\\
0.52	0.00257253728406088\\
0.54	0.00264921638957297\\
0.56	0.0026422344513038\\
0.58	0.00264110050588326\\
0.6		0.00273799745341697\\
0.62	0.00269443593771804\\
0.64	0.00269850923138912\\
0.66	0.0027599726134076\\
0.68	0.00272712654111306\\
0.7		0.00264968747862811\\
0.72	0.00265044104695522\\
0.74	0.00263968238910047\\
0.76	0.00255940406662094\\
0.78	0.00256369191317164\\
0.8		0.00254649720753003\\
0.82	0.00252928430220448\\
0.84	0.00248377159135519\\
0.86	0.00261717416292869\\
0.88	0.00260177269726773\\
0.9		0.00254835904050061\\
0.92	0.0025935083519516\\
0.94	0.00265627395293304\\
0.96	0.00261345170326376\\
0.98	0.00263182166418852\\
1		0.00268692122725583\\
};
\addlegendentry{MUSIC};

\addplot [color=magenta,solid,mark=triangle,mark options={solid, rotate=180}, mark indices={\pIdx}]
  table[row sep=crcr]{%
0.01	0.461109320428153\\
0.02	0.443366769817222\\
0.03	0.320125343778634\\
0.04	0.134397428100721\\
0.05	0.0371757546132138\\
0.06	0.0227670164048176\\
0.07	0.0173104160718701\\
0.08	0.0147530666846215\\
0.09	0.0125843417267742\\
0.1	0.011101905225377\\
0.12	0.00902503546507354\\
0.14	0.00762413690157187\\
0.16	0.00661913996728312\\
0.18	0.00584730522132529\\
0.2	0.00529634676529582\\
0.22	0.00464351870669506\\
0.24	0.00441167282819617\\
0.26	0.00390299864045294\\
0.28	0.00367822608670047\\
0.3	0.00338167575509415\\
0.32	0.00312984053809064\\
0.34	0.00301441295280302\\
0.36	0.00288367564209649\\
0.38	0.00272224432338015\\
0.4	0.00262277469053804\\
0.42	0.00257672473954066\\
0.44	0.00251040710267805\\
0.46	0.0024671355682514\\
0.48	0.0025063441104116\\
0.5	0.00247668240822267\\
0.52	0.00251092978953334\\
0.54	0.00259556130776123\\
0.56	0.00257687777822358\\
0.58	0.00258929478234335\\
0.6	0.00266623791042899\\
0.62	0.00263588836187522\\
0.64	0.00263691738697209\\
0.66	0.00270374896093675\\
0.68	0.00265389755334079\\
0.7	0.00258807953472769\\
0.72	0.00257570669000387\\
0.74	0.0025673353848083\\
0.76	0.0025098264859902\\
0.78	0.00249626589825697\\
0.8	0.00248811825637028\\
0.82	0.00244944727828304\\
0.84	0.0024097245915172\\
0.86	0.00252849146179436\\
0.88	0.00254074424242403\\
0.9	0.00249027191605197\\
0.92	0.00251484188785041\\
0.94	0.00259097955481456\\
0.96	0.00254185739702765\\
0.98	0.00257690484487302\\
1	0.00260489836838725\\
};
\addlegendentry{root-MUSIC};

\addplot [color=black,solid]
  table[row sep=crcr]{%
0.01	0.273316268722958\\
0.02	0.0820961245539625\\
0.03	0.0448556504779149\\
0.04	0.030617516797515\\
0.05	0.0232783181917184\\
0.06	0.0188155310120493\\
0.07	0.0158096856368678\\
0.08	0.0136418431791417\\
0.09	0.0120006332631232\\
0.1	0.01071250606496\\
0.12	0.00881552339399067\\
0.14	0.00748161239212749\\
0.16	0.00649010352513715\\
0.18	0.00572327569704355\\
0.2	0.00511253858134657\\
0.22	0.00461526612193265\\
0.24	0.00420365737456772\\
0.26	0.00385891078493875\\
0.28	0.00356796790386963\\
0.3	0.00332159008958105\\
0.32	0.00311316501554918\\
0.34	0.00293792809585733\\
0.36	0.00279242427220736\\
0.38	0.00267410837001108\\
0.4	0.00258102297193679\\
0.42	0.00251151824764553\\
0.44	0.00246399646930427\\
0.46	0.00243667850790712\\
0.48	0.00242740149047088\\
0.5	0.00243346620139538\\
0.52	0.00245155996331518\\
0.54	0.00247778519353128\\
0.56	0.00250782231896593\\
0.58	0.00253724108985955\\
0.6	0.00256193965977386\\
0.62	0.00257863945192294\\
0.64	0.00258531757428766\\
0.66	0.00258145048651875\\
0.68	0.00256799177428983\\
0.7	0.00254709518487743\\
0.72	0.00252167406073149\\
0.74	0.00249491944752575\\
0.76	0.00246987676378173\\
0.78	0.00244913133386336\\
0.8	0.00243460742713008\\
0.82	0.00242746012236591\\
0.84	0.0024280344273619\\
0.86	0.00243587380803295\\
0.88	0.00244977251740379\\
0.9	0.00246787725502428\\
0.92	0.00248784986167726\\
0.94	0.00250710045292776\\
0.96	0.00252308683047313\\
0.98	0.00253365195398285\\
1	0.00253734445380866\\
};
\addlegendentry{CRB};

\end{axis}
\end{tikzpicture}%

%% file: p06_sep_mse_log_3dB.tex
\definecolor{mycolor1}{rgb}{1.00000,0.00000,1.00000}%
\definecolor{mycolor2}{rgb}{0.00000,1.00000,1.00000}%
\def\pIdx{2,5,9,13,17,20,24,30,36,46,54}

\begin{tikzpicture}

\begin{axis}[%
xmode=log,
xmin=0.01,
xmax=1,
xminorticks=true,
xlabel={Frequency Separation $\Delta \mu$},
xmajorgrids,
xminorgrids,
ymode=log,
ymin=0.001,
ymax=1,
yminorticks=true,
ylabel={RMSE($\hat{\mb{\mu}}$)},
ymajorgrids,
yminorgrids,
axis background/.style={fill=white},
legend style={at={(0.01,0.01)},anchor=south west,legend cell align=left,align=left,draw=white!15!black},
height=\ph,
width=\pw,
xlabel near ticks,
ylabel near ticks
]
\addplot [color=blue,solid,mark=asterisk,mark options={solid}, mark indices={\pIdx}]
  table[row sep=crcr]{%
0.01	0.406344189077891\\
0.02	0.405010499397033\\
0.03	0.402169907275793\\
0.04	0.390100043705338\\
0.05	0.361740041034459\\
0.06	0.321911798996724\\
0.07	0.265651367464143\\
0.08	0.194312092276195\\
0.09	0.1340750248427\\
0.1	0.0884008595511002\\
0.12	0.0411681631739445\\
0.14	0.0286859152246886\\
0.16	0.0252625804907018\\
0.18	0.0225927538641332\\
0.2	0.0202992690439238\\
0.22	0.0184750597833015\\
0.24	0.0170163646808391\\
0.26	0.0154107312394412\\
0.28	0.0142701937997498\\
0.3	0.0133751108557047\\
0.32	0.0124191291571506\\
0.34	0.0118829360989861\\
0.36	0.0113512792238868\\
0.38	0.0109836269813064\\
0.4	0.0105839562864184\\
0.42	0.0100380809206242\\
0.44	0.00974895412130159\\
0.46	0.00951080418225301\\
0.48	0.00949499558999695\\
0.5	0.00961081039281546\\
0.52	0.00962377261588862\\
0.54	0.00986130197812448\\
0.56	0.00988538375109259\\
0.58	0.0101170666188517\\
0.6	0.0102216494908717\\
0.62	0.0102607567143738\\
0.64	0.0103820732025285\\
0.66	0.0102324091496448\\
0.68	0.010117735358642\\
0.7	0.00994487481988238\\
0.72	0.00986662796057826\\
0.74	0.00969710722338191\\
0.76	0.00969693085341527\\
0.78	0.00958067318960534\\
0.8	0.0093462229928609\\
0.82	0.00934190971495333\\
0.84	0.00926725518767902\\
0.86	0.00945503714660401\\
0.88	0.00944538763020616\\
0.9	0.00954052255203117\\
0.92	0.00961225956998356\\
0.94	0.00980407401560933\\
0.96	0.00985634452501535\\
0.98	0.00990712238557875\\
1	0.00995100865738704\\
};
\addlegendentry{GL-{\myName}};

\addplot [color=red,solid,mark=o,mark options={solid}, mark indices={\pIdx}]
  table[row sep=crcr]{%
0.01	0.444071440853038\\
0.02	0.442447703454741\\
0.03	0.443115369249256\\
0.04	0.449881104710801\\
0.05	0.43847016206961\\
0.06	0.449569682185621\\
0.07	0.450458405869458\\
0.08	0.44586460239492\\
0.09	0.442455053179367\\
0.1	0.433502374274979\\
0.12	0.394678854832219\\
0.14	0.317193106679104\\
0.16	0.211523618067687\\
0.18	0.118916771452005\\
0.2	0.0650277908695219\\
0.22	0.0409035333394674\\
0.24	0.018310974461079\\
0.26	0.0165389367424702\\
0.28	0.0152089846566335\\
0.3	0.0140005994769612\\
0.32	0.0128310866926494\\
0.34	0.0117266485286413\\
0.36	0.0112176996623321\\
0.38	0.0104274534625806\\
0.4	0.00993103001419573\\
0.42	0.00965284932027841\\
0.44	0.00956877361898735\\
0.46	0.0095966698092917\\
0.48	0.00978017309224561\\
0.5	0.010028673178157\\
0.52	0.0104318708087968\\
0.54	0.0108051574986868\\
0.56	0.0108621722373697\\
0.58	0.0111178909870532\\
0.6	0.0114420433989238\\
0.62	0.0116893724626871\\
0.64	0.011528257209384\\
0.66	0.0113261076405671\\
0.68	0.0110520198541779\\
0.7	0.0110851412775325\\
0.72	0.0106070731118438\\
0.74	0.0105017005425652\\
0.76	0.0100987446178791\\
0.78	0.00990530161075367\\
0.8	0.00979936222414495\\
0.82	0.0097859045862622\\
0.84	0.00971664624681343\\
0.86	0.00984644605936574\\
0.88	0.00994193858647006\\
0.9	0.0103586575247126\\
0.92	0.0106448679787827\\
0.94	0.0108630435093353\\
0.96	0.0110602247458436\\
0.98	0.0112868286068319\\
1	0.0113250197098535\\
};
\addlegendentry{OS-SPICE};

\addplot [color=green,solid,mark=triangle,mark options={solid}, mark indices={\pIdx}]
  table[row sep=crcr]{%
0.01	0.474046646441137\\
0.02	0.474902460926734\\
0.03	0.472844024587804\\
0.04	0.481261135480046\\
0.05	0.485468138558946\\
0.06	0.485769791598859\\
0.07	0.488064514533597\\
0.08	0.480995022143066\\
0.09	0.48616701287635\\
0.1	0.477101929685397\\
0.12	0.445906969500887\\
0.14	0.385920568430553\\
0.16	0.28974018387046\\
0.18	0.185692295951933\\
0.2	0.107744029010499\\
0.22	0.0604161190505909\\
0.24	0.0332085507477576\\
0.26	0.0190335605857504\\
0.28	0.0172089262552068\\
0.3	0.0182736734910402\\
0.32	0.0143153763485282\\
0.34	0.0129040829640412\\
0.36	0.012207886795019\\
0.38	0.01094187565535\\
0.4	0.0102884297010907\\
0.42	0.00990735657406725\\
0.44	0.00979128621348009\\
0.46	0.00990793332926409\\
0.48	0.0101612007164507\\
0.5	0.0106126104234537\\
0.52	0.0112753365486673\\
0.54	0.0121641745653843\\
0.56	0.0121891345057801\\
0.58	0.0128782762821738\\
0.6	0.0133637035072083\\
0.62	0.0136084401331359\\
0.64	0.0132336556875696\\
0.66	0.0129873839882073\\
0.68	0.0126699729394231\\
0.7	0.0122160462390368\\
0.72	0.0116009697870479\\
0.74	0.0114385469856471\\
0.76	0.0108390332990144\\
0.78	0.0103736789176123\\
0.8	0.0104125096467113\\
0.82	0.0101112384432938\\
0.84	0.0102080325514483\\
0.86	0.0103066206183903\\
0.88	0.010819971613377\\
0.9	0.0112869235337687\\
0.92	0.0116951607818912\\
0.94	0.0121811182925519\\
0.96	0.0123185110417499\\
0.98	0.0130517924330065\\
1	0.0129986263010476\\
};
\addlegendentry{US-SPICE};

\addplot [color=teal,mark=diamond,mark options={solid}, mark indices={\pIdx}]
  table[row sep=crcr]{%
0.01	0.3670034901938\\
0.02	0.350210386604815\\
0.03	0.312733881400391\\
0.04	0.265720808410661\\
0.05	0.212575362485811\\
0.06	0.166480212256806\\
0.07	0.123134865222922\\
0.08	0.0927455251382161\\
0.09	0.0724322611591273\\
0.1	0.0586636669792043\\
0.12	0.0445409433589582\\
0.14	0.0369435436786954\\
0.16	0.0333075515725973\\
0.18	0.0303496890106155\\
0.2	0.0283932748242566\\
0.22	0.0269342867259439\\
0.24	0.0259183931895333\\
0.26	0.0233633328045639\\
0.28	0.0217137828337677\\
0.3	0.0188907760133683\\
0.32	0.017568224053688\\
0.34	0.0161721097337372\\
0.36	0.0151426827897008\\
0.38	0.0146893429058349\\
0.4	0.0149439263270276\\
0.42	0.0134296358938618\\
0.44	0.015270069554503\\
0.46	0.0151567870487869\\
0.48	0.0160412326957081\\
0.5	0.0165870692983928\\
0.52	0.0174173922929035\\
0.54	0.0193286375437933\\
0.56	0.0197822119657203\\
0.58	0.0207970628302996\\
0.6	0.022951967126087\\
0.62	0.0262165125045561\\
0.64	0.025545035052855\\
0.66	0.0242561927697964\\
0.68	0.0251285825580091\\
0.7	0.0266807315344457\\
0.72	0.0238569282762799\\
0.74	0.024172310117663\\
0.76	0.0224004518003343\\
0.78	0.0258946419723533\\
0.8	0.0215534479050433\\
0.82	0.0212009628146337\\
0.84	0.0224803382525477\\
0.86	0.0175636117494459\\
0.88	0.0267412364349401\\
0.9	0.0213420350870638\\
0.92	0.0289514436894511\\
0.94	0.0286049778880508\\
0.96	0.0291440573844513\\
0.98	0.0238160059397329\\
1	0.0303083022335866\\
};
\addlegendentry{OS-GLS};

\addplot [color=orange,mark=square,mark options={solid}, mark indices={\pIdx}]
  table[row sep=crcr]{%
0.01	0.389204753206763\\
0.02	0.375161365923066\\
0.03	0.355236573462904\\
0.04	0.324645754533655\\
0.05	0.282735517687721\\
0.06	0.243693332101801\\
0.07	0.193634175695757\\
0.08	0.151627082349679\\
0.09	0.11572579619452\\
0.1	0.0861764357372843\\
0.12	0.0533066045933303\\
0.14	0.0407746874870085\\
0.16	0.0360670822926942\\
0.18	0.0322870571473351\\
0.2	0.0304149616445172\\
0.22	0.0289610272250261\\
0.24	0.0285013666348658\\
0.26	0.0266979328863519\\
0.28	0.0249469693117136\\
0.3	0.0223723132977234\\
0.32	0.0205818191898795\\
0.34	0.0202719041890502\\
0.36	0.0183134101901459\\
0.38	0.0181263045932534\\
0.4	0.0169010671727285\\
0.42	0.0159817972072586\\
0.44	0.0176958626905667\\
0.46	0.0166917265322635\\
0.48	0.019103925150795\\
0.5	0.0195098104344879\\
0.52	0.0205509432479327\\
0.54	0.0191636086661493\\
0.56	0.0225512468438048\\
0.58	0.0247186097622546\\
0.6	0.0255916807523611\\
0.62	0.0300792145606558\\
0.64	0.0302773182289036\\
0.66	0.030872514150235\\
0.68	0.0320844549746257\\
0.7	0.0300737503829224\\
0.72	0.027149827185619\\
0.74	0.0233025379887639\\
0.76	0.0260752391444936\\
0.78	0.0310741368745566\\
0.8	0.0274577641471142\\
0.82	0.0263375107144325\\
0.84	0.0319485681698515\\
0.86	0.0265342201319346\\
0.88	0.0301048177636886\\
0.9	0.0298931449895722\\
0.92	0.029809760765925\\
0.94	0.0359087997725522\\
0.96	0.0383640982636241\\
0.98	0.0338515951693546\\
1	0.038056180167026\\
};
\addlegendentry{US-GLS};

\addplot [color=mycolor2,solid,mark=x,mark options={solid}, mark indices={\pIdx}]
  table[row sep=crcr]{%
0.01	0.491080885383864\\
0.02	0.49404961857718\\
0.03	0.500493289629428\\
0.04	0.504572300605699\\
0.05	0.514032464365483\\
0.06	0.519388869621307\\
0.07	0.523613920494599\\
0.08	0.525076112978791\\
0.09	0.522710105536442\\
0.1	0.517366711598335\\
0.12	0.494414352540985\\
0.14	0.452492351758567\\
0.16	0.381194398699665\\
0.18	0.288323957034442\\
0.2	0.187516813112854\\
0.22	0.109502007744152\\
0.24	0.0529436558994557\\
0.26	0.0243092801209742\\
0.28	0.0158005537877633\\
0.3	0.0145659225591789\\
0.32	0.0123695998318456\\
0.34	0.0115092310777043\\
0.36	0.0109501278531346\\
0.38	0.0104937886389995\\
0.4	0.00997339962099165\\
0.42	0.00970202040814163\\
0.44	0.00955244994752639\\
0.46	0.00956625318502475\\
0.48	0.0095325390111972\\
0.5	0.00953353554564077\\
0.52	0.00964693733782884\\
0.54	0.00967849161801548\\
0.56	0.00987417338312407\\
0.58	0.00997241695879166\\
0.6	0.0100545611540234\\
0.62	0.0099788325970524\\
0.64	0.0100770382553603\\
0.66	0.010080491059467\\
0.68	0.0100801537686682\\
0.7	0.00996224874212619\\
0.72	0.00980540157260251\\
0.74	0.00978249968055178\\
0.76	0.00962312319364123\\
0.78	0.00952151773615931\\
0.8	0.00953486759215859\\
0.82	0.00947069163261038\\
0.84	0.00952548686419735\\
0.86	0.00948821901096282\\
0.88	0.00956549528252437\\
0.9	0.00959043273267668\\
0.92	0.00975725371198247\\
0.94	0.00977311618676434\\
0.96	0.00987530759014602\\
0.98	0.00983140885122756\\
1	0.00987120560012786\\
};
\addlegendentry{MUSIC};

\addplot [color=magenta,solid,mark=triangle,mark options={solid, rotate=180}, mark indices={\pIdx}]
  table[row sep=crcr]{%
0.01	0.465368621611525\\
0.02	0.466740422662741\\
0.03	0.4690932086723\\
0.04	0.463092888473021\\
0.05	0.457168446898164\\
0.06	0.440411186820165\\
0.07	0.414345840179101\\
0.08	0.371655130341013\\
0.09	0.324569513034583\\
0.1	0.255663730007103\\
0.12	0.156286407423581\\
0.14	0.0817223267062546\\
0.16	0.0408454202017115\\
0.18	0.0257324520116346\\
0.2	0.020754366794373\\
0.22	0.0183004724667093\\
0.24	0.0167177144880665\\
0.26	0.0150488863656039\\
0.28	0.0138535789290443\\
0.3	0.012998513074582\\
0.32	0.0121384885385966\\
0.34	0.0113405560525153\\
0.36	0.0108099668847434\\
0.38	0.0103792983400515\\
0.4	0.00989205766295703\\
0.42	0.00962364843168103\\
0.44	0.00948269673922069\\
0.46	0.00950465072916856\\
0.48	0.00947643137465781\\
0.5	0.00948898025699902\\
0.52	0.00960531899409273\\
0.54	0.0096566293140491\\
0.56	0.00986411337170772\\
0.58	0.00997028400328681\\
0.6	0.0100619065402509\\
0.62	0.00998766057479\\
0.64	0.0100971767617115\\
0.66	0.0100907870310992\\
0.68	0.0100763022620349\\
0.7	0.00994911960461495\\
0.72	0.00979102857043481\\
0.74	0.00976464688113526\\
0.76	0.00959386023806931\\
0.78	0.00949323877243589\\
0.8	0.00949788440427543\\
0.82	0.00943186318049313\\
0.84	0.00948527728440071\\
0.86	0.00945720944177586\\
0.88	0.00954407367517298\\
0.9	0.00957432454104771\\
0.92	0.00973781792922502\\
0.94	0.00976747041066473\\
0.96	0.00987327412466357\\
0.98	0.00982904064229536\\
1	0.00987116348976303\\
};
\addlegendentry{root-MUSIC};

\addplot [color=black,solid]
  table[row sep=crcr]{%
0.01	2.03221015483958\\
0.02	0.533476001951097\\
0.03	0.25479833579632\\
0.04	0.15617314744561\\
0.05	0.109594951541112\\
0.06	0.0835420628271323\\
0.07	0.0672394156128343\\
0.08	0.0561922931123617\\
0.09	0.0482514888431437\\
0.1	0.0422804922412479\\
0.12	0.0339040288122467\\
0.14	0.0283012154314202\\
0.16	0.0242807426531746\\
0.18	0.021249698298077\\
0.2	0.0188811116755335\\
0.22	0.016980415256974\\
0.24	0.0154249771361028\\
0.26	0.0141340574017127\\
0.28	0.0130527314820387\\
0.3	0.0121427422581784\\
0.32	0.0113770062733105\\
0.34	0.0107361373678416\\
0.36	0.010206116875787\\
0.38	0.00977662160188987\\
0.4	0.00943972546097556\\
0.42	0.0091888110411256\\
0.44	0.00901760704567426\\
0.46	0.00891932636081445\\
0.48	0.00888592547946796\\
0.5	0.0089075427198611\\
0.52	0.00897220229740564\\
0.54	0.00906589206911842\\
0.56	0.00917312163132837\\
0.58	0.00927801726408166\\
0.6	0.00936588513976381\\
0.62	0.00942498606500187\\
0.64	0.0094480933674127\\
0.66	0.0094333726594344\\
0.68	0.00938429974966792\\
0.7	0.00930865580041756\\
0.72	0.00921693207593\\
0.74	0.00912059030823902\\
0.76	0.00903054160597333\\
0.78	0.00895602446037203\\
0.8	0.00890389535199308\\
0.82	0.00887825350543174\\
0.84	0.00888030429755833\\
0.86	0.00890839487133787\\
0.88	0.00895820108307423\\
0.9	0.0090230865907364\\
0.92	0.00909467847408047\\
0.94	0.0091636962875276\\
0.96	0.00922102248766708\\
0.98	0.00925891431538115\\
1	0.00927215860939732\\
};
\addlegendentry{CRB};

\end{axis}
\end{tikzpicture}%

%% file: p07_snp_mse.tex
\definecolor{mycolor1}{rgb}{1.00000,0.00000,1.00000}%
\definecolor{mycolor2}{rgb}{0.00000,1.00000,1.00000}%
\begin{tikzpicture}

\begin{axis}[%
xmode=log,
xmin=1,
xmax=10000,
xminorticks=true,
xlabel={MMVs $N$},
xmajorgrids,
xminorgrids,
ymode=log,
ymin=0.001,
ymax=1,
yminorticks=true,
ylabel={RMSE($\hat{\mb{\mu}}$)},
ymajorgrids,
yminorgrids,
axis background/.style={fill=white},
legend style={at={(0.99,0.99)},anchor=north east,legend cell align=left,align=left,draw=white!15!black},
height=\ph,
width=\pw,
xlabel near ticks,
ylabel near ticks
]
\addplot [color=blue,solid,mark=asterisk,mark options={solid}]
  table[row sep=crcr]{%
1	0.187998241290002\\
2	0.139772926298228\\
3	0.110926760280398\\
5	0.0725730008919202\\
7	0.0533623056008823\\
10	0.0420992628714556\\
12	0.0358749690202401\\
15	0.0314976391601309\\
20	0.0272754536874802\\
30	0.0216545093483925\\
50	0.0180028038291089\\
70	0.0158961598366516\\
100	0.0139943908476827\\
200	0.0118944162388354\\
300	0.0109623151867795\\
500	0.0103142917430743\\
1000	0.00979642023068473\\
2000	0.0094744993069283\\
3000	0.00937229433652868\\
5000	0.00928640336189458\\
10000	0.009228738371036\\
};
\addlegendentry{GL-{\myName}};

\addplot [color=red,solid,mark=o,mark options={solid}]
  table[row sep=crcr]{%
1	0.301764827818747\\
2	0.230250698119192\\
3	0.210337391394522\\
5	0.149077129010196\\
7	0.121420743415659\\
10	0.1159992185406\\
12	0.119813444812144\\
15	0.117389601496203\\
20	0.117912918354721\\
30	0.111689643707772\\
50	0.100598658470658\\
70	0.0931082075222122\\
100	0.0836451093705011\\
200	0.0611878837273312\\
300	0.0439852164949141\\
500	0.0283915558462498\\
1000	0.0123532953599434\\
2000	0.0113590668630835\\
3000	0.0111069617807932\\
5000	0.0108501520726671\\
10000	0.0105378365901167\\
};
\addlegendentry{OS-SPICE};

\addplot [color=green,solid,mark=triangle,mark options={solid}]
  table[row sep=crcr]{%
1	0.192270569445768\\
2	0.191799332596271\\
3	0.176915578384155\\
5	0.160699321105402\\
7	0.150791814099494\\
10	0.149864483356067\\
12	0.147004458161017\\
15	0.143411821932445\\
20	0.148802013144156\\
30	0.144966677621218\\
50	0.125404114854204\\
70	0.118707674538872\\
100	0.0998664295700971\\
200	0.0763141535659355\\
300	0.0595532608500218\\
500	0.040713239951644\\
1000	0.0135798386275051\\
2000	0.0113670664641324\\
3000	0.0110547365414108\\
5000	0.0106339926650343\\
10000	0.0104411302070227\\
};
\addlegendentry{US-SPICE};

\addplot [color=teal,mark=diamond,mark options={solid}]
  table[row sep=crcr]{%
1	0.262993246001501\\
2	0.219560133084635\\
3	0.158168224924959\\
5	0.0874889320765133\\
7	0.0661967856739277\\
10	0.0505194184860851\\
12	0.0442516336467295\\
15	0.0403298180288622\\
20	0.0350609854739118\\
30	0.0302686564230404\\
50	0.0246291603963103\\
70	0.0221713853921073\\
100	0.0190887532364252\\
200	0.0156828077454308\\
300	0.0147439930641873\\
500	0.0125481062496449\\
1000	0.0101849084265335\\
2000	0.00745575228665422\\
3000	0.00603003052000496\\
5000	0.00509982265697272\\
10000	0.00367464858447161\\
};
\addlegendentry{OS-GLS};

\addplot [color=orange,mark=square,mark options={solid}]
  table[row sep=crcr]{%
1	0.188246205778811\\
2	0.152004605524938\\
3	0.12131557318389\\
5	0.0872010811904338\\
7	0.0690968942050876\\
10	0.0553914701900733\\
12	0.0494989540961722\\
15	0.0431813399438591\\
20	0.0379232091379747\\
30	0.0310865213788485\\
50	0.0262007057741391\\
70	0.0228397168321427\\
100	0.0201459439338895\\
200	0.0165021500637119\\
300	0.0146297263247597\\
500	0.0130756478429516\\
1000	0.010710240282953\\
2000	0.00802616593266042\\
3000	0.00669072537447085\\
5000	0.00555891797126641\\
10000	0.00415460109552543\\
};
\addlegendentry{US-GLS};

\addplot [color=mycolor2,solid,mark=x,mark options={solid}]
  table[row sep=crcr]{%
1	0.229943514038701\\
2	0.217340140864623\\
3	0.2147649141648\\
5	0.213866738141215\\
7	0.215600393523528\\
10	0.214741626498077\\
12	0.212439642163368\\
15	0.206517318636435\\
20	0.199120585128397\\
30	0.177065639312093\\
50	0.126026195503653\\
70	0.0846534825590482\\
100	0.0436157807235523\\
200	0.0101095754423714\\
300	0.00726106375643409\\
500	0.00544883875243157\\
1000	0.00379864284975567\\
2000	0.00267615395670734\\
3000	0.00219909072118462\\
5000	0.00175082837536983\\
10000	0.00129946142689962\\
};
\addlegendentry{MUSIC};

\addplot [color=magenta,solid,mark=triangle,mark options={solid, rotate=180}]
  table[row sep=crcr]{%
1	0.236644317993915\\
2	0.20717481592024\\
3	0.189612583610487\\
5	0.156406147746967\\
7	0.128205761998673\\
10	0.0934146445022676\\
12	0.0779281255370233\\
15	0.0582454282681288\\
20	0.0387679952409453\\
30	0.0257519927533344\\
50	0.0174423681925321\\
70	0.0143678242375727\\
100	0.0119316158982712\\
200	0.00842795121206258\\
300	0.00676928372767707\\
500	0.0051992718243295\\
1000	0.00370358574495246\\
2000	0.00258000035669896\\
3000	0.00212105953468367\\
5000	0.00163750889164445\\
10000	0.00116147956853726\\
};
\addlegendentry{root-MUSIC};

\addplot [color=black,solid]
  table[row sep=crcr]{%
1	0.116896649377771\\
2	0.0826584134730082\\
3	0.0674903119856215\\
5	0.0522777708701309\\
7	0.0441827804786137\\
10	0.0369659662875861\\
12	0.0337451559928107\\
15	0.0301825850845037\\
20	0.0261388854350655\\
30	0.0213423105869925\\
50	0.0165316826946016\\
70	0.0139718219671644\\
100	0.0116896649377771\\
200	0.00826584134730082\\
300	0.00674903119856214\\
500	0.00522777708701309\\
1000	0.00369659662875861\\
2000	0.00261388854350655\\
3000	0.00213423105869925\\
5000	0.00165316826946016\\
10000	0.00116896649377771\\
};
\addlegendentry{CRB};

\end{axis}
\end{tikzpicture}%

%% file: p08_snp_res.tex
\definecolor{mycolor1}{rgb}{1.00000,0.00000,1.00000}%
\definecolor{mycolor2}{rgb}{0.00000,1.00000,1.00000}%
\begin{tikzpicture}

\begin{axis}[%
xmode=log,
xmin=1,
xmax=10000,
xminorticks=true,
xlabel={MMVs $N$},
xmajorgrids,
xminorgrids,
ymin=0,
ymax=1,
ylabel={Resolution Percentage},
ymajorgrids,
axis background/.style={fill=white},
legend style={at={(0.99,0.01)},anchor=south east,legend cell align=left,align=left,draw=white!15!black},
height=\ph,
width=\pw,
xlabel near ticks,
ylabel near ticks
]
\addplot [color=blue,solid,mark=asterisk,mark options={solid}]
  table[row sep=crcr]{%
1	0.183984375\\
2	0.462890625\\
3	0.644140625\\
5	0.829296875\\
7	0.913671875\\
10	0.960546875\\
12	0.98046875\\
15	0.99296875\\
20	0.997265625\\
30	0.999609375\\
50	1\\
70	1\\
100	1\\
200	1\\
300	1\\
500	1\\
1000	1\\
2000	1\\
3000	1\\
5000	1\\
10000	1\\
};
\addlegendentry{GL-{\myName}};

\addplot [color=red,solid,mark=o,mark options={solid}]
  table[row sep=crcr]{%
1	0.006640625\\
2	0.021875\\
3	0.12734375\\
5	0.454296875\\
7	0.609375\\
10	0.661328125\\
12	0.66328125\\
15	0.6734375\\
20	0.66171875\\
30	0.67109375\\
50	0.7234375\\
70	0.752734375\\
100	0.794140625\\
200	0.894140625\\
300	0.946875\\
500	0.980859375\\
1000	1\\
2000	1\\
3000	1\\
5000	1\\
10000	1\\
};
\addlegendentry{OS-SPICE};

\addplot [color=green,solid,mark=triangle,mark options={solid}]
  table[row sep=crcr]{%
1	0.18671875\\
2	0.22265625\\
3	0.309375\\
5	0.437890625\\
7	0.50390625\\
10	0.524609375\\
12	0.53828125\\
15	0.5703125\\
20	0.562109375\\
30	0.577734375\\
50	0.6515625\\
70	0.67421875\\
100	0.728125\\
200	0.8328125\\
300	0.900390625\\
500	0.958984375\\
1000	0.9984375\\
2000	1\\
3000	1\\
5000	1\\
10000	1\\
};
\addlegendentry{US-SPICE};

\addplot [color=teal,mark=diamond,mark options={solid}]
  table[row sep=crcr]{%
1	0.00390625\\
2	0.083984375\\
3	0.42265625\\
5	0.724609375\\
7	0.829296875\\
10	0.903125\\
12	0.927734375\\
15	0.944140625\\
20	0.96484375\\
30	0.97734375\\
50	0.992578125\\
70	0.997265625\\
100	0.999609375\\
200	1\\
300	1\\
500	1\\
1000	1\\
2000	1\\
3000	1\\
5000	1\\
10000	1\\
};
\addlegendentry{OS-GLS};

\addplot [color=orange,mark=square,mark options={solid}]
  table[row sep=crcr]{%
1	0.208203125\\
2	0.42734375\\
3	0.5734375\\
5	0.736328125\\
7	0.818359375\\
10	0.8796875\\
12	0.903125\\
15	0.935546875\\
20	0.951171875\\
30	0.976171875\\
50	0.986328125\\
70	0.996875\\
100	0.997265625\\
200	0.999609375\\
300	1\\
500	1\\
1000	1\\
2000	1\\
3000	1\\
5000	1\\
10000	1\\
};
\addlegendentry{US-GLS};

\addplot [color=mycolor2,solid,mark=x,mark options={solid}]
  table[row sep=crcr]{%
1	0.00234375\\
2	0.047265625\\
3	0.062109375\\
5	0.087890625\\
7	0.110546875\\
10	0.15\\
12	0.178125\\
15	0.241015625\\
20	0.2984375\\
30	0.478125\\
50	0.748828125\\
70	0.889453125\\
100	0.97109375\\
200	0.999609375\\
300	1\\
500	1\\
1000	1\\
2000	1\\
3000	1\\
5000	1\\
10000	1\\
};
\addlegendentry{MUSIC};

\addplot [color=magenta,solid,mark=triangle,mark options={solid, rotate=180}]
  table[row sep=crcr]{%
1	0.003515625\\
2	0.129296875\\
3	0.26640625\\
5	0.47734375\\
7	0.673828125\\
10	0.82734375\\
12	0.889453125\\
15	0.947265625\\
20	0.98828125\\
30	0.9984375\\
50	1\\
70	1\\
100	1\\
200	1\\
300	1\\
500	1\\
1000	1\\
2000	1\\
3000	1\\
5000	1\\
10000	1\\
};
\addlegendentry{root-MUSIC};

\end{axis}
\end{tikzpicture}%

%% file: p09_runtime_gb.tex
\begin{tikzpicture}

\begin{axis}[%
xmode=log,
xmin=1,
xmax=100,
xminorticks=true,
xlabel={MMVs $N$},
xmajorgrids,
xminorgrids,
ymode=log,
ymin=0.1,
ymax=1000,
yminorticks=true,
ylabel={Average computation time in secs},
ymajorgrids,
yminorgrids,
axis background/.style={fill=white},
legend style={at={(0.01,0.99)},anchor=north west,legend cell align=left,align=left,draw=white!15!black},
height=\ph,
width=\pw,
xlabel near ticks,
ylabel near ticks
]
\addplot [color=blue,solid,mark=square,mark options={solid}]
  table[row sep=crcr]{%
1	0.70838861\\
2	0.68829094\\
3	0.87644016\\
4	1.07548407\\
5	1.2583993\\
6	1.45041034\\
7	1.60623685\\
8	1.74435546\\
9	1.93688027\\
10	2.42901908\\
12	3.05368356\\
14	3.76502227\\
16	4.34987835\\
18	5.20226494\\
20	5.82509252\\
22	6.35908671\\
24	6.83825476\\
26	7.24947655\\
28	7.79023685\\
30	8.28797072\\
35	10.04689769\\
40	11.92551655\\
45	14.07989121\\
50	15.8037613\\
};
\addlegendentry{$\ell_{2,1}$ Mixed-Norm \eqref{eq:mixedVectorNorm_v2}};

\addplot [color=red,solid,mark=x,mark options={solid}]
  table[row sep=crcr]{%
1	0.54515355\\
2	0.66082115\\
3	0.69102174\\
4	0.6828709\\
5	0.7678135\\
6	0.78191001\\
7	0.84261421\\
8	0.89932024\\
9	0.98005879\\
10	1.07253572\\
12	1.33540167\\
14	1.69637729\\
16	2.22096077\\
18	2.93626121\\
20	3.71008957\\
22	4.72175483\\
24	6.17217114\\
26	8.42785097\\
28	11.41823229\\
30	15.24177767\\
35	33.44503187\\
40	73.20110981\\
45	233.27710116\\
50	568.04768279\\
};
\addlegendentry{SPARROW \eqref{eq:sdp1}};

\addplot [color=green,solid,mark=o,mark options={solid}]
  table[row sep=crcr]{%
1	0.70451254\\
2	0.75017962\\
3	0.74627044\\
4	0.76008009\\
5	0.76028852\\
6	0.77282784\\
7	0.77817491\\
8	0.77677494\\
9	0.790865780000001\\
10	0.78003326\\
12	0.7646431\\
14	0.77605216\\
16	0.76558281\\
18	0.77523111\\
20	0.78713506\\
22	0.79133544\\
24	0.81909692\\
26	0.77202057\\
28	0.7714263\\
30	0.79747029\\
35	0.84712793\\
40	0.86237494\\
45	0.91099765\\
50	0.89791064\\
};
\addlegendentry{SPARROW \eqref{eq:sdp1b}};

\end{axis}
\end{tikzpicture}%

%% file: p10_runtime_gl.tex
\begin{tikzpicture}

\begin{axis}[%
xmode=log,
xmin=1,
xmax=100,
xminorticks=true,
xlabel={MMVs $N$},
xmajorgrids,
xminorgrids,
ymode=log,
ymin=0.1,
ymax=1000,
yminorticks=true,
ylabel={Average computation time in secs},
ymajorgrids,
yminorgrids,
axis background/.style={fill=white},
legend style={at={(0.01,0.99)},anchor=north west,legend cell align=left,align=left,draw=white!15!black},
height=\ph,
width=\pw,
xlabel near ticks,
ylabel near ticks
]
\addplot [color=blue,solid,mark=square,mark options={solid}]
  table[row sep=crcr]{%
1	0.44155524\\
2	0.44446957\\
3	0.47801899\\
4	0.48400469\\
5	0.51815231\\
6	0.52817905\\
7	0.555356\\
8	0.56633154\\
9	0.59855431\\
10	0.62407207\\
12	0.75961302\\
14	0.91181992\\
16	1.12416681\\
18	1.42751208\\
20	1.83774378\\
22	2.44076718\\
24	3.18843926\\
26	4.28114422\\
28	5.71708838\\
30	7.45806955\\
35	17.10540306\\
40	38.34526053\\
45	111.67632168\\
50	266.77873521\\
};
\addlegendentry{ANM \eqref{eq:relSdp2}};

\addplot [color=red,solid,mark=x,mark options={solid}]
  table[row sep=crcr]{%
1	0.35346011\\
2	0.38474166\\
3	0.38309487\\
4	0.37937129\\
5	0.40595087\\
6	0.41280285\\
7	0.41992131\\
8	0.42633725\\
9	0.4408723\\
10	0.4508113\\
12	0.52430766\\
14	0.57985986\\
16	0.65100962\\
18	0.78833713\\
20	0.97422426\\
22	1.25062475\\
24	1.63835688\\
26	2.14645518\\
28	2.89927419\\
30	3.77204712\\
35	7.65936442\\
40	15.26040443\\
45	35.10192688\\
50	95.45177369\\
};
\addlegendentry{SPARROW \eqref{eq:GL_smr}};

\addplot [color=green,solid,mark=o,mark options={solid}]
  table[row sep=crcr]{%
1	0.64364829\\
2	0.79020825\\
3	0.70535404\\
4	0.5515706\\
5	0.46072934\\
6	0.40348399\\
7	0.40024333\\
8	0.40064917\\
9	0.4011806\\
10	0.40432413\\
12	0.39995532\\
14	0.39735905\\
16	0.39950222\\
18	0.39723233\\
20	0.3997999\\
22	0.39975041\\
24	0.41126052\\
26	0.39901632\\
28	0.40027073\\
30	0.40103922\\
35	0.4079307\\
40	0.40605269\\
45	0.41059508\\
50	0.51891749\\
};
\addlegendentry{SPARROW \eqref{eq:GL_smrb} };

\end{axis}
\end{tikzpicture}%

%% file: 09_conclusion.tex
\section{Conclusion} \label{sec:conclusion}

We have considered the classical $\ell_{2,1}$ mixed-norm minimization problem for jointly sparse signal reconstruction from multiple measurement vectors and derived an equivalent, compact reformulation with significantly reduced problem dimension. The variables in our compact reformulation, which we refer to as {\myName} (SPARse ROW norm reconstruction), represent the row-norms of the jointly sparse signal representation. Our {\myName} reformulation shows that the signal support is fully encoded in the sample covariance matrix, instead of the instantaneous measurement vectors as might be expected from classical sparse reconstruction formulations. 

In relation to existing techniques for gridless sparse recovery, we furthermore presented a gridless {\myName} implementation for the special case of uniform linear sampling. The gridless {\myName} implementation is based on semidefinite programming and we have established exact equivalence between the gridless {\myName} formulation and the recently proposed atomic norm minimization problem for multiple measurement vectors. However, in contrast to atomic norm minimization, our gridless {\myName} implementation shows reduced problem size, resulting in significantly reduced computational complexity. Additionally, we have established theoretical links between the {\myName} formulation and the SPICE method.

In our numerical evaluation we have demonstrated that {\myName} provides a viable supplement to classical subspace-based methods, such as MUSIC, especially in the non-asymptotic regime of low signal-to-noise ratio and low number of measurement vectors.

%% file: 10_appendix.tex
% \newpage \clearpage

\appendix[Equivalence of {\myNameB} and Anm]
% \section*{Appendix - Equivalence of {\myName} and ANM}

Consider the GL-{\myName} formulation \eqref{eq:GL_smr} and the ANM formulation \eqref{eq:relSdp2}. The problems are equivalent in the sense that both problems yield the same optimal function values and the minimizers are related by
\begin{align}
  \opt{\mb{u}} &= \frac{1}{\sqrt{N}} \opt{\mb{v}} \label{eq:optU1} \\
  \opt{\mb{U}}_N &= \sqrt{N} \opt{\mb{V}}_N + \frac{1}{\lambda} \opt{\mb{Z}}^\tH \opt{\mb{Z}}, \label{eq:optU2} 
\end{align}
for an appropriate $M \times N$ matrix $\mb{Z}$. 

To see the equivalence, consider the reformulation 
\begin{subequations}
\label{eq:GL_smr1}
\begin{align}
  \min_{\mb{u}, \mb{U}_N} \; & \;\; \frac{\lambda}{2} \tr\big( \mb{U}_N \big) + \frac{\lambda N}{2 M} \tr\big( \toep(\mb{u}) \big) \\
  \tst & \; \mtx{ \mb{U}_N / \sqrt{N} & \mb{Y}^\tH \\ \mb{Y} & \sqrt{N} \toep(\mb{u}) + \lambda \sqrt{N} \mb{I}_M } \succeq \mb{0}  \label{eq:GL_smr1b} \\
  & \; \; \toep(\mb{u})\succeq \mb{0} ,
\end{align}	
\end{subequations}
of the GL-{\myName} formulation \eqref{eq:GL_smr}, where the objective function in \eqref{eq:GL_smr1} is scaled by $\lambda N /2$ and the constraints \eqref{eq:GL_smr0b} and \eqref{eq:GL_smr1b} have identical Schur complements. Inserting \eqref{eq:optU1} and \eqref{eq:optU2} into problem \eqref{eq:GL_smr1} results in
\begin{subequations}
\label{eq:GL_smr2}
\begin{align}
  \smash{\min_{\substack{\mb{v}, \mb{V}_N, \\\mb{Z}}}} & \;
  \frac{\lambda \sqrt{N}}{2} \tr\big( \mb{V}_N \big) + \frac{1}{2} \tr\big( \mb{Z}^\tH \mb{Z} \big) +
  \frac{\lambda \sqrt{N}}{2 M} \tr\big( \toep(\mb{v}) \big) \label{eq:GL_smr2a} \\
  \tst & \; \mtx{ \mb{V}_N + \frac{1}{\lambda \sqrt{N}} \mb{Z}^\tH \mb{Z} & \mb{Y}^\tH \\
				  \mb{Y} & \toep(\mb{v}) + \lambda \sqrt{N} \mb{I}_M } \succeq \mb{0}  \label{eq:GL_smr2b} \\
  & \; \; \toep(\mb{v})\succeq \mb{0} .
\end{align}	
\end{subequations}
% The semidefinite constraint \eqref{eq:GL_smr2b} can be partitioned as
% \begin{align}
%   &\mtx{ \mb{V}_N + \frac{1}{\lambda \sqrt{N}} \mb{Z}^\tH \mb{Z} & \mb{Y}^\tH \nonumber \\
% 		\mb{Y} & \toep(\mb{v}) + \lambda \sqrt{N} \mb{I}_M } \\ =
% %   &\mtx{ \mb{V}_N & \mb{Z}^\tH \!\!- \!\!\mb{Y}^\tH\\ \mb{Z} \!\!-\!\! \mb{Y} & \toep(\mb{v}) } +
% % 			\mtx{ \frac{1}{\lambda \sqrt{N}} \mb{Z}^\tH \mb{Z} & \mb{Z}^\tH \\
% % 				  \mb{Z} & \lambda \sqrt{N} \mb{I}_M } \nonumber \\ =
% &\mtx{ \mb{V}_N & \mb{Z}^\tH\!\! -\!\! \mb{Y}^\tH \\ \mb{Z}\!\! - \!\!\mb{Y} & \toep(\mb{v}) } +
%  \lambda \sqrt{N} \mtx{\frac{1}{\lambda \sqrt{N}} \mb{Z}^\tH \\ \mb{I}_M } \!\!
%   \mtx{\frac{1}{\lambda \sqrt{N}} \mb{Z}^\tH \\ \mb{I}_M }^\tH 
%   \succeq \mb{0} \label{eq:GL_smrEq} .
% \end{align}
% By application of \eqref{eq:GL_smrEq}, 
Problem \eqref{eq:GL_smr2} can be equivalently written as
\begin{subequations}
\label{eq:GL_smr3}
\begin{align}
  \smash{\min_{\substack{\mb{v}, \mb{V}_N, \\\mb{Z}}}} & \;
  \frac{1}{2}\tr \big( \mb{Z}^\tH \mb{Z} \big) + \frac{\lambda \sqrt{N}}{2} \Big( \tr\big( \mb{V}_N \big) +
  \frac{1}{M} \tr\big( \toep(\mb{v}) \big) \Big) \\
  \tst & \; \mtx{ \mb{V}_N & \mb{Z}^\tH\!\! - \!\!\mb{Y}^\tH \\ \mb{Z}\!\! - \!\!\mb{Y} & \toep(\mb{v}) } \!+\!
			 \lambda \sqrt{N} \mtx{\frac{1}{\lambda \sqrt{N}} \mb{Z}^\tH \\ \mb{I}_M } \!\! 
  \mtx{\frac{1}{\lambda \sqrt{N}} \mb{Z}^\tH \\ \mb{I}_M }^\tH \!\!  \succeq \mb{0} \label{eq:canbedropted}\\
  & \; \; \toep(\mb{v})\succeq \mb{0} \label{eq:psdoftoepv}
\end{align}	
\end{subequations}
which in turn is equivalent to 
\begin{subequations}
\label{eq:GL_smr4}
\begin{align}
  \smash{\min_{\substack{\mb{v}, \mb{V}_N, \\\mb{Z}}}} & \;
  \frac{1}{2}\tr \mb{Z}^\tH \mb{Z} + \frac{\lambda \sqrt{N}}{2} \Big( \tr\big( \mb{V}_N \big) +
  \frac{1}{M} \tr\big( \toep(\mb{v}) \big) \Big) \\
  \tst & \; \mtx{ \mb{V}_N & \mb{Z}^\tH\!\! - \!\!\mb{Y}^\tH \\ \mb{Z}\!\! - \!\!\mb{Y} & \toep(\mb{v}) }  \succeq \mb{0} \label{eq:canbedropted2}\\
  & \; \; \toep(\mb{v})\succeq \mb{0}
\end{align}	
\end{subequations}
To prove the equivalence of \eqref{eq:GL_smr3} and \eqref{eq:GL_smr4} we first remark that any optimal point of \eqref{eq:GL_smr4} is clearly feasible for \eqref{eq:GL_smr3}. Reversely, for any
optimal solution $(\opt{\mb{U}}_N, \opt{\mb{u}})$ of problem \eqref{eq:GL_smr1} we can always find a partition \eqref{eq:optU2} which, due to the equivalence, is optimal for \eqref{eq:GL_smr3} and which satisfies condition \eqref{eq:canbedropted2}, i.e., is feasible for \eqref{eq:GL_smr4}. To prove the last statement it suffices to show that we can find w.l.o.g. a partition \eqref{eq:optU2} such that
\begin{align}\label{eq:alwayspsd}
 & \mtx{ \frac{1}{\lambda \sqrt{N}} \opt{\mb{Z}} & \mb{I}_M }
  \mtx{ \opt{\mb{V}}_N & \opt{\mb{Z}}^\tH\!\! - \!\!\mb{Y}^\tH \\ \opt{\mb{Z}}\!\! - \!\!\mb{Y} & \toep(\opt{\mb{v}}) }
 \mtx{\frac{1}{\lambda \sqrt{N}} \opt{\mb{Z}}^\tH \\ \mb{I}_M } \nonumber \\ 
 =& \frac{1}{\lambda^2 N} \opt{\mb{Z}} \opt{\mb{V}}_N \opt{\mb{Z}}^\tH + \frac{2}{\lambda \sqrt{N}} \opt{\mb{Z}}\opt{\mb{Z}}^\tH - \frac{1}{\lambda \sqrt{N}}\mb{Y} \opt{\mb{Z}}^\tH \nonumber \\
 & - \frac{1}{\lambda \sqrt{N}}\opt{\mb{Z}} \mb{Y}^\tH + \toep(\opt{\mb{v}})
 \succeq \mb{0}
\end{align}
with which \eqref{eq:psdoftoepv} is achieved, e.g., for $\hat{\mb{Z}} = \mb{0}$.

Introducing the change of variable $\mb{Y}_0 = \mb{Z} - \mb{Y}$ in \eqref{eq:GL_smr4} we arrive at ANM formulation \eqref{eq:relSdp2}, which completes the prove. 

%% file: 00_main_ieee.bbl
\begin{thebibliography}{10}
\providecommand{\url}[1]{#1}
\csname url@samestyle\endcsname
\providecommand{\newblock}{\relax}
\providecommand{\bibinfo}[2]{#2}
\providecommand{\BIBentrySTDinterwordspacing}{\spaceskip=0pt\relax}
\providecommand{\BIBentryALTinterwordstretchfactor}{4}
\providecommand{\BIBentryALTinterwordspacing}{\spaceskip=\fontdimen2\font plus
\BIBentryALTinterwordstretchfactor\fontdimen3\font minus
  \fontdimen4\font\relax}
\providecommand{\BIBforeignlanguage}[2]{{%
\expandafter\ifx\csname l@#1\endcsname\relax
\typeout{** WARNING: IEEEtran.bst: No hyphenation pattern has been}%
\typeout{** loaded for the language `#1'. Using the pattern for}%
\typeout{** the default language instead.}%
\else
\language=\csname l@#1\endcsname
\fi
#2}}
\providecommand{\BIBdecl}{\relax}
\BIBdecl

\bibitem{Tibshirani:Lasso}
R.~Tibshirani, ``Regression shrinkage and selection via the {LASSO},''
  \emph{Journal of the Royal Statistical Society. Series B (Methodological)},
  vol.~58, pp. 267--288, 1996.

\bibitem{Chen98atomicdecomposition}
S.~S. Chen, D.~L. Donoho, and M.~A. Saunders, ``Atomic decomposition by basis
  pursuit,'' \emph{SIAM Journal On Scientific Computing}, vol.~20, pp. 33--61,
  1998.

\bibitem{1614066}
D.~Donoho, ``Compressed sensing,'' \emph{IEEE Transactions on Information
  Theory}, vol.~52, no.~4, pp. 1289--1306, April 2006.

\bibitem{1580791}
E.~Cand{\`{e}}s, J.~Romberg, and T.~Tao, ``Robust uncertainty principles: exact
  signal reconstruction from highly incomplete frequency information,''
  \emph{IEEE Transactions on Information Theory}, vol.~52, no.~2, pp. 489--509,
  Feb 2006.

\bibitem{1542412}
E.~Cand{\`{e}}s and T.~Tao, ``Decoding by linear programming,'' \emph{IEEE
  Transactions on Information Theory}, vol.~51, no.~12, pp. 4203--4215, Dec
  2005.

\bibitem{candes:2688127}
E.~J. Cand\`{e}s, J.~K. Romberg, and T.~Tao, ``Stable signal recovery from
  incomplete and inaccurate measurements,'' \emph{Comm. Pure Appl. Math.},
  vol.~59, no.~8, pp. 1207--1223, August 2006.

\bibitem{candes:12380704}
E.~J. Cand{\`{e}}s and J.~Romberg, ``Quantitative robust uncertainty principles
  and optimally sparse decompositions,'' vol.~6, no.~2, pp. 227--254, 2006.

\bibitem{Donoho02optimallysparse}
D.~L. Donoho and M.~Elad, ``Optimally sparse representation in general
  (nonorthogonal) dictionaries via \texorpdfstring{$\ell^{1}$}{l1}
  minimization,'' vol. 100, no.~5.\hskip 1em plus 0.5em minus 0.4em\relax
  National Acad Sciences, 2003, pp. 2197--2202.

\bibitem{tropp2010}
J.~Tropp, J.~Laska, M.~Duarte, J.~Romberg, and R.~Baraniuk, ``Beyond {N}yquist:
  Efficient sampling of sparse bandlimited signals,'' \emph{IEEE Transactions
  on Information Theory}, vol.~56, no.~1, pp. 520--544, Jan 2010.

\bibitem{donoho1992}
D.~L. Donoho, ``Superresolution via sparsity constraints,'' \emph{SIAM Journal
  on Mathematical Analysis}, vol.~23, no.~5, pp. 1309--1331, 1992.

\bibitem{258082}
S.~Mallat and Z.~Zhang, ``Matching pursuits with time-frequency dictionaries,''
  \emph{IEEE Transactions on Signal Processing}, vol.~41, no.~12, pp.
  3397--3415, Dec 1993.

\bibitem{4385788}
J.~Tropp and A.~Gilbert, ``Signal recovery from random measurements via
  orthogonal matching pursuit,'' \emph{IEEE Transactions on Information
  Theory}, vol.~53, no.~12, pp. 4655--4666, Dec 2007.

\bibitem{Needell2009301}
D.~Needell and J.~Tropp, ``{CoSaMP}: Iterative signal recovery from incomplete
  and inaccurate samples,'' \emph{Applied and Computational Harmonic Analysis},
  vol.~26, no.~3, pp. 301 -- 321, 2009.

\bibitem{candes2012b}
E.~J. Cand{\`{e}}s and C.~Fernandez{-}Granda, ``Towards a mathematical theory
  of super-resolution,'' \emph{CoRR}, vol. abs/1203.5871, 2012.

\bibitem{candes2012}
------, ``Super-resolution from noisy data,'' \emph{CoRR}, vol. abs/1211.0290,
  2012.

\bibitem{chandrasekaran2012}
V.~Chandrasekaran, B.~Recht, P.~A. Parrilo, and A.~S. Willsky, ``The convex
  geometry of linear inverse problems,'' \emph{Foundations of Computational
  Mathematics}, vol.~12, no.~6, pp. 805--849, 2012.

\bibitem{gongguo2013}
G.~Tang, B.~Bhaskar, P.~Shah, and B.~Recht, ``Compressed sensing off the
  grid,'' \emph{IEEE Transactions on Information Theory}, vol.~59, no.~11, pp.
  7465--7490, Nov 2013.

\bibitem{bhaskar2011}
B.~N. Bhaskar and B.~Recht, ``Atomic norm denoising with applications to line
  spectral estimation,'' in \emph{Communication, Control, and Computing
  (Allerton), 2011 49th Annual Allerton Conference on}, Sept 2011, pp.
  261--268.

\bibitem{6552292}
G.~Tang, B.~N. Bhaskar, and B.~Recht, ``Near minimax line spectral
  estimation,'' vol.~61, no.~1, Jan 2015, pp. 499--512.

\bibitem{10.2307/3647556}
Y.~L. Ming~Yuan, ``Model selection and estimation in regression with grouped
  variables,'' \emph{Journal of the Royal Statistical Society. Series B
  (Statistical Methodology)}, vol.~68, no.~1, pp. 49--67, 2006.

\bibitem{Tropp2006589}
J.~A. Tropp, ``Algorithms for simultaneous sparse approximation. {P}art {II}:
  Convex relaxation,'' \emph{Signal Processing}, vol.~86, no.~3, pp. 589 --
  602, 2006.

\bibitem{turlach2005simultaneous}
B.~A. Turlach, W.~N. Venables, and S.~J. Wright, ``Simultaneous variable
  selection,'' \emph{Technometrics}, vol.~47, no.~3, pp. 349--363, 2005.

\bibitem{Kowalski2009303}
M.~Kowalski, ``Sparse regression using mixed norms,'' \emph{Applied and
  Computational Harmonic Analysis}, vol.~27, no.~3, pp. 303 -- 324, 2009.

\bibitem{tropp2006algorithms}
J.~A. Tropp, A.~C. Gilbert, and M.~J. Strauss, ``Algorithms for simultaneous
  sparse approximation. {P}art {I}: Greedy pursuit,'' \emph{Signal Processing},
  vol.~86, no.~3, pp. 572--588, 2006.

\bibitem{1453780}
S.~Cotter, B.~Rao, K.~Engan, and K.~Kreutz-Delgado, ``Sparse solutions to
  linear inverse problems with multiple measurement vectors,'' \emph{IEEE
  Transactions on Signal Processing}, vol.~53, no.~7, pp. 2477--2488, July
  2005.

\bibitem{6408167}
Y.~Jin and B.~Rao, ``Support recovery of sparse signals in the presence of
  multiple measurement vectors,'' \emph{IEEE Transactions on Information
  Theory}, vol.~59, no.~5, pp. 3139--3157, May 2013.

\bibitem{4014378}
J.~Chen and X.~Huo, ``Theoretical results on sparse representations of
  multiple-measurement vectors,'' \emph{IEEE Transactions on Signal
  Processing}, vol.~54, no.~12, pp. 4634--4643, Dec 2006.

\bibitem{Lai2011402}
M.-J. Lai and Y.~Liu, ``The null space property for sparse recovery from
  multiple measurement vectors,'' \emph{Applied and Computational Harmonic
  Analysis}, vol.~30, no.~3, pp. 402 -- 406, 2011.

\bibitem{6145474}
M.~Davies and Y.~Eldar, ``Rank awareness in joint sparse recovery,'' \emph{IEEE
  Transactions on Information Theory}, vol.~58, no.~2, pp. 1135--1146, Feb
  2012.

\bibitem{7313018}
Y.~Li and Y.~Chi, ``Off-the-grid line spectrum denoising and estimation with
  multiple measurement vectors,'' \emph{IEEE Transactions on Signal
  Processing}, vol.~64, no.~5, pp. 1257--1269, March 2016.

\bibitem{yang2014a}
Z.~Yang and L.~Xie, ``Exact joint sparse frequency recovery via optimization
  methods,'' \emph{IEEE Transactions on Signal Processing}, vol.~PP, no.~99,
  pp. 1--1, 2016.

\bibitem{yang2014b}
------, ``On gridless sparse methods for line spectral estimation from complete
  and incomplete data,'' \emph{CoRR}, vol. abs/1407.2490, 2014.

\bibitem{Krim:TwoDecades}
H.~Krim and M.~Viberg, ``Two decades of array signal processing research: the
  parametric approach,'' \emph{IEEE Signal Processing Magazine}, vol.~13,
  no.~4, pp. 67--94, Jul 1996.

\bibitem{vanTrees2002}
H.~L. van Trees, \emph{Optimum Array Processing: Part IV of Detection,
  Estimation, and Modulation Theory}.\hskip 1em plus 0.5em minus 0.4em\relax
  New York: John Wiley \& Sons, Inc., 2002.

\bibitem{1143830}
R.~Schmidt, ``Multiple emitter location and signal parameter estimation,''
  \emph{IEEE Transactions on Antennas and Propagation}, vol.~34, no.~3, pp.
  276--280, Mar 1986.

\bibitem{17564}
P.~Stoica and N.~Arye, ``{MUSIC}, maximum likelihood, and cramer-rao bound,''
  \emph{IEEE Transactions on Acoustics, Speech, and Signal Processing},
  vol.~37, no.~5, pp. 720--741, May 1989.

\bibitem{Malioutov:LassoDoa}
D.~Malioutov, M.~\c{C}etin, and A.~Willsky, ``A sparse signal reconstruction
  perspective for source localization with sensor arrays,'' \emph{IEEE
  Transactions on Signal Processing}, vol.~53, no.~8, pp. 3010--3022, 2005.

\bibitem{5466152}
M.~M. Hyder and K.~Mahata, ``Direction-of-arrival estimation using a mixed
  \texorpdfstring{$\ell_{2,0}$}{l20} norm approximation,'' \emph{IEEE
  Transactions on Signal Processing}, vol.~58, no.~9, pp. 4646--4655, Sept
  2010.

\bibitem{kim2010compressive}
J.~Kim, O.~K. Lee, and J.~C. Ye, ``Compressive {MUSIC:} {A} missing link
  between compressive sensing and array signal processing,'' \emph{CoRR}, vol.
  abs/1004.4398, 2010.

\bibitem{clean}
J.~A. H\"{o}gbom, ``Aperture synthesis with a non-regular distribution of
  interferometer baselines,'' \emph{Astronomy and Astrophysics Supplement
  Series}, vol.~15, pp. 417--426, Jun. 1974.

\bibitem{558475}
I.~Gorodnitsky and B.~Rao, ``Sparse signal reconstruction from limited data
  using focuss: a re-weighted minimum norm algorithm,'' \emph{IEEE Transactions
  on Signal Processing}, vol.~45, no.~3, pp. 600--616, Mar 1997.

\bibitem{1687-6180-2012-111}
L.~Blanco and M.~Najar, ``Sparse covariance fitting for direction of arrival
  estimation,'' \emph{EURASIP Journal on Advances in Signal Processing}, vol.
  2012, no.~1, p. 111, 2012.

\bibitem{6494328}
J.~Zheng and M.~Kaveh, ``Sparse spatial spectral estimation: A covariance
  fitting algorithm, performance and regularization,'' \emph{IEEE Transactions
  on Signal Processing}, vol.~61, no.~11, pp. 2767--2777, June 2013.

\bibitem{5599897}
P.~Stoica, P.~Babu, and J.~Li, ``New method of sparse parameter estimation in
  separable models and its use for spectral analysis of irregularly sampled
  data,'' \emph{IEEE Transactions on Signal Processing}, vol.~59, no.~1, pp.
  35--47, Jan 2011.

\bibitem{5617289}
------, ``{SPICE}: A sparse covariance-based estimation method for array
  processing,'' \emph{IEEE Transactions on Signal Processing}, vol.~59, no.~2,
  pp. 629--638, Feb 2011.

\bibitem{2014arXiv1406.7698S}
P.~Stoica, D.~Zachariah, and J.~Li, ``Weighted {SPICE}: A unifying approach for
  hyperparameter-free sparse estimation,'' \emph{Digital Signal Processing},
  vol.~33, pp. 1--12, 2014.

\bibitem{6553252}
C.~Rojas, D.~Katselis, and H.~Hjalmarsson, ``A note on the {SPICE} method,''
  \emph{IEEE Transactions on Signal Processing}, vol.~61, no.~18, pp.
  4545--4551, Sept 2013.

\bibitem{babu2014connection}
P.~Babu and P.~Stoica, ``Connection between {SPICE} and square-root {LASSO} for
  sparse parameter estimation,'' \emph{Signal Processing}, vol.~95, pp. 10--14,
  2014.

\bibitem{yuejie2011}
Y.~Chi, L.~Scharf, A.~Pezeshki, and A.~Calderbank, ``Sensitivity to basis
  mismatch in compressed sensing,'' \emph{IEEE Transactions on Signal
  Processing}, vol.~59, no.~5, pp. 2182--2195, May 2011.

\bibitem{herman2010}
M.~A. Herman and T.~Strohmer, ``General deviants: An analysis of perturbations
  in compressed sensing,'' \emph{IEEE Journal of Selected topics in signal
  processing}, vol.~4, no.~2, pp. 342--349, 2010.

\bibitem{S98guide}
J.~Sturm, ``Using {SeDuMi} 1.02, a {MATLAB} toolbox for optimization over
  symmetric cones,'' \emph{Optimization Methods and Software}, vol. 11--12, pp.
  625--653, 1999.

\bibitem{Cara1}
C.~Carath\'{e}odory, ``\"{U}ber den {V}ariabilit\"{a}tsbereich der
  {F}ourierschen {K}onstanten von positiven harmonischen {F}unktionen,''
  \emph{Rendiconti del Circolo Matematico di Palermo (1884-1940)}, vol.~32,
  no.~1, p. 193–217, 1911.

\bibitem{Cara2}
C.~Carath\'{e}odory and L.~Fej\'{e}r, ``\"{U}ber den {Z}usammenhang der
  extremen von harmonischen {F}unktionen mit ihren {K}oeffizienten und \"{u}ber
  den {P}icard-{L}andauschen {S}atz,'' \emph{Rendiconti del Circolo Matematico
  di Palermo (1884-1940)}, vol.~32, no.~1, p. 218–239, 1911.

\bibitem{Toeplitz}
O.~Toeplitz, ``{Z}ur {T}heorie der quadratischen und bilinearen {F}ormen von
  unendlich vielen {V}er\"{a}nderlichen,'' \emph{Mathematische Annalen},
  vol.~70, no.~3, pp. 351--376, 1911.

\bibitem{prony1795}
G.~de~Prony, ``Essai exp{\'e}rimental et analytique: sur les lois de la
  dilatabilit{\'e} des fluides {\'e}lastiques et sur celles de la force
  expansive de la vapeur de l'eau et de la vapeur de l'alcool {\`a}
  diff{\'e}rentes temp{\'e}ratures,'' \emph{Journal de l'{\'E}cole
  Polytechnique}, vol.~1, no.~22, pp. 24--76, 1795.

\bibitem{56027}
Y.~Hua and T.~K. Sarkar, ``Matrix pencil method for estimating parameters of
  exponentially damped/undamped sinusoids in noise,'' \emph{IEEE Transactions
  on Acoustics, Speech, and Signal Processing}, vol.~38, no.~5, pp. 814--824,
  May 1990.

\bibitem{1456696}
D.~W. Tufts and R.~Kumaresan, ``Estimation of frequencies of multiple
  sinusoids: Making linear prediction perform like maximum likelihood,''
  \emph{Proceedings of the IEEE}, vol.~70, no.~9, pp. 975--989, Sept 1982.

\bibitem{6810450}
G.~Tang, B.~Bhaskar, and B.~Recht, ``Sparse recovery over continuous
  dictionaries-just discretize,'' in \emph{Signals, Systems and Computers, 2013
  Asilomar Conference on}, Nov 2013, pp. 1043--1047.

\bibitem{qin2013efficient}
Z.~Qin, K.~Scheinberg, and D.~Goldfarb, ``Efficient block-coordinate descent
  algorithms for the group {LASSO},'' \emph{Mathematical Programming
  Computation}, vol.~5, no.~2, pp. 143--169, 2013.

\bibitem{wright2015}
S.~J. Wright, ``Coordinate descent algorithms,'' \emph{Mathematical
  Programming}, vol. 151, no.~1, pp. 3--34, 2015.

\bibitem{stoica2001stochastic}
P.~Stoica, G.~Larsson, and A.~B. Gershman, ``The stochastic {CRB} for array
  processing: a textbook derivation,'' \emph{Signal Processing Letters, IEEE},
  vol.~8, no.~5, pp. 148--150, 2001.

\bibitem{6375850}
E.~T. Northardt, I.~Bilik, and Y.~I. Abramovich, ``Spatial compressive sensing
  for direction-of-arrival estimation with bias mitigation via expected
  likelihood,'' \emph{IEEE Transactions on Signal Processing}, vol.~61, no.~5,
  pp. 1183--1195, March 2013.

\bibitem{Stoica20121580}
P.~Stoica and P.~Babu, ``{SPICE} and {LIKES}: Two hyperparameter-free methods
  for sparse-parameter estimation,'' \emph{Signal Processing}, vol.~92, no.~7,
  pp. 1580 -- 1590, 2012.

\bibitem{grant2008}
M.~Grant and S.~Boyd, ``Graph implementations for nonsmooth convex programs,''
  in \emph{Recent Advances in Learning and Control}, ser. Lecture Notes in
  Control and Information Sciences, V.~Blondel, S.~Boyd, and H.~Kimura,
  Eds.\hskip 1em plus 0.5em minus 0.4em\relax Springer-Verlag Limited, 2008,
  pp. 95--110.

\bibitem{grant2014}
------, ``{CVX}: Matlab software for disciplined convex programming, version
  2.1,'' \url{http://cvxr.com/cvx}, Mar. 2014.

\end{thebibliography}
